\documentclass[iicol,sn-mathphys-num]{sn-jnl}

\usepackage{caption}
\usepackage{graphicx}%
\usepackage{multirow}%
\usepackage{amsmath,amssymb,amsfonts}%
\usepackage{amsthm}%
\usepackage{mathrsfs}%
\usepackage[title]{appendix}%
\usepackage{xcolor}%
\usepackage{textcomp}%
\usepackage{manyfoot}%
\usepackage{booktabs}%
\usepackage{algorithm}%
\usepackage{algorithmicx}%
\usepackage{algpseudocode}%
\usepackage{listings}%
\usepackage{natbib}%
\usepackage{tikz}
\usepackage{tcolorbox}
\usepackage{pgfplots}
\usepackage{pgfplotstable}
\usepackage{csquotes}
\usepackage{tablefootnote}
\usepackage{hyperref}

\theoremstyle{thmstyleone}%
\theoremstyle{thmstyletwo}%

\theoremstyle{thmstylethree}%

\begin{document}

\title[Article Title]{From Demographics to Survey Anchors: Evaluating LLM Agents for Modeling Retirement Attitudes}


\author*[1]{\fnm{Rubén} \sur{Garzón}}\email{ruben.garzon@axa.com}
\author[2]{\fnm{Pauline} \sur{Baron}}
\author[3]{\fnm{Vincent} \sur{Grari}}
\author[4]{\fnm{Jonne} \sur{Kamphorst}}
\author[5]{\fnm{Michael} \sur{Bernstein}}
\author[2]{\fnm{Marcin} \sur{Detyniecki}}

\affil[1]{\orgdiv{AI Research}, \orgname{AXA Group Operations}, 
  \orgaddress{\city{Madrid}, \country{Spain}}}

\affil[2]{\orgdiv{AI Research}, \orgname{AXA Group Operations}, 
  \orgaddress{\city{Paris}, \country{France}}}

\affil[3]{\orgdiv{AI Research}, \orgname{AXA Group Operations}, 
  \orgaddress{\city{Lausanne}, \country{Switzerland}}}

\affil[4]{\orgdiv{CDSP \& CEE}, \orgname{Sciences Po}, 
  \orgaddress{\city{Paris}, \country{France}}}

\affil[5]{\orgdiv{Computer Science Department}, \orgname{Stanford University}, 
  \orgaddress{\city{Stanford}, \state{California}, \country{United States}}}

\abstract{Large language models (LLM) agents may offer tools to predict human responses to surveys. A common technique for defining these agents uses only demographics, for example country, age, gender, employment status, income, education and marital status. We compare the predictive accuracy of demographic agents to that of survey agents defined with a larger set of in-domain survey responses. We test both approaches in predicting responses to the multidisciplinary, cross-national Survey of Health, Ageing and Retirement in Europe (SHARE), focusing on five variables from three policy-relevant constructs around personal finance. In these three constructs, we observe that, compared to survey agents trained on broader data, demographics-only agents (1)~exhibited a central tendency bias, skewing answers toward population means, and 
(2)~were unrealistically accurate, failing to reproduce the incorrect answers and ``don't know'' responses typical of human respondents. 
These performance differences are further substantiated through the replication
of a hierarchical regression analysis from prior retirement planning research. Agents based solely on demographic information reproduce the outcome that financial risk tolerance, future time perspective, and knowledge of retirement planning each are predictive of retirement savings. However, only the survey-anchored agents succeed in reproducing the interaction among these three factors.
These findings suggest caution in using only demographics to define LLM agents for predicting survey responses.}




\maketitle
\section{Introduction}\label{sec1}

An emerging field of research explores how large language models (LLMs), in the form of generative agents~\cite{park2023generative,park2024generative}, can be used for simulating human behaviors across a variety of contexts, such as social science, education, politics and behavioral economics \cite{xu2025classroom,park2023generative,park2024generative,xie2024can}. Such work holds particular promise for policymaking and research, as it could enable the generation of collective insights based on synthetic user data \cite{agnew2024illusion, hamalainen_evaluating_llms, hwang_human_subjects_research_age_of_genai,kapania_simulacrum_2025} and the testing of public interventions in a more accessible, faster and less costly way.

A common method for constructing these AI agent simulations is to use only demographic variables \cite{argyle2023out,horton2023largelanguagemodelssimulated,ashokkumar2024predicting}. For example, a demographic agent might be defined with a series of demographics such as, ``Ideologically, I describe myself as liberal. Politically, I am a strong Democrat. Racially, I am white. I am female. Financially, I am poor. In terms of my age, I am
old''~\cite{argyle2023out}. The LLM is then prompted with this demographic profile and a survey question, e.g., ``In 2024, I voted for...'' and a list of candidates. These \textit{demographic agents} are an attractive way to create agents because the data can be easily drawn from openly-available census or survey data. However, research has begun to question the use of demographic agents, suggesting that they may under-represent variation within groups~\cite{wang2025large} and may be less accurate than other, richer, sources of data~\cite{park2024generative}. Yet, they remain a popular choice for their ease of construction.


In this paper, we test the empirical accuracy of this popular approach of demographic agents by comparing them to the accuracy of richer, \textit{survey-anchored
agents} built from each respondent's full question–answer history (approximately 175 items per individual, spanning health, employment, finances, and social context).
We pursue this investigation in the context of a broad, multinational survey focused on retirement savings.

This task is a useful proxy for the accuracy of such agents, since it focuses on a policy-relevant outcome and allows comparison across many national and cultural contexts.
It is also an intrinsically important prediction goal, as European governments face ageing populations and the limits of their social security apparatuses, so are exploring mechanisms to promote private savings---for which rapid simulated feedback might help with exploration and early iteration.
Specifically, our work draws on a large-scale structured dataset from the \textit{Survey of Health, Ageing and Retirement in Europe (SHARE)}, which is a multidisciplinary and cross-national research infrastructure for studying the effects of health, social, economic and environmental policies over the life-course of European citizens. It provides
data about almost 70k individuals of over 50 years old from 28 European countries. This paper uses data from SHARE Wave 9 (2021-2022)~\cite{SHAREwave9, SHAREwave9methodology,borsch2013data}. 

We build survey-anchored agents from SHARE by prompting a local LLM with each respondent's full question–answer history (approximately 175 items per agent).

We test these survey agents, and demographic agents representing the same individuals but only their demographic variables, on five dependent variables comprising three core constructs that impact retirement saving practices—future time perspective, financial risk tolerance, and financial knowledge \cite{jacobs2005influence}—and then compare the simulation outputs to the empirical results at three different levels:
\begin{itemize}
    \item \textit{Individual level}: do the survey agents, and the demographic agents, replicate the same answers to the SHARE survey as the individuals each agent was based on?
    \item \textit{Country level}: we obtained additional questions related to our constructs from the EU Barometer survey on Financial Literacy (2023) \cite{doi/10.2874/956514}. Do the agents match the distribution of responses when aggregated to a country level (the EU barometer does not provide participant answers but only per country statistics)?
    \item \textit{Study level}: do the agents replicate prior findings correlating future time perspective, financial risk tolerance, and financial knowledge with retirement savings? \cite{jacobs2005influence} (Unlike the previous two levels, which use European SHARE-based agents, here we build agents from the 2004 General Social Survey (GSS)\cite{marsden2016overview}, as the original study was conducted with US participants.)

\end{itemize}

Our results 
 raise concerns about the usage of demographic agents in survey responses. First, compared to survey-anchored agents, they  
exhibited central-tendency bias: on subjective questions, their responses converged toward population means and under-represent the variance and tails of the distribution. Second, on our measurements, they were hyper-reasoners compared to survey-anchored agents: on objective financial literacy and numeracy questions, they over-represented correct answers and produced fewer errors and ``don't know'' responses, thus failing to replicate observed human error rates.
In comparison, the survey-anchored agents shifted response distributions closer to empirical human data, capturing greater heterogeneity. This result supports prior work that richer source data can improve the accuracy of generative agent simulations~\cite{park2024generative}. 

We observe a similar pattern in psychometric reliability. We replicate a hierarchical regression analysis from prior retirement research, which tests whether psychological predictors (future time perspective, financial knowledge, risk tolerance) and their interactions explain savings behavior. 
Demographics-only agents exhibit central tendency bias across multiple diagnostic dimensions—producing internally consistent but undifferentiated response profiles that fail to recover the higher-order interactions central to the original study's findings.
Survey-anchored agents mitigate these pathologies by generating more heterogeneous, individuated responses and are the only condition to recover significant two- and three-way interactions. However, they are not without limitations: important discrepancies in effect magnitude and direction persist, they still under-represent extreme values, exhibit hyper-accuracy on verifiable questions, and show weaker alignment on questions involving temporal relations awareness---just to a lesser extent than demographic agents, suggesting that additional prompting strategies or model finetuning may be required to address these shortcomings in the long run \cite{suh2025language}. Our results must be interpreted with caution, as future work must establish whether they generalize to a broader set of questions beyond our context---yet the trends are striking.

The paper is structured as follows: Section~\ref{sec:related_work} provides a literature review of LLM-powered simulation of human behavior and the determinants that influence retirement saving (in particular future time perspective, financial risk tolerance and financial knowledge). Section~\ref{sec:constr_synth} details the methodology for constructing our generative agents using the SHARE dataset. Section~\ref{sec:expe} describes the construct and results of the three studies. We conclude with a discussion (Section~\ref{sec:discu}) and a summary of findings (Section~\ref{sec:conclu}).

\section{Related Work}
\label{sec:related_work}

This section offers background information relevant to the paper. It begins by outlining the opportunities, challenges, and limitations of current methods used to evaluate large language models (LLMs) for simulating human attitudes. It then examines the various factors that influence individuals' attitudes towards retirement planning and saving, which is the application of interest.

\subsection{Simulating Human Attitudes with Large Language Models}

Recent work investigates LLMs as simulators of human judgment and survey responses~ \cite{horton2023largelanguagemodelssimulated,manning2024automated,binz2025foundation,dillion2023can,harding2024ai,grossmann2023ai}. In particular, one strand evaluates the use of LLMs for generating synthetic research data in studies that usually require human participation 
\cite{hamalainen_evaluating_llms,hwang_human_subjects_research_age_of_genai,kapania_simulacrum_2025,agnew2024illusion,tseng2024two}. Persona- or interview-anchored prompting can reproduce subgroup-specific answer patterns and even individual attitudes across domains~\cite{argyle2023out,park2024generative}. LLMs and surveys can also synergistically improve one another's abilities \cite{kim2023ai}.

A popular method for defining simulated agents uses only demographic variables~\cite{argyle2023out, horton2023largelanguagemodelssimulated, ashokkumar2024predicting}. In this approach, agents are constructed with prompts specifying characteristics such as country, gender, age, and ideology. The LLM is then queried with a survey question, and the model simulates how a person with that profile would respond. These demographic agents are attractive for their simplicity and scalability, since demographic data is often publicly available. However, growing evidence suggests limitations: demographic agents may misportray groups and flatten the variation within population subgroups~\cite{wang2025large} and are often less accurate than agents conditioned on richer context~\cite{park2024generative}. This central tension—between accessibility and realism—motivates our comparison of demographic agents to survey-anchored agents built from in-domain response histories.


At the same time, a growing body of work has documented systematic distortions when LLMs are used as stand-ins for human subjects. Evaluations across classic economic, psycholinguistic, and social psychology experiments report hyper-accuracy on numeracy tasks and sensitivity to response formats~\cite{aher2023using}. From a causal-inference angle, \cite{gui_challenge_2023} further show that LLM-simulated experiments can violate unconfoundedness, as ambiguous prompts let treatment variations shift unspecified variables that should remain constant. In economic decision tasks, models often appear more internally consistent and less heterogeneous than humans, with notable framing effects~\cite{chen2023emergence}. Analyses from cognitive psychology likewise find human-like biases alongside gaps in causal reasoning~\cite{binz2023using}. 
Large-scale benchmarking efforts have reinforced these concerns: \cite{wang_sociobench_2025} introduced SocioBench, derived from the International Social Survey Programme with over 480,000 respondent records across 30+ countries, finding that LLMs achieve only 30–40\% accuracy in complex survey scenarios with significant variation across demographic subgroups. Similarly, \cite{zhao_large_2025} proposed LLM-S³, a comprehensive evaluation suite that systematically demonstrates how context and prompt design impact simulation fidelity.

A related strand of work has focused specifically on the representativeness of LLM-generated responses. Several studies indicate that large language models tend to underestimate the variability of human responses~\cite{bisbee2024synthetic,park2024diminished}, and public-opinion studies suggest that conditioning on media or persona signals can align outputs with population surveys but that misalignment persists for some groups~\cite{chu2023language,santurkar2023whose}. In a comprehensive position paper, \cite{anthis_llm_2025} argued that five tractable challenges---diversity, bias, sycophancy, alienness, and generalization---must be addressed for LLM social simulations to fulfill their promise.
These findings have motivated efforts to improve simulation fidelity through richer agent grounding. \cite{suh2025language} showed that fine-tuning LLMs on scaled survey data can reduce the LLM-human gap by up to 46\% and achieve strong generalization to unseen surveys and subpopulations. \cite{toubia2025twin} created ``digital twins'' of 2,058 U.S. participants using survey data from 500 questions and evaluated them at both the individual and group levels, demonstrating that extensive respondent-level context can improve alignment. Both \cite{anthis_llm_2025} and \cite{suh2025language} identify context-rich prompting and fine-tuning with social science datasets as promising directions.

Taken together, this literature suggests two empirical regularities: first, demographics-only agents tend to flatten within-group variation and produce overly systematic response patterns; second, richer sources of respondent information—whether from interviews~\cite{park2024generative}, survey batteries~\cite{toubia2025twin}, or fine-tuning corpora~\cite{suh2025language}—can mitigate these distortions. This motivates our central hypothesis: that anchoring LLM agents in structured, in-domain survey data (approximately 175 question–answer pairs per respondent from the SHARE study) will yield more faithful simulations of human attitudes than agents defined by demographics alone. 


\subsection{Research on retirement savings}

This work situates its measurement question in the context of financial planning. 
One strand of work seeks to encourage long-term savings \cite{10.1145/2702123.2702408}. Another strand of work builds tools more broadly for digital financial literacy \cite{10.1145/3290605.3300620,10.1145/3706599.3719934,10.1145/3334480.3382898}, as well as accessible financial tools for older adults \cite{10.1145/3706598.3713427, maqbool2018understanding}. Recently, generative AI has been envisioned as a tool to help people perform financial planning \cite{10.1145/3706599.3720145}.

In the broader behavioral sciences, researchers have sought to understand the reasons why individuals do not carefully plan and save for retirement. An extensive part of literature \cite{wiatrowski1993factors, bassett1998workers, sterns2003self} has therefore examined the influence of several determinants on retirement savings practices such as demographic and socio-economic factors. Variables such as age, gender \cite{behling1983scarce}, educational level, occupation, income level, health status but also country type of social model \cite{rey2018influence} have been evidenced to impact retirement individuals' savings practices.
While demographic and socio-economic factors have been explored in many studies, a lesser deal of attention has been paid to the influence of psychological determinants on retirement planning and savings practices such as cognitive and behavioral factors \cite{gough2011retirement}. Yet, it has been argued that psychological factors have a more direct effect on savings decisions than demographics, the latter being mediated through the psyche \cite{hershey2004psychological}. While there is no exhaustive list nor a unique classification of psychological determinants, some have been found to substantially impact retirement contributions such as heuristics and bias that individuals have when making (or failing to make) decisions about retirement, e.g.,  procrastination \cite{thaler2004save}, inertia, loss aversion (i.e., tendency of individuals to weigh losses about twice as much as gains), myopia or present bias (i.e., hypersensitivity to short-term losses) \cite{benartzi2007heuristics}. The "peer effect" is another factor that has been evidenced; it describes the phenomenon through which rational but unsophisticated investors may seek expert advice for help but paradoxically tend to ask people who do not necessarily qualify as experts, such as relatives \cite{benartzi1999risk}. 
Others showed the influence of retirement goal clarity \cite{moen1996life, stawski2007goal, petkoska2009understanding}. Besides these indicators, research has broadly demonstrated the influence of three specific dimensions as predictors of retirement savings practices: \textit{future time perspective }(i.e. extent to which one focuses on the future rather than the past), \textit{financial risk tolerance} (i.e., one's attitude towards financial risk), and \textit{financial knowledge} or \textit{literacy} (i.e., one's proficiency in financial matters)  \cite{hershey2000psychological, yang2012determinants, rey2015determinants, banks2007understanding, alessie2011financial, rey2015determinants, chou2014social}. 
Higher levels of each of these three dimensions has been evidenced to be associated with more aggressive retirement saving proﬁles \cite{jacobs2005influence}. 

\section{Constructing Survey-Anchored  Agents}
\label{sec:constr_synth}
Our goal is to predict broad, cross-national survey results using generative agent simulations. We will compare the most common approach---demographic agents---to agents anchored in a larger set of survey responses. In both cases, we create the agents from the same underlying dataset, called the SHARE survey. So, the individuals underlying the agents are identical: the only difference is an ablation of what information from the survey is used to define each agent. Only for the third study, we use a different dataset (US based, General Social Survey \cite{marsden2016overview}), but use the same methodology.

To build \emph{demographic agents}, we follow the principle from prior work (e.g,.~\cite{argyle2023out}). In particular, given the subject of our study, we focus on seven demographic variables: country, age, gender, employment status, income, education and marital status (see \ref{app:demographics7_share_items} for details).

In contrast, we build the \emph{survey-anchored} generative agents using the same individuals' full question–answer history on the SHARE survey ($\approx175$ items per agent). Unlike interview-anchored \emph{generative agents}~\cite{park2024generative} that rely on free-form transcripts, our agents are constructed solely from structured survey data. It is worth noting that the survey-anchored agent also incorporated the same seven demographic variables as the demographics-only agents (age, country, gender, employment status, income, education and marital status).

For any evaluation item, we hold it out from the context and prompt the model to infer how the \emph{same} respondent would answer the unseen question. This setup enables the downstream assessments reported in this paper: individual-level fidelity, country-level distributional alignment, and construct-level relationships relevant to retirement saving.

\subsection{The SHARE dataset}

The Survey of Health, Ageing and Retirement in Europe (SHARE) is a rich, multidisciplinary, cross-national panel study that provides a unique resource for research on ageing. The dataset tracks a representative sample of individuals aged 50 and older across Europe and Israel, with the same individuals being followed over time to allow for the analysis of dynamic changes in their lives. The study began in 2004 with 11 participating countries and has since expanded to 28 countries in its most recent wave, Wave 9 \cite{SHAREwave9, borsch2013data}, which surveyed almost 70k participants. Data collection is conducted on a biennial basis, with a new wave of the survey taking place approximately every two years. 

SHARE collects a comprehensive range of data on the health, socioeconomic status, and social networks of older Europeans. The survey is structured into several core modules that cover a broad spectrum of topics (see Appendix materials for exhaustive list of sections).
Key areas of data collection include detailed information on physical and mental health, employment history, retirement, income, and wealth. The study also gathers data on social support, family networks, and housing conditions, as well as providing a retrospective look at life history events. Unlike a qualitative interview, which relies on open-ended questions and free-writing to capture detailed narratives, the SHARE survey employs a Computer-Assisted Personal Interviewing (CAPI) format with a structured questionnaire. This design utilizes a standardized, closed-list of answers for each question. This format, while not capturing the full scope of a respondent's personal story, ensures consistency and allows for the aggregation of data into statistical insights.

We used Wave 9 \cite{SHAREwave9, SHAREwave9methodology,borsch2013data} of the SHARE dataset, whose interviews happened between 2021 and 2022. This wave includes data from 69448 individuals. We filtered the dataset down to only participants from France, Germany and Spain; our aim was to ensure a sufficient number of instances---at least around 1000 survey-anchored agents for each research question---while considering  the cost and duration required to execute studies in each additional country. Each individual could potentially answer more than 3000 questions, but not all of them apply to the participant or was not answered for different reasons.

\subsection{Creation of generative agent simulations}

We created a generative agent for each individual in the SHARE study. For each participant, we extracted questions/answers from almost all sections of the SHARE study and used this to create its corresponding survey-anchored generative agent.\footnote{We excluded sections Interviewer and Linking; see Appendix materials for complete list of sections.} 
In addition to the questions and answers, we added explicitly the age (calculated assuming the interviews took place in 2021) and the country as the first information visible to the Large Language Model.

This data was structured as part of the prompt, for example, as:
\begin{tcolorbox}[colback=gray!5, colframe=gray!80, title=Survey Information in LLM Prompt]
\enquote{Country}: \enquote{France}, \enquote{Age}: \enquote{58}, \enquote{Note sex of respondent from observation (ask if unsure)}: \enquote{Female}, \enquote{How often do you feel full of energy these days?}: \enquote{Sometimes}, ...
\end{tcolorbox}

We then queried the LLM using the following prompt:
\begin{tcolorbox}[colback=gray!5, colframe=gray!80, title=LLM Prompt]
[SURVEY INPUT]

\enquote{The text above contains answers from a person to a survey of health, ageing and retirement in Europe. Analyzing those questions and answers, try to predict how this same person would answer to the following question: }

[QUESTION INPUT]
\end{tcolorbox}

[SURVEY INPUT] is the set of questions/answers for that participant obtained from the SHARE dataset as described above, and [QUESTION INPUT] is the question of interest and also the multiple-choice options provided as answers: e.g., \enquote{What are the chances that you will live to age XX or more} \enquote{[0,10,20,30,40, ...,90,100]}? See Appendix materials for more detailed example of questions/answers provided as [SURVEY INPUT].  During our studies, the question of interest ([QUESTION INPUT]) varied and came from different sources: the SHARE dataset (Study 1, see \ref{subsec:EXP1}), the Eurobarometer survey on Financial Literacy (2023) \cite{doi/10.2874/956514} (Study 2, see \ref{subsec:EXP2}), or previous research by \citet{jacobs2005influence} for replication (Study 3 in section \ref{subsec:EXP3}). The SHARE dataset was used to construct the survey-anchored agents in Studies 1 and 2, and the GSS dataset was used in Study 3 (to replicate US based participants); these demographic and attitudinal responses served as the foundational characteristics that the LLM used to instantiate each persona. Study 1 evaluates in-dataset fidelity on SHARE, while Study 2 and 3 are explicitly designed as out-of-dataset generalization tests rather than evaluations of SHARE prediction accuracy. 

The Eurobarometer~\cite{doi/10.2874/956514} and \citet{jacobs2005influence} instruments offered additional questions closely related to our topic of interest—retirement attitudes—that were not available in SHARE. In studies 2 (see \ref{subsec:EXP2}) and 3 (see \ref{subsec:EXP3}), we therefore tested whether agents anchored to SHARE or GSS respondents' profiles could generalize to answer these related but distinct questions from external datasets, evaluating whether this prompting setup elicited stable, respondent-specific patterns that extend beyond the specific items in the original dataset.

Due to the SHARE dataset usage agreement, which requires data privacy, it is not possible to use cloud (frontier) models. So, we restricted our attention to local language models. We employed quantized version of Qwen 3 14b~\cite{yang2025qwen3} with Ollama~\cite{ollama2023} and vLLM \cite{kwon2023efficient}. We used the default hyperparameters provided by Ollama for all models (see Appendix), and performed no hyperparameter tuning. Context length varied only due to differences in prompt length between demographics-only and survey-anchored agents. We additionally ran exploratory tests with other local LLMs using their default Ollama settings (see Section~\ref{subsubsection:robustness}).

\begin{table*}[]
\centering
\begin{tabular}{|l|p{6.5cm}|l|}
\hline
\textbf{Question Code} & \textbf{Question Text} & \textbf{Dimension} \\
\hline
SHARE-FTP01 & What are the chances that you will live to age XX or more?\protect\footnotemark & Future Time Perspective\\
\hline
SHARE-FTP02 & Thinking about your work generally and not just your present job, what are the chances that you will be working full-time after you reach age 63? & Future Time Perspective\\
\hline
SHARE-FTP03 & In planning your saving and spending, which of the following time periods is most important to you? & Future Time Perspective\\
\hline
SHARE-FRT01 & Which of the statements on the card comes closest to the amount of financial risk that you are willing to take when you save or make investments? & Risk Tolerance\\
\hline
SHARE-FK01 & Let's say you have 2000€ in a savings account. The account earns 10\% interest each year. How much would you have in the account at the end of two years? & Financial Literacy\\
\hline
\end{tabular}
\caption{Questions from the SHARE survey related to the dimensions of interest in our work (pension savings and retirement planning).}
\label{tab:SHARE_questions}

\end{table*}


\section{Studies}
\label{sec:expe}

To evaluate how well each agent reproduces human attitudes, we conduct three evaluations—at the individual, country, and study levels. Each study focuses on the same three measurement constructs: Future Time Perspective (FTP), Financial Risk Tolerance (FRT), and Financial Knowledge (FK) dimensions. 



\footnotetext{Age XX follows the SHARE questionnaire specification and is defined as a target age conditional on the respondent's current age (e.g., XX = 80 for respondents aged 67).}
\begin{itemize}
\item \emph{Individual level} (Section~\ref{subsec:EXP1}): This evaluation assesses whether each agent reproduces its matched respondent's held-out SHARE answers on a subset of FTP, FRT, and FK items (Table~\ref{tab:SHARE_questions}). Each target question is withheld from the agent's context; a prediction is generated and compared with the respondent's observed answer. As a baseline we use a supervised random-forest model trained on SHARE (a separate model is trained for each target question). We consider the random forest as difficult baseline to match LLM performance, since the random forest is directly trained on the target prediction data; the LLM agents, in contrast, are not given any supervision on the specific prediction task. Performance is reported using Pearson's Correlation \(r\) for numerical items and weighted \(F_1\) for categorical items. Moreover, the total variation distance (TVD) is calculated for each item (See Appendix for a detailed description of the Total Variation Distance (TVD) calculation, including its application to numerical variables.)
    \item \emph{Country level} (Section~\ref{subsec:EXP2}): We aggregate the agents' responses on a subset of the  Eurobarometer survey \cite{doi/10.2874/956514}  items aligned to the same FTP, FRT, and FK constructs. 
    We then compare the country-level responses repartition —specifically, the percentage of agents selecting each answer option—against the corresponding proportions observed in the official Eurobarometer data. One constraint of our research is the age discrepancy between our agents (based on the SHARE questions/answers), who are all over 50, and the Eurobarometer study participants, who are aged 18 and above. 
    \item \emph{Study level} (Section~\ref{subsec:EXP3}): We conceptually reproduce the same study from \cite{jacobs2005influence} using a population of agents based on participants of the General Social Survey \cite{marsden2016overview} of 2004. We compute agent-level FTP, FRT and FK scale scores across the answers of our agents, and replicate a hierarchical regression of retirement savings on these three scales (including pairwise and three-way interactions), comparing estimated coefficients and model fit to the original study \cite{jacobs2005influence}. 
\end{itemize}

\subsection{Individual-level evaluation (SHARE)}
\label{subsec:EXP1}

\subsubsection{Methodology}
A leave-one-item-out evaluation was conducted on the SHARE questions in Table~\ref{tab:SHARE_questions}. For each respondent and target question, the model context comprised all remaining question--answer pairs for that respondent. The target question's text and answer were withheld. The model then received the held-out item question (and, for categorical items, the corresponding answer set) and produced a predicted response. It's crucial to emphasize that eliminating only the target question, while keeping other closely related questions from the same section, might enhance performance.  Identifying these associated queries would require a subjective, expert-driven process. Ultimately, the supervised learning method gains from the same kind of assistance.

Categorical items used the response categories defined by the SHARE survey. Numerical items were elicited on a continuous \([0,100]\) scale to limit discretization artifacts associated with coarse grids~\cite{chen2023emergence}. Only for the question about subjective survival expectations (\texttt{SHARE-FTP01}), the threshold age was individualized to each respondent (e.g., a 67-year-old was queried about living to at least age 80). This was precisely derived from the SHARE questionnaire's design, with target ages determined according to the survey's guidelines.

Performance was computed at the respondent--item level. 

\subsubsection{Overall Performance: Survey-Anchored vs. Demographics-Only Agents}
\label{subsubsec:overall_performance}

Table \ref{tab:SHARE_results} summarizes performance across all five SHARE questions. Survey-anchored agents consistently outperform demographics-only agents in terms of Total Variation Distance (TVD), indicating closer alignment with the empirical distribution of responses. Across all five questions except one (SHARE-FTP01), survey-anchored agents achieve substantially lower TVD values (more than 30\% improvement).

\begin{table*}[ht]
    \centering
    \footnotesize
    \renewcommand{\arraystretch}{1.2}
    \setlength{\tabcolsep}{3pt}
    \resizebox{\textwidth}{!}{
    \begin{tabular}{|p{3.2cm}|c|p{3.5cm}|p{2.5cm}|p{2.5cm}|}
    \hline
        \textbf{Question} & \textbf{N} & \textbf{Supervised ML} & \textbf{Demographic agents} & \textbf{Survey-anchored agents} \\
    \hline
        \multirow{3}{3.2cm}{SHARE-FTP01: Expect to live to age XX?} & \multirow{3}{*}{1857} 
        & Train Corr: 0.66 / Test Corr: \textbf{0.54} & Corr: 0.44 & Corr: 0.50 \\
        & & Train TVD: 0.44 / Test TVD: 0.45 & TVD: 0.47 & TVD: \textbf{0.44} \\
        & & Ntrees: 20, max depth: 5 & & \\
    \hline
        \multirow{3}{3.2cm}{SHARE-FTP02: Expect to be working full time after 63?} & \multirow{3}{*}{986} 
        & Train Corr: 0.74 / Test Corr: \textbf{0.44} & Corr: 0.08 & Corr: 0.28 \\
        & & Train TVD: 0.59 / Test TVD: 0.60 & TVD: 0.49 & TVD: \textbf{0.33} \\
        & & Ntrees: 10, max depth: 5 & & \\
    \hline
        \multirow{3}{3.2cm}{SHARE-FTP03: In planning your saving and spending, which time period is most important?} & \multirow{3}{*}{2000} 
        & Train F1: 0.33 / Test F1: 0.27 & Weighted F1: 0.22 & Weighted F1: \textbf{0.29} \\
        & & Train TVD: 0.42 / Test TVD: 0.42 & TVD: 0.41 & TVD: \textbf{0.28} \\
        & & Ntrees: 10, max depth: 5 & & \\
    \hline
        \multirow{3}{3.2cm}{SHARE-FRT01: Amount of risk you are willing to take?} & \multirow{3}{*}{2000} 
        & Train F1: 0.62 / Test F1: 0.59 & Weighted F1: 0.53 & Weighted F1: \textbf{0.60} \\
        & & Train TVD: 0.26 / Test TVD: 0.26 & TVD: 0.16 & TVD: \textbf{0.11} \\
        & & Ntrees: 10, max depth: 7 & & \\
    \hline
        \multirow{3}{3.2cm}{SHARE-FK01: 2000€ savings account, 10\% interest, after 2 years I will have?} & \multirow{3}{*}{1100} 
        & Train F1: 0.51 / Test F1: \textbf{0.42} & Weighted F1: 0.23 & Weighted F1: 0.28 \\
        & & Train TVD: 0.22 / Test TVD: \textbf{0.23} & TVD: 0.58 & TVD: 0.37 \\
        & & Ntrees: 20, max depth: 7 & & \\
    \hline
    \end{tabular}
    }
    \caption{Results to questions from the SHARE survey. Ground truth are the answers given by real survey participants. For each question, the first subrow shows Correlation/F1 scores (higher is better), the second subrow shows TVD scores (lower is better), and the third subrow shows hyperparameters for the supervised ML model. Survey-anchored agents enhance the distribution's closeness to ground truth, improving TVD over demographics.}
    \label{tab:SHARE_results}
\end{table*}

We extend the analysis to 10 non-retirement SHARE items that capture subjective self-assessments of personal states, abilities, or experiences rather than objective indicators. \footnote{Although the SHARE Wave 9 questionnaire comprises approximately 1,000 questions across 23 modules, most are unsuitable for evaluating LLM-based agent simulation. Many items record verifiable biographical facts (e.g., year of birth, education level, number of children, diagnosed conditions) whose answers are uniquely determined by each respondent's life history. Others involve physical performance measurements collected with calibrated instruments (e.g., grip strength, walking speed, peak flow), and a large share consist of detailed financial or asset inventories requiring precise numerical recall.} Our selected items, by contrast, elicit evaluative judgments mostly along ordinal or Likert-type scales — spanning self-rated health, perceived memory, job satisfaction, physical activity frequency, computer skills, health literacy, financial strain, depressive mood, and life satisfaction — making them well-suited to test whether LLM agents can reproduce the distributional patterns of subjective human judgments. Table \ref{tab:SHARE_results_beyond_retirement_attitudes} indicates that, for all but three questions (CF820, MH002 and AC012), survey-anchored agents replicate the observed response distributions more accurately than agents relying solely on demographics. When examining items by domain, survey-anchored agents achieve a comparable average reduction in TVD for non-retirement questions (mean $\Delta$TVD = 0.142; 26.68\% reduction) compared to retirement-related questions (mean $\Delta$TVD = 0.116; 27.64\% reduction), though these averages are computed over a small number of items (10 and 5, respectively).

\begin{table*}[ht]
    \centering
    \renewcommand{\arraystretch}{1.3}
    \setlength{\tabcolsep}{4pt}
    \resizebox{\textwidth}{!}{
    \begin{tabular}{|p{7cm}|p{1.8cm}|p{3.6cm}|p{3.0cm}|p{3.0cm}|}
    \hline
        \textbf{\large Question} & \textbf{N (Number of agents)} & \textbf{Supervised ML} & \textbf{Demographic agents} & \textbf{Survey-anchored agents} \\
    \hline
        \textbf{PH003\_HealthGen2:} Would you say your health is? & 1099 & Train F1 0.45 TVD 0.34 Test F1 0.41 TVD 0.35 Ntrees 20, max depth 7 & Weighted F1 0.30 TVD 0.50 & Weighted F1 \textbf{0.45} TVD \textbf{0.18} \\
    \hline
        \textbf{CF103\_Memory:} How would you rate your memory at the present time? & 1085 & Train F1 0.4 TVD 0.45 Test F1 0.38 TVD 0.45 Ntrees 5, max depth 7 & Weighted F1 0.35 TVD 0.34 & Weighted F1 \textbf{0.42} TVD \textbf{0.20} \\
    \hline
        \textbf{EP026\_SatJob:} All things considered, I am satisfied with my job. & 1095 & Train F1 0.74 TVD 0.10 Test F1 \textbf{0.58} TVD \textbf{0.11} Ntrees 50, max depth 5 & Weighted F1 0.31 TVD 0.45 & Weighted F1 0.51 TVD 0.28 \\
    \hline
        \textbf{BR015\_PartInVigSprtsAct:} How often do you engage in vigorous physical activity? & 1099 & Train F1 0.51 TVD 0.20 Test F1 \textbf{0.49} TVD \textbf{0.20} Ntrees 10, max depth 7 & Weighted F1 0.25 TVD 0.57 & Weighted F1 0.36 TVD 0.47\\
    \hline
        \textbf{IT003\_PC\_skills:} How would you rate your computer skills? & 1084 & Train F1 0.39 TVD 0.30 Test F1 \textbf{0.30} TVD \textbf{0.32} Ntrees 5, max depth 7 & Weighted F1 0.25 TVD 0.41 & Weighted F1 0.28 TVD 0.34\\
    \hline
        \textbf{HC889\_HealthLiteracy:} How often do you need help reading medical instructions? & 1096 & Train F1 0.74 TVD 0.17 Test F1 \textbf{0.74} TVD 0.17 Ntrees 5, max depth 7 & Weighted F1 0.28 TVD 0.67 & Weighted F1 0.67 TVD \textbf{0.15}\\
    \hline
        \textbf{CO007\_AbleMakeEndsMeet:} Is your household able to make ends meet? & 1089 & Train F1 0.51 TVD 0.18 Test F1 \textbf{0.45} TVD \textbf{0.19} Ntrees 10, max depth 7 & Weighted F1 0.25 TVD 0.45 & Weighted F1 0.43 TVD 0.28 \\
    \hline
        \textbf{CF820\_MemoryChange:} Compared to last interview, is your memory better, same, or worse? & 1088 & Train F1 0.66 TVD 0.24 Test F1 \textbf{0.64} TVD \textbf{0.26} Ntrees 10, max depth 7 & Weighted F1 0.59 TVD 0.26 & Weighted F1 0.59 TVD 0.30 \\
    \hline
        \textbf{MH002\_Depression:} In the last month, have you been sad or depressed? & 1096 & Train F1 0.74 TVD 0.06 Test F1 \textbf{0.70} TVD \textbf{0.08} Ntrees 50, max depth 7 & Weighted F1 0.50 TVD 0.29 & Weighted F1 0.61 TVD 0.31\\
    \hline
        \textbf{AC012\_HowSat:} On a scale from 0 to 10, how satisfied are you with your life? & 1001 & Train Corr 0.69 TVD 0.28 Test Corr \textbf{0.61} TVD 0.29 Ntrees 20, max depth 5 & Corr. 0.31 TVD \textbf{0.26} & Corr. 0.49 TVD 0.27\\
    \hline
    \end{tabular}
    }
   \caption{Results to questions from the SHARE survey beyond retirement attitudes. Ground truth are the answers given by real survey participants. Survey-anchored agents enhance the distribution's closeness to ground truth, improving TVD over demographics in all questions except three (CF820,MH002,AC012).}
    \label{tab:SHARE_results_beyond_retirement_attitudes}
\end{table*}

To evaluate the statistical significance of these performance differences, we employ a participant-level bootstrap procedure with 5{,}000 resamples \cite{efron1994introduction,koehn2004statistical,berg2012empirical}. Resampling at the participant level accounts for non-independence in the data, as the same respondents contribute responses to multiple questions.

We construct a participant-level dataset containing ground-truth survey responses and corresponding simulated responses from survey-anchored and demographics-only agents for all 15 evaluation questions ($N = 5{,}461$ SHARE participants). In each bootstrap iteration, participants are resampled with replacement, preserving their full response profiles across all questions they answered. For each question, we compute the Total Variation Distance (TVD) between agent-generated and ground-truth response distributions for both agent types. We then compute the mean difference in TVD (survey-anchored minus demographics-only) across the 15 questions. Statistical significance is assessed using the 95\% bootstrap confidence interval of this mean difference, rejecting the null hypothesis of no difference when the interval excludes zero.

Table~\ref{tab:boostrap_results} reports bootstrapped TVD estimates for both agent types. On average across the 15 questions, survey-anchored agents achieve substantially lower TVD than demographics-only agents (mean $\Delta\text{TVD} = -0.142$, corresponding to a mean 29.02\% reduction). Survey-anchored agents produce response distributions that are closer to the ground truth for all questions except two (\textit{CF820\_MemoryChange, MH002\_Depression}). For \textit{AC012\_HowSat}, both types of agents perform comparably. In 12 out of the 15 questions, the decrease in TVD is greater than 5\% in favor of survey-anchored agents.

The bootstrap analysis (Table~\ref{tab:bootstrap_summary}) yields a mean difference of $\Delta\text{TVD} = -0.142$ (95\% CI: [$-0.1516$, $-0.1309$]), where the negative value indicates superior performance by survey-anchored agents. The confidence interval excludes zero, indicating a statistically significant difference ($p < 0.05$). Results are consistent across 1{,}000 and 5{,}000 bootstrap iterations, indicating robustness to the choice of resampling size.


\begin{table*}[]
\label{tab:tvd_demog7_vs_survey}
\begin{tabular}{lcccccc}
\toprule
Question Name & Type & Demog7 & Survey & $\Delta$ TVD & \% Change & Better \\
              &      &        &        & (S $-$ D)    &           & Method \\
\midrule
PH003\_HealthGen2        & Cat. & 0.514 & \textbf{0.132}$^\dagger$ & -0.382 & -74.3\% & Survey-anch. \\
CF103\_Memory            & Cat. & 0.321 & \textbf{0.190}$^\dagger$ & -0.131 & -40.8\% & Survey-anch. \\
EP026\_SatJob            & Cat. & 0.456 & \textbf{0.287}$^\dagger$ & -0.169 & -37.1\% & Survey-anch. \\
BR015\_PartInVigSprtsAct & Cat. & 0.606 & \textbf{0.463}$^\dagger$ & -0.143 & -23.6\% & Survey-anch. \\
IT003\_PC\_skills        & Cat. & 0.408 & \textbf{0.340}$^\dagger$ & -0.068 & -16.7\% & Survey-anch. \\
HC889\_HealthLiteracy    & Cat. & 0.638 & \textbf{0.145}$^\dagger$ & -0.493 & -77.3\% & Survey-anch. \\
CO007\_AbleMakeEndsMeet  & Cat. & 0.455 & \textbf{0.263}$^\dagger$ & -0.192 & -42.2\% & Survey-anch. \\
CF820\_MemoryChange      & Cat. & \textbf{0.248}$^\dagger$ & 0.280 & +0.032 & +12.9\% & Demo-only \\
MH002\_Depression        & Cat. & \textbf{0.293} & 0.307 & +0.014 & +4.8\%  & Demo-only \\
AC012\_HowSat            & Num. & \textbf{0.248} & 0.251 & +0.003 & +1.21\%  & Demo-only \\
SHARE-FTP01              & Num. & 0.469 & \textbf{0.436}$^\dagger$ & -0.033 & -7.0\%  & Survey-anch. \\
SHARE-FTP02              & Num. & 0.494 & \textbf{0.334}$^\dagger$ & -0.161 & -32.39\% & Survey-anch. \\
SHARE-FRT01              & Cat. & 0.160 & \textbf{0.108}$^\dagger$ & -0.052 & -32.5\% & Survey-anch. \\
SHARE-FTP03              & Cat. & 0.435 & \textbf{0.284}$^\dagger$ & -0.151 & -34.71\% & Survey-anch. \\
SHARE-FK01              & Cat. & 0.581 & \textbf{0.374}$^\dagger$ & -0.207 & -35.6\% & Survey-anch. \\
\midrule
\textbf{Mean}            &      & 0.422 & \textbf{0.280} & -0.142 & -29.02\% & Survey-anch. \\
\bottomrule
\end{tabular}
  \small
  \caption{Bootstrap resampling: Comparison of Total Variation Distance between 
  demographics-only and survey-anchored agents across 15 questions ($N$ = 5,461). 
  Survey-anchored agents show lower TVD in 12 out of 15 questions. 
  $^\dagger$indicates $|\Delta \mathrm{TVD}| > 5\%$. 
  \% Change = (Survey $-$ Demo)/Demo $\times$ 100.}
\label{tab:boostrap_results}
\end{table*}

\begin{table}[htbp]
\centering
\small
\begin{tabular}{@{}lc@{}}
\toprule
Metric & Value \\
\midrule
Participants & 5,461 \\
Questions & 15 \\
Bootstrap Iterations & 5,000 \\
\midrule
Mean TVD (Demo) & 0.4217 \\
Mean TVD (Survey) & 0.2796 \\
\midrule
$\Delta$ TVD & $-0.1421$ \\
95\% CI & [$-0.1516$, $-0.1309$] \\
$p$-value & $< 0.05$ \\
\bottomrule
\end{tabular}
\caption{Bootstrap resampling analysis of TVD differences between demographics-only (Demo) and survey-anchored (Survey) agents. Negative $\Delta$ TVD indicates survey-anchored agents are closer to ground truth ($N$ = 5,461; 5,000 bootstrap iterations).}
\label{tab:bootstrap_summary}
\end{table}

\subsubsection{Central Tendency Bias}

Prior work has identified phenomena such as central tendency bias and reduced variance in LLM survey simulations \cite{bisbee2024synthetic,kaiser2025simulating,wang2025large,murthy2025one}. We assess whether these limitations also apply to demographics-only agents and survey-anchored agents in our setting. We observe consistent evidence for this pattern across several questions.

\begin{figure*}[]
    \centering
    \begin{minipage}[t]{0.37\textwidth}
        \centering
        \includegraphics[width=\textwidth]{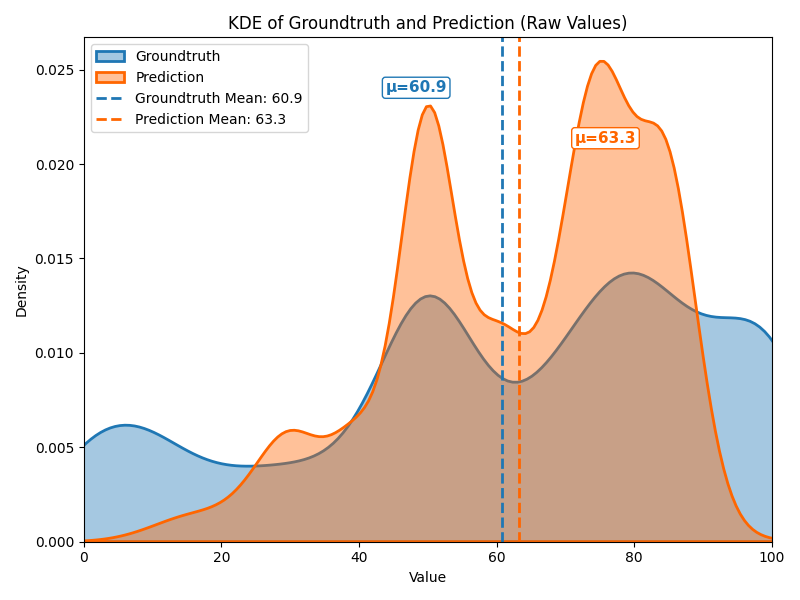}
        \caption*{\small (a) Demographics-only agents.}
        \label{fig:ex009_demog_kde}
    \end{minipage}
    \hspace{0.05\textwidth}
    \begin{minipage}[t]{0.37\textwidth}
        \centering
        \includegraphics[width=\textwidth]{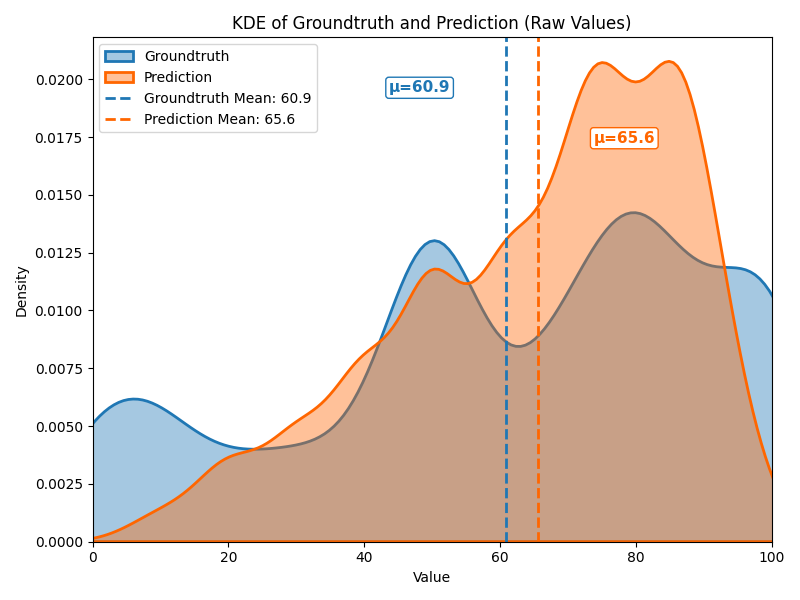}
        \caption*{\small (b) Survey-anchored agents.}
        \label{fig:ex009_full_kde}
    \end{minipage}
    
    \vspace{0.3cm}
    
    \begin{minipage}[t]{0.37\textwidth}
        \centering
        \includegraphics[width=\textwidth]{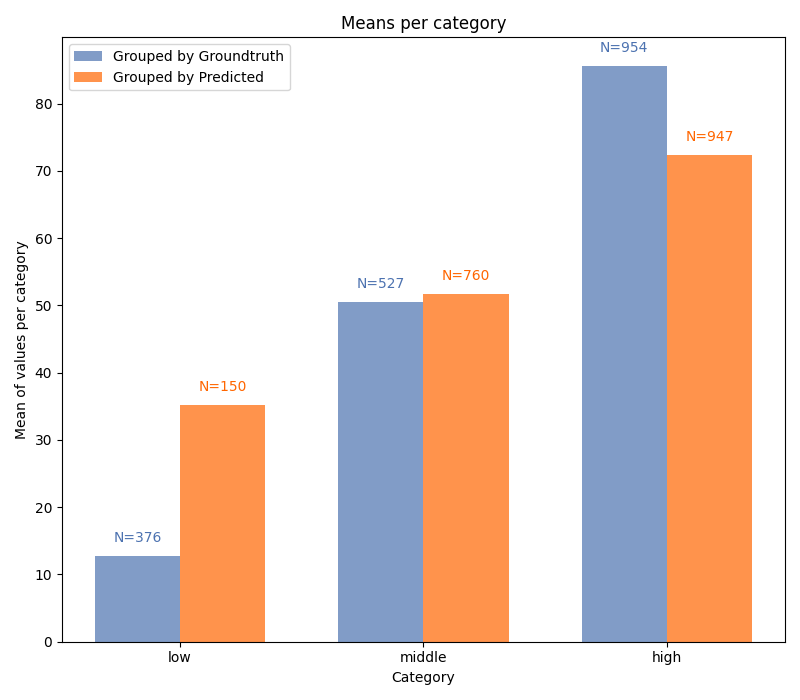}
        \caption*{\small (c) Demographics-only agents.}
        \label{fig:ex009_demog_barplot}
    \end{minipage}
    \hspace{0.05\textwidth}
    \begin{minipage}[t]{0.37\textwidth}
        \centering
        \includegraphics[width=\textwidth]{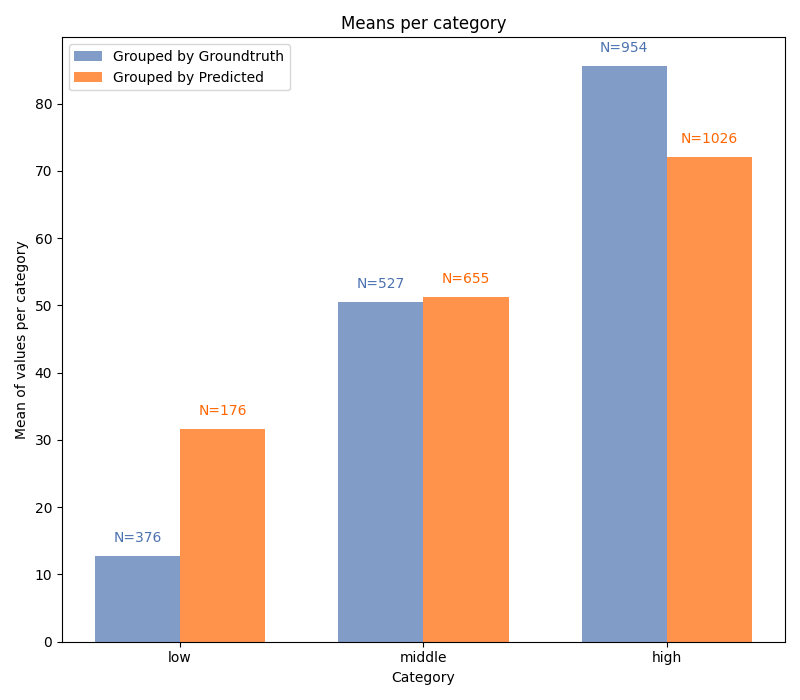}
        \caption*{\small (d) Survey-anchored agents.}
        \label{fig:ex009_full_barplot}
    \end{minipage}
    
    \vspace{0.3cm}
    
    \begin{minipage}[t]{0.37\textwidth}
        \centering
        \includegraphics[width=\textwidth]{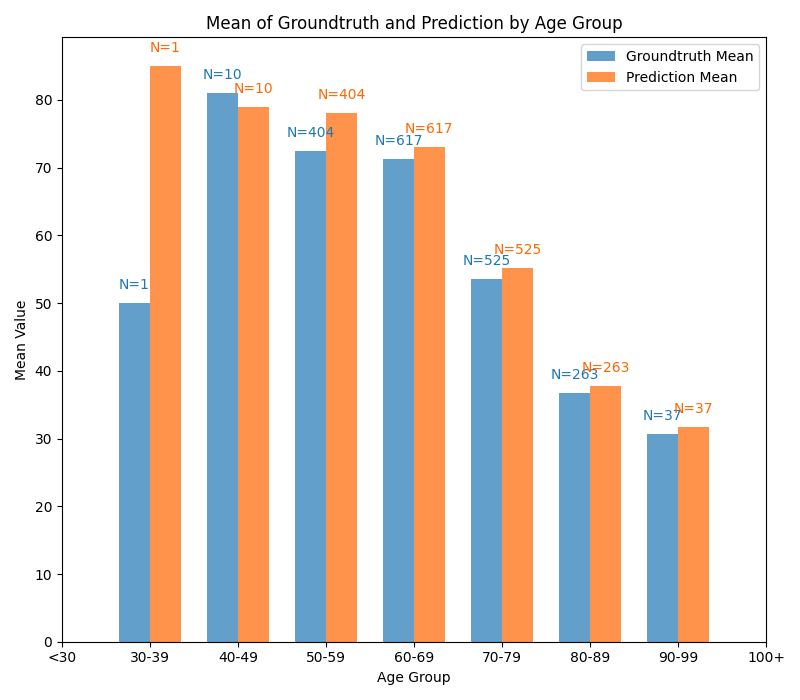}
        \caption*{\small (e) Demographics-only agents.}
        \label{fig:ex009_demog_byages}
    \end{minipage}
    \hspace{0.05\textwidth}
    \begin{minipage}[t]{0.37\textwidth}
        \centering
        \includegraphics[width=\textwidth]{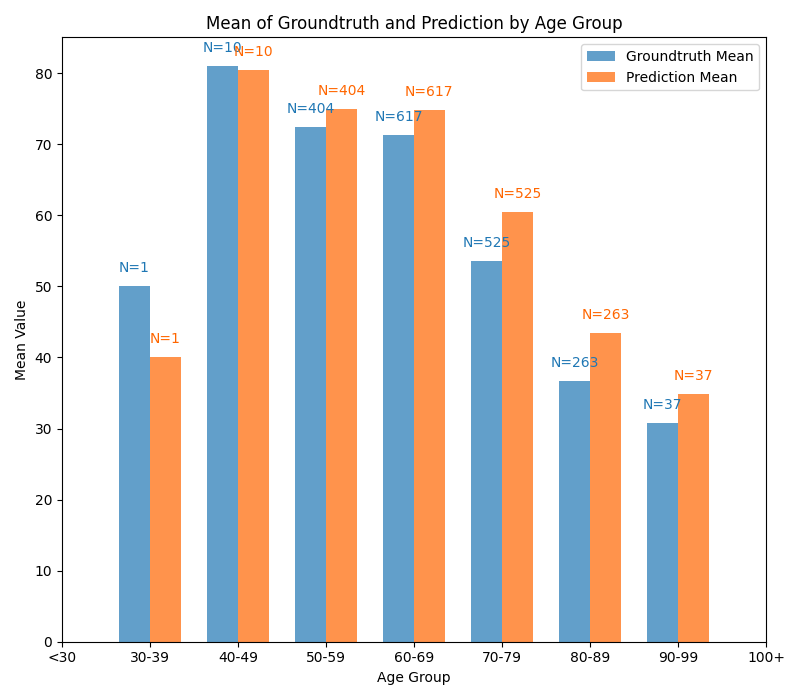}
        \caption*{\small (f) Survey-anchored agents.}
        \label{fig:ex009_full_byages}
    \end{minipage}
    
    \caption{\small SHARE-FTP01 – \enquote{\textit{What are the chances that you will live to age XX or more?}} 
\newline 
Comparison of predictions from demographics-only agents (left) and survey-anchored agents (right). 
(a,b) Kernel density estimates show overestimation and focal-point heaping (demographics-only) versus smooth upward shift (survey-anchored).  \newline
(c,d) Validation of LLM categorization through preserved group means: participants were categorized into three groups using either ground truth responses (blue) or LLM predictions (orange), with category means computed from actual human values in both cases. While LLMs effectively identify \textit{central tendencies}, they struggle to capture \textbf{extreme responses}. \newline
(e,f) Mean values across age groups. LLMs correctly reproduce the \textit{monotonic decline} in subjective life expectancy with increasing age.}  
    \label{fig:ex009_figures}
\end{figure*}

\paragraph{Subjective life expectancy (SHARE-FTP01).}
Figure \ref{fig:ex009_figures} compares predictions for \enquote{\textit{What are the chances that you will live to age XX or more?}} Both agent types overestimate subjective survival probability relative to the groundtruth. The demographics-only distribution appears more multimodal, with visible clustering around focal values (30, 50, 80) (Figure \ref{fig:ex009_figures}a,b). When we examine subgroup means (Figure \ref{fig:ex009_figures}c,d), a clear pattern emerges. Participants’ responses, which range from 0 to 100, were divided into three categories: Low values(0–33.3), Middle values (33.3–66.7), and High values (66.7–100). These categories were defined either using the ground-truth human responses or the LLM’s predicted responses, and in both cases, the mean for each category was always computed from the actual human response values (see \ref{appendix:Validation_LLM_Categorization} for further details). While LLMs effectively identify central tendencies (the Middle category matches closely), they systematically \textbf{overestimate means in the Low category and underestimate means in the High category}—failing to capture extreme responses. This pattern is consistently reproduced across subsequent questions.

An analysis across age groups (Figure \ref{fig:ex009_figures}e,f) shows that both agent types capture the expected decline in subjective life expectancy with increasing age. Both types of agents systematically produce more optimistic estimates than human respondents, consistent with findings that instruction-finetuned models tend to yield higher scores on well-being measures \cite{li2022gpt}.

\begin{figure*}[]
    \centering

    \begin{minipage}[t]{0.45\textwidth}
        \centering
        \includegraphics[width=\textwidth]{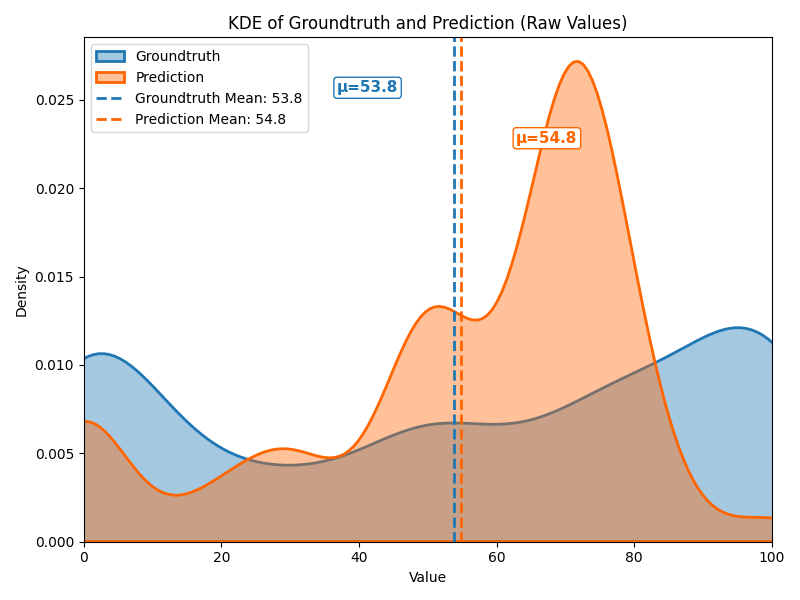}
        \caption*{(a) Demographics-only agents.}
        \label{fig:ex025_demog_kde}
    \end{minipage}
    \hfill
    \begin{minipage}[t]{0.45\textwidth}
        \centering
        \includegraphics[width=\textwidth]{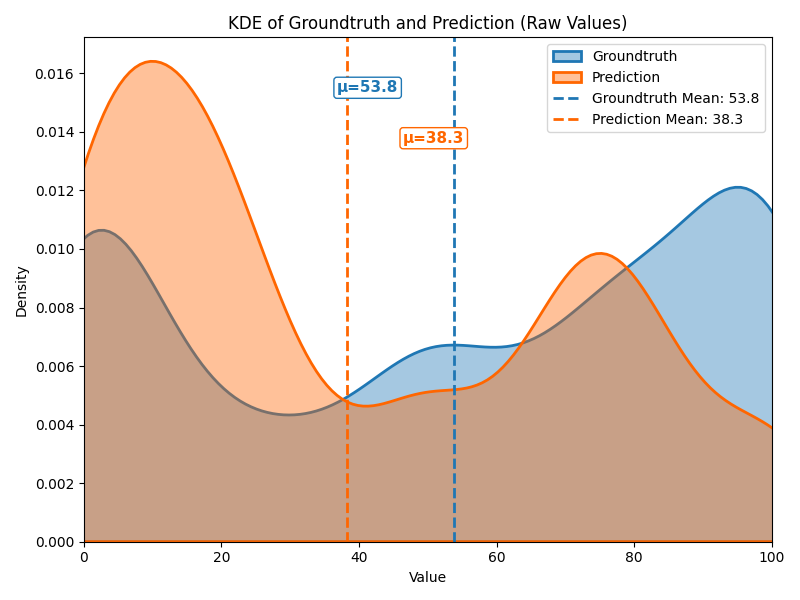}
        \caption*{(b) Survey-anchored agents.}
        \label{fig:ex025_full_kde}
    \end{minipage}

    \vspace{0.5cm}

    \begin{minipage}[t]{0.45\textwidth}
        \centering
        \includegraphics[width=\textwidth]{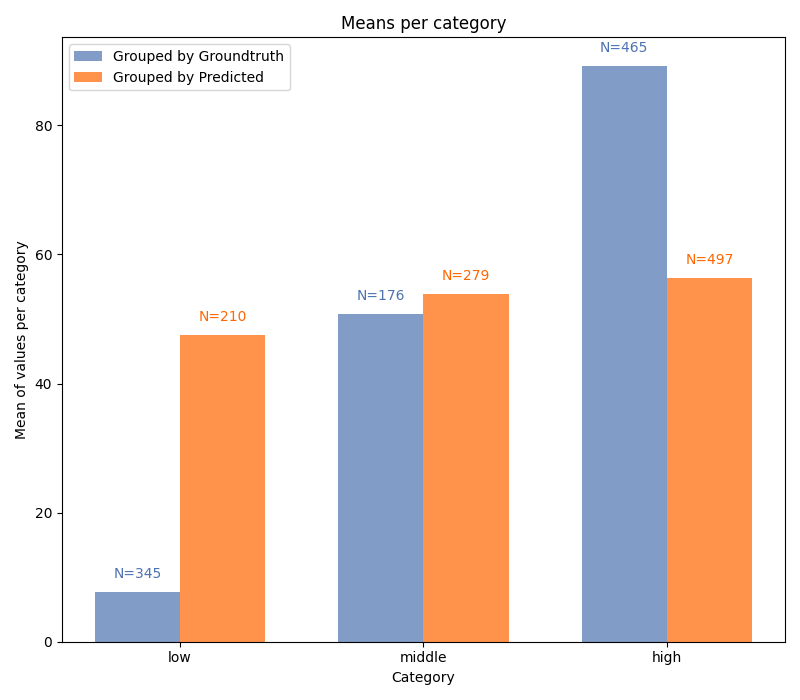}
        \caption*{(c) Demographics-only agents.}
        \label{fig:ex025_demog_barplot}
    \end{minipage}
    \hfill
    \begin{minipage}[t]{0.45\textwidth}
        \centering
        \includegraphics[width=\textwidth]{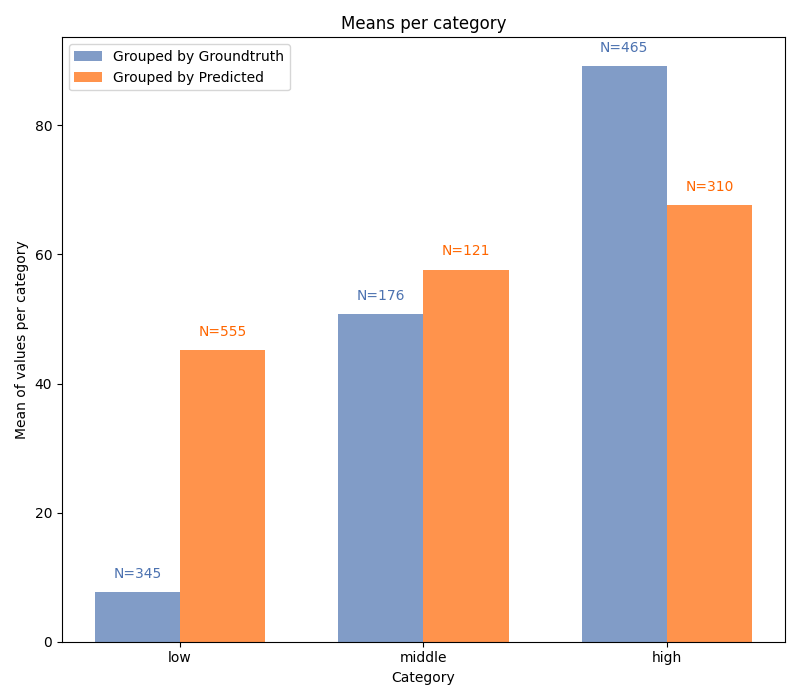}
        \caption*{(d) Survey-anchored agents.}
        \label{fig:ex025_full_barplot}
    \end{minipage}

    \vspace{0.5cm}

    \caption{
    SHARE-FTP02 – "\textit{What are the chances that you will be working full-time after you reach age 63?}" Comparison of predictions from demographics-only agents (left) and survey-anchored agents (right), evaluated against human responses.\newline
    Kernel density plots (a,b) reveal that survey-anchored agents (b) approximate the bimodal distribution of human responses, capturing distinct clusters representing respondents with low versus high expectations for post-retirement work. In contrast, demographics-only agents (a) \textbf{do not position these clusters} at the distributional extremes, instead concentrating predictions in the middle range.\newline
    (c,d) Validation of LLM categorization fidelity through preserved group means. Participants were categorized into three groups using either ground truth human responses (blue) or LLM predictions (orange), with category means computed from actual human values in both cases. Both (c,d) recover category-level means more accurately in the center (c), though they overestimate and underestimate at the extremes. Survey-anchored agents (d) \textbf{outperform demographics-only agents} (c) specially for the category of high expectations.
    }
    \label{fig:ex025_figures}
\end{figure*}

\paragraph{Post-retirement work expectations (SHARE-FTP02).}
For \enquote{\textit{What are the chances that you will be working full-time after you reach age 63?}}, the central tendency bias is more pronounced. Figure \ref{fig:ex025_figures}(a,b) shows kernel density plots: survey-anchored agents approximate the \textbf{bimodal distribution} of human responses, capturing distinct clusters of respondents with low versus high expectations. In contrast, demographics-only agents \textbf{do not position these clusters} at the distributional extremes, instead concentrating predictions in the middle range.

\begin{figure*}[]
    \centering
    \begin{minipage}[t]{0.45\textwidth}
        \centering
        \includegraphics[width=\linewidth]{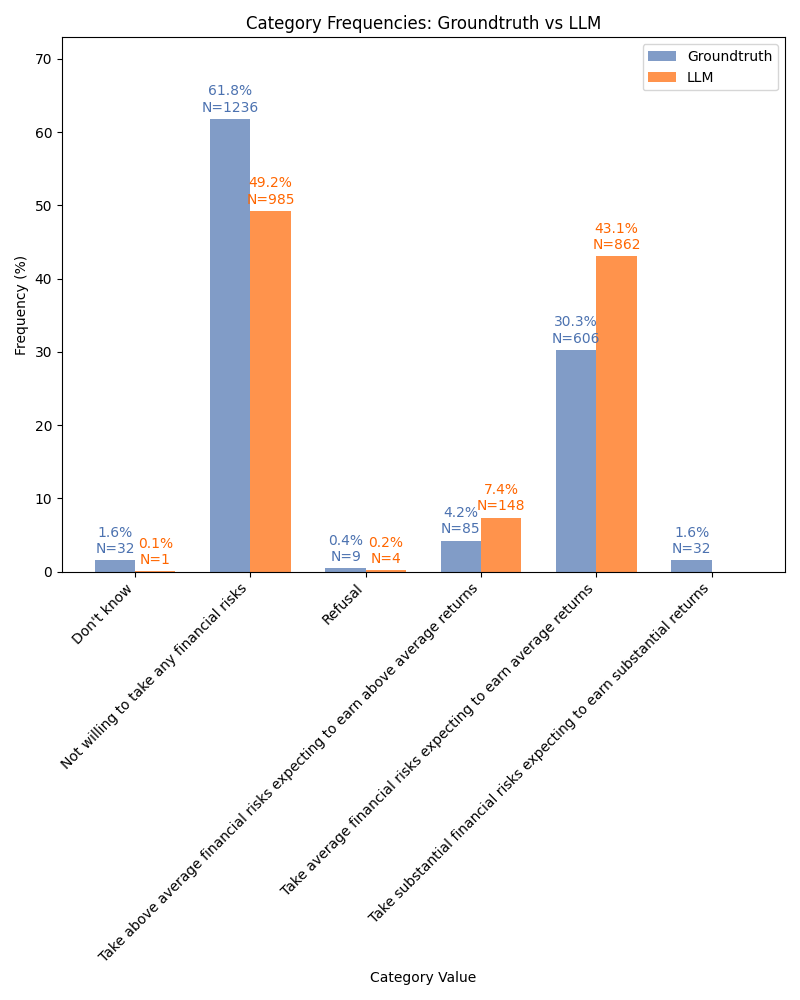}
        \caption*{(a) Demographics-only agents.}
        \label{fig:ex110_demog}
    \end{minipage}
    \hspace{0.04\textwidth}
    \begin{minipage}[t]{0.45\textwidth}
        \centering
        \includegraphics[width=\linewidth]{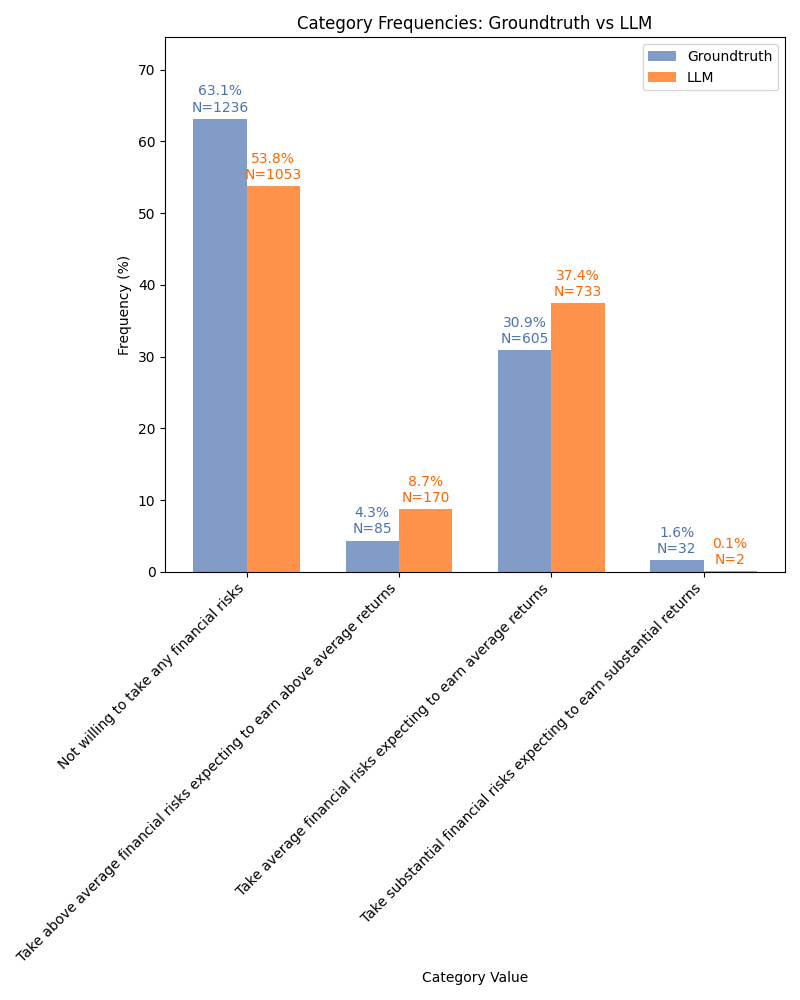}
        \caption*{(b) Full survey-anchored agents.}
        \label{fig:ex110_full}
    \end{minipage}
    \caption{
        SHARE-FRT-01 — \textit{\enquote{Which amount of risk are you willing to take when you save or make investments?}}.
        Demographics-only agents (a) exhibit more pronounced central tendency bias than survey-anchored agents (b), which produce distribution closer to real partipants. Both agents tend to be less risk-averse than humans.
    }
    \label{fig:ex110_figures}
\end{figure*}

\paragraph{Financial risk tolerance (SHARE-FRT01).}
In the question related to risk tolerance, demographics-only agents (Figure \ref{fig:ex110_figures}a) over-estimate the less frequent answers by human respondents and under-estimate the most frequent answers, again exhibiting central tendency bias. Survey-anchored agents (Figure \ref{fig:ex110_figures}b) have the same problem but yield an answer distribution closer to that of real participants. Both types of agents enhance the performance of the supervised machine learning baseline (Table \ref{tab:SHARE_results}) and exhibit less risk-averse behavior than human respondents. The survey-anchored agents never generated Don't know or Refusal responses, and the demographics-only agents produced such responses only rarely.

\subsubsection{Hyper-Accuracy on Objective Questions}
\label{subsubsec:hyper_accuracy}

A second key hypothesis is that LLM agents exhibit \textit{hyper-accuracy}\cite{aher2023using}—performing unrealistically well on objective questions by failing to reproduce human error patterns. This limitation is particularly pronounced for demographics-only agents.

\begin{figure*}[htbp]
    \centering
    \begin{minipage}[t]{0.45\textwidth}
        \centering
        \includegraphics[width=\linewidth]{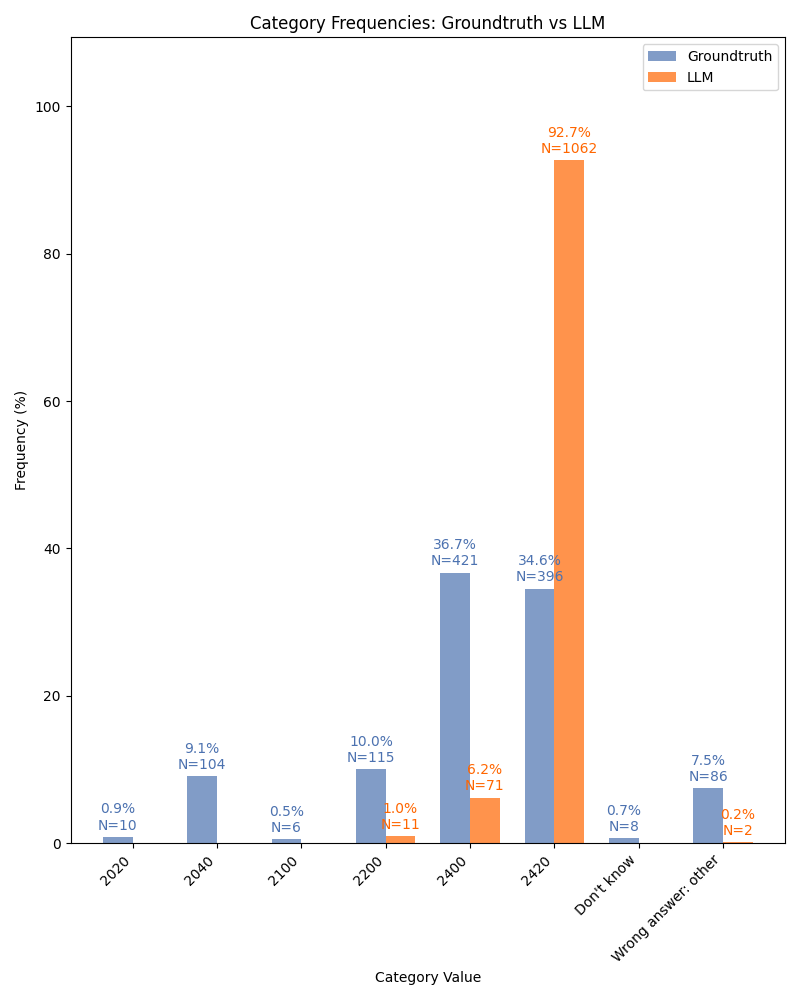}
        \caption*{(a) Demographics-only agents.}
    \end{minipage}
    \hspace{0.04\textwidth}
    \begin{minipage}[t]{0.45\textwidth}
        \centering
        \includegraphics[width=\linewidth]{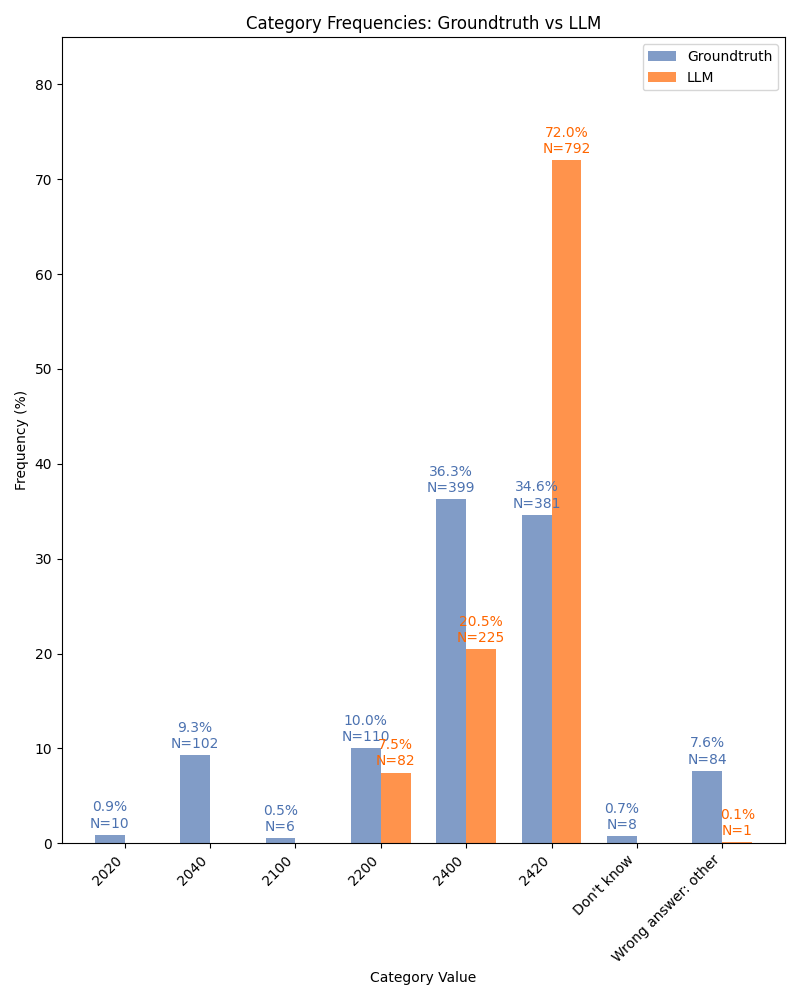}
        \caption*{(b) Survey-anchored agents.}
    \end{minipage}

    \caption{SHAREFK-01 – \enquote{\textit{Starting today with 2000€ in savings account, with 10\% interest, how much will I have in 2 years?}} (Correct answer: 2420€). Left: Demographics-only agents are hyper accurate and miss most human errors. Right: Full survey-anchored agents capture slightly more errors (like using simple instead of compound interest) but still display hyper accuracy and overlook most mistakes.}
    \label{fig:cf015_figures}
\end{figure*}

For the compound interest question (SHARE-FK01: \textit{\enquote{Let's say you have 2000€ in a savings account. The account earns 10\% interest each year. How much would you have in the account at the end of two years?}}), both agent types exhibit hyper-accuracy, but demographics-only agents are especially affected (Figure \ref{fig:cf015_figures}).

Demographics-only agents (Figure \ref{fig:cf015_figures}a) return the correct answer (€2420) at nearly 100\% frequency, completely missing the variability in human responses. This outcome likely reflects LLM alignment toward producing accurate and truthful responses, which reduces their ability to reproduce human-like error patterns and limits their validity for attitude simulation \cite{lyman2025balancing}.

Survey-anchored agents (Figure \ref{fig:cf015_figures}b) capture somewhat more errors—including the common mistake of applying simple interest rather than compound interest—but still display substantial hyper-accuracy. They fail to reproduce the ``don't know'' responses and counter-intuitive errors present in human data.

\subsubsection{Temporal Reasoning Limitations}
\label{subsubsec:temporal_reasoning}

\begin{figure*}[]
    \centering
    \begin{minipage}[t]{0.45\textwidth}
        \centering
        \includegraphics[width=\linewidth]{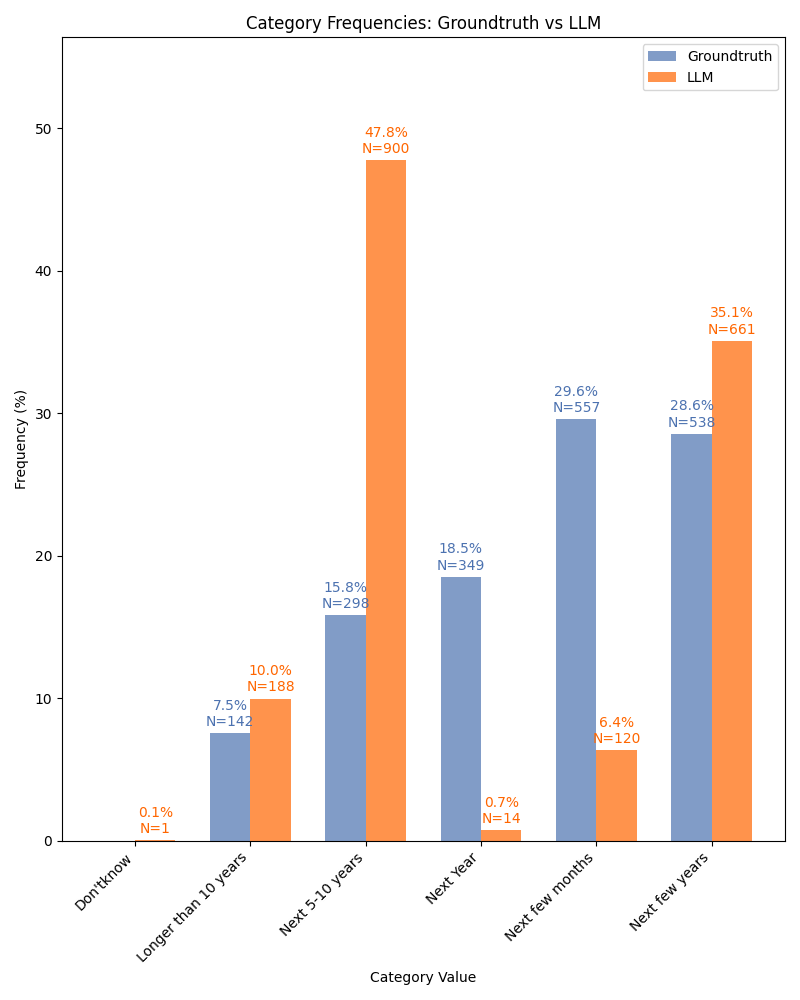}
        \caption*{(a) Demographics-only agents.}
        \label{fig:ex111_demog}
    \end{minipage}
    \hspace{0.04\textwidth}
    \begin{minipage}[t]{0.45\textwidth}
        \centering
        \includegraphics[width=\linewidth]{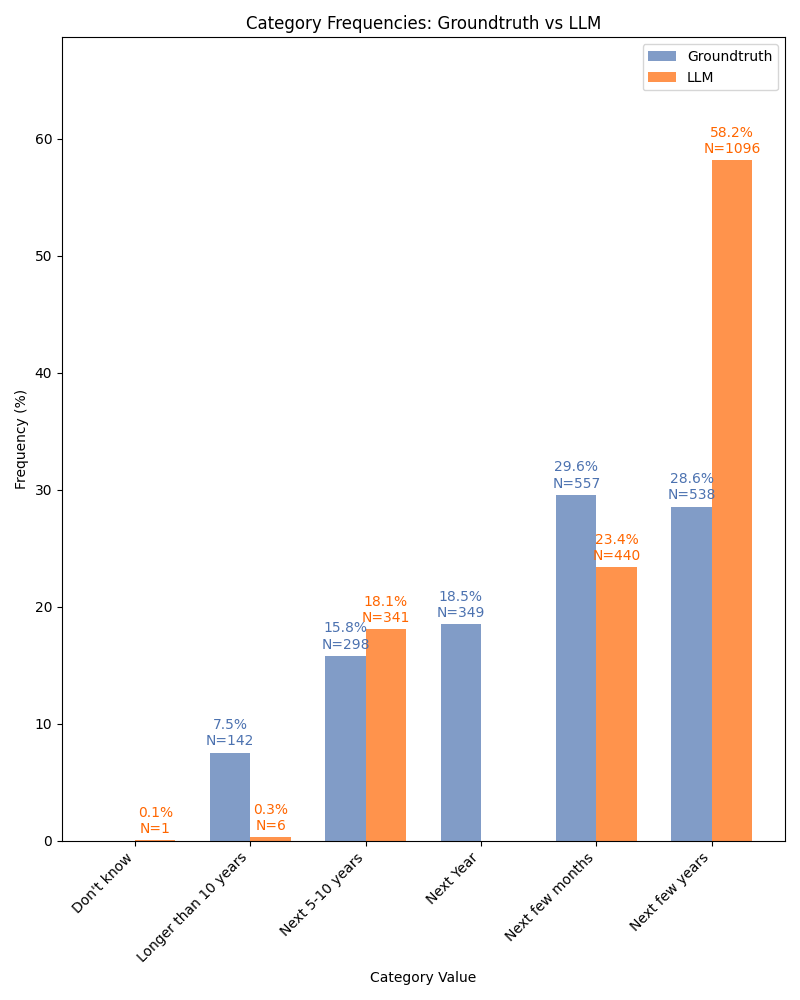}
        \caption*{(b) Survey-anchored agents.}
        \label{fig:ex111_full}
    \end{minipage}
    \caption{
        SHARE-FTP-03 — \enquote{\textit{In planning your saving and spending, which of the following time period is most important to you?}}.
        Comparison of predicted response frequencies from demographics-only agents (left) and survey-anchored agents (right) against human responses.
        Demographics-only agents (a) produce narrow predictions, largely concentrated on \enquote{Next few years} and \enquote{Next 5-10 years}, and fail to reflect the diversity of human responses.
        Survey-anchored agents (b) improve distributional alignment and better reflect the prevalence of \enquote{Next few months} and \enquote{Next few years} answers.
        However, they overpredict this option and substantially underpredict the \enquote{Next year} response, which is also frequent in the ground truth.
        Both models show limited sensitivity to differences between nearby time categories.
    }
    \label{fig:ex111_figures}
\end{figure*}

Beyond central tendency bias and hyper-accuracy, we observe a third systematic limitation: difficulties with temporal reasoning. For the time horizon question (SHARE-FTP03: \textit{\enquote{In planning your saving and spending, which of the following time period is most important to you?}}), both agent types show limited sensitivity to differences between nearby time categories (Figure \ref{fig:ex111_figures}).

Demographics-only agents (Figure \ref{fig:ex111_figures}a) produce narrow predictions largely concentrated on \enquote{Next few years} and \enquote{Next 5-10 years} failing to reflect the diversity of human responses. Survey-anchored agents (Figure \ref{fig:ex111_figures}b) improve distributional alignment but still overpredict \enquote{Next few years} and substantially underpredict \enquote{Next year}—which is the third most frequent human response.

These results may suggest that LLMs face systematic challenges in distinguishing between time intervals, a limitation we explore further in the country-level evaluation (Section~\ref{subsec:EXP2}).

\subsubsection{Robustness and Generalization}
\label{subsubsection:robustness}

\begin{figure*}[ht]
    \centering
    \begin{minipage}[t]{0.45\textwidth}
        \centering
        \includegraphics[width=\linewidth]{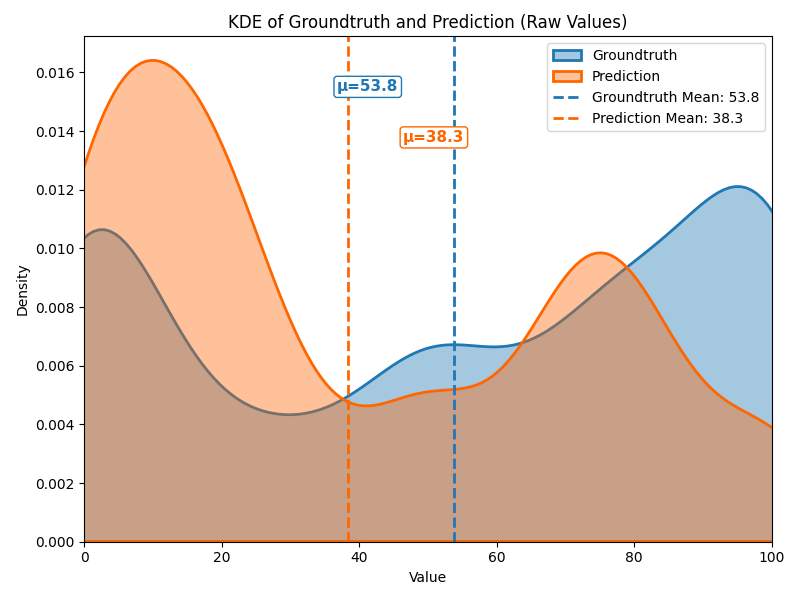}
        \caption*{(a) [Answer from 0 to 100, 0 (you are certain you will not reach that age) and 100 (you are certain you will live to that age or more.)]}
        \label{fig:subfig1}
    \end{minipage}
    \hspace{0.04\textwidth}
    \begin{minipage}[t]{0.45\textwidth}
        \centering
        \includegraphics[width=\linewidth]{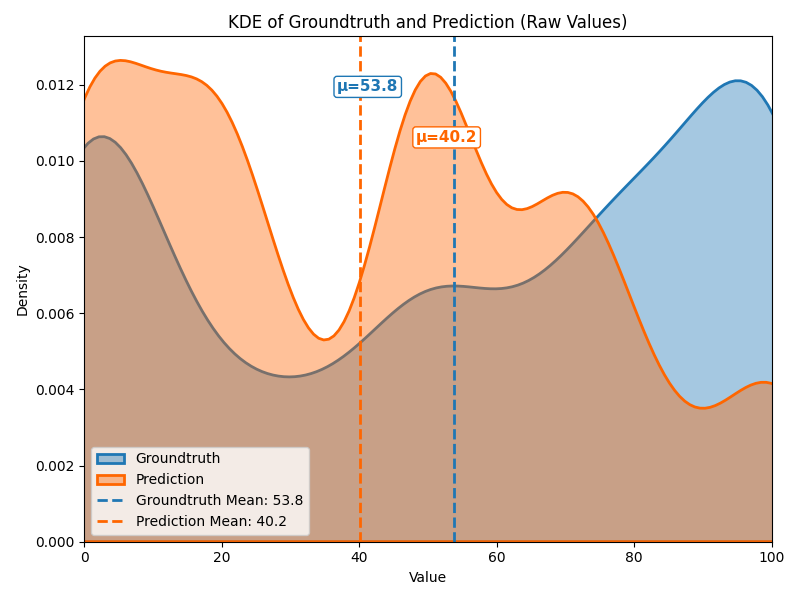}
        \caption*{(b) Chances that you will live to age XX? [0,10,20,30,40,50,60,70,80,90,100]}
        \label{fig:subfig2}
    \end{minipage}

    \vspace{0.02\textwidth}

    \begin{minipage}[t]{0.45\textwidth}
        \centering
        \includegraphics[width=\linewidth]{compl_agentscont_ex025_kde_raw_gt_vs_pred.png}
        \caption*{(c) [Answer from 0 to 100, (0 means you will not be working full-time after that age) and 100 (you will not be working full-time after that age).]}
        \label{fig:subfig3}
    \end{minipage}
    \hspace{0.04\textwidth}
    \begin{minipage}[t]{0.45\textwidth}
        \centering
        \includegraphics[width=\linewidth]{compl_agentsdiscr_ex025_kde_raw_gt_vs_pred.png}
        \caption*{(d) Chances to work full-time after 63? [0,10,20,30,40,50,60,70,80,90,100]}
        \label{fig:subfig4}
    \end{minipage}

    \caption{Asking the LLM to answer a continuous interval (a,c) vs discrete set of choices (b,d) changes the distribution of the answers. Continuous answer outperforms discrete choices.}
    \label{fig:sensitivy_format_answers}
\end{figure*}

\paragraph{Prompt sensitivity.}
On numerical questions, model performance proved highly sensitive to the presentation format of response options. Figure \ref{fig:sensitivy_format_answers} compares results when models were instructed to select from discrete values (0, 10, 20, …, 100) versus providing an open response on the interval 0–100. In the open-response setting (Figure \ref{fig:sensitivy_format_answers}a,c), predictions yield smoother distributions but underestimate the mean. In the discrete-response setting (Figure \ref{fig:sensitivy_format_answers}b,d), predictions become more concentrated around focal points, yet the underestimation bias persists. These differences highlight that model behavior depends not only on question content but also on response format.

\begin{figure*}[h]
\centering
\begin{tikzpicture}
\begin{axis}[
    xlabel={Pearson Correlation},
    ylabel={MMLU PRO},
    title={Model Performance Comparison},
    grid=major,
    width=12cm,
    height=8cm,
    enlargelimits=true,
    scatter/classes={
        a={mark=*,blue}
    },
    legend pos=north west,
    every axis plot/.append style={only marks}
]
\addplot[
    mark=*,
    mark size=4pt,
    color=blue,
    nodes near coords,
    point meta=explicit symbolic,
    every node near coord/.append style={font=\small, anchor=south west}
] coordinates {
    (0.343, 0.478) [Gemma3:4b]
    (0.368, 0.563) [Gemma3:12b]
    (0.394, 0.565) []
    (0.506, 0.665) [Qwen3:14b]
    (0.467, 0.775) [Llama3.3:70b]
};
\node[font=\small, anchor=south west, yshift=8pt,color=blue] at (axis cs:0.394, 0.565) {Llama3.1:8b};
\end{axis}
\end{tikzpicture}
\caption{Pearson Correlation on question SHARE-FTP01 vs Performance on MMLU PRO for various models. Exploratory preliminary tests indicate a positive relationship: higher MMLU-PRO performance tend to exhibit stronger correspondence with survey-anchored agents performance.}
\label{fig:comparison_llms_mmlu_fpt01}
\end{figure*}
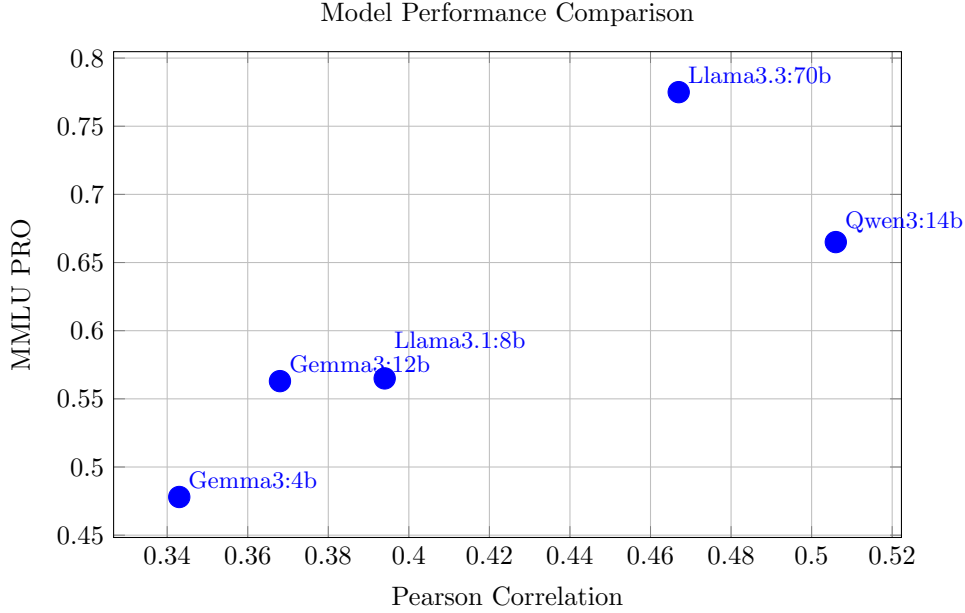

\paragraph{Model comparison.}
We conducted main experiments with Qwen3-14B and ran exploratory tests with additional local LLMs. Figure \ref{fig:comparison_llms_mmlu_fpt01} shows the association between model accuracy on MMLU-PRO and their alignment with responses to SHARE-FTP01. The results indicate a positive relationship: models with higher MMLU-PRO performance tend to exhibit stronger correspondence with survey-based behavioral measures. Detailed tests with Gemma3 \cite{team2025gemma} and LLaMA 3.1 \cite{grattafiori2024llama} reported similar results to Qwen3 (see Appendix \ref{app:resulst_gemma3_llama3}).

\paragraph{Multiple runs.}
For each question and participant, we queried the LLM a single time. We verified that conducting multiple runs with majority voting would not have materially changed outcomes (see Appendix \ref{app:robustness_experiments} for details).


\subsection{Country-level evaluation: Eurobarometer Experiments}
\label{subsec:EXP2}

The previous experiment observed accuracy rates
at the level of individual agents. But many
descriptive studies focus on averages over large
groups, rather than individual-level accuracy. So,
this second experiment test how well demographic
vs. survey agents can predict the results of
country-level surveys (we aggregated Eurobarometer data from three countries: Germany, France and Spain.


\subsubsection{Methodology}

The Eurobarometer serves as the polling tool utilized by the European Commission to track public opinion trends in Europe. We selected questions from Flash Eurobarometer 525~\cite{doi/10.2874/956514} titled ``Monitoring the level of financial literacy in the EU'' (Table \ref{tab:EUROBAR_questions}). The survey targeted 1000 individuals aged 18 and over in each country. Several items from the SHARE Cognitive Function section were excluded from survey-anchored agent profiles to prevent information leakage (see Appendix \ref{appendix:excluded_questions_information_leakage}). We used data from France, Germany and Spain.

\begin{table*}[]
\centering
\begin{tabular}{|l|p{7.5cm}|l|}
\hline
\textbf{Question Code} & \textbf{Question Text} & \textbf{Dimension} \\
\hline
EUBAR-FTP01 & If you lost your main source of income today, how long could you continue to cover your living expenses, without borrowing any money or moving house? & Future Time Perspective\\
\hline
EUBAR-FTP02 & Overall, how confident are you that you will have enough money to live comfortably throughout your retirement years? & Future Time Perspective\\
\hline
EUBAR-FRT01 & To what extent do you agree or disagree with the following statement? Before I buy something, I carefully consider whether I can afford it. & Financial Risk Tolerance
\\
\hline
EUBAR-FK01 & Imagine that someone puts €100 into a savings account with a guaranteed interest rate of 2\% per year. They don't make any further payments into this account and they don't withdraw any money. How much would be in the account at the end of five years? & Financial Knowledge\\
\hline
EUBAR-FK02 & Now imagine the following situation. You are going to be given a gift of €1,000 in one year and, over that year, inflation stays at 2\%. In one year's time, with the €1,000, will you be able to buy: & Financial Knowledge\\
\hline\end{tabular}
\caption{Selected questions from Flash Eurobarometer 525 and associated dimensions of interest.}
\label{tab:EUROBAR_questions}
\end{table*}

\subsubsection{Results: Replication of Individual-Level Findings}

Although the age groups differ (Eurobarometer 18+ vs. SHARE 50+), country-level analysis corroborated key findings from the individual-level experiment.

\begin{figure*}[]
   \centering
    \begin{minipage}[t]{0.45\textwidth}
        \centering
        \includegraphics[width=\linewidth]{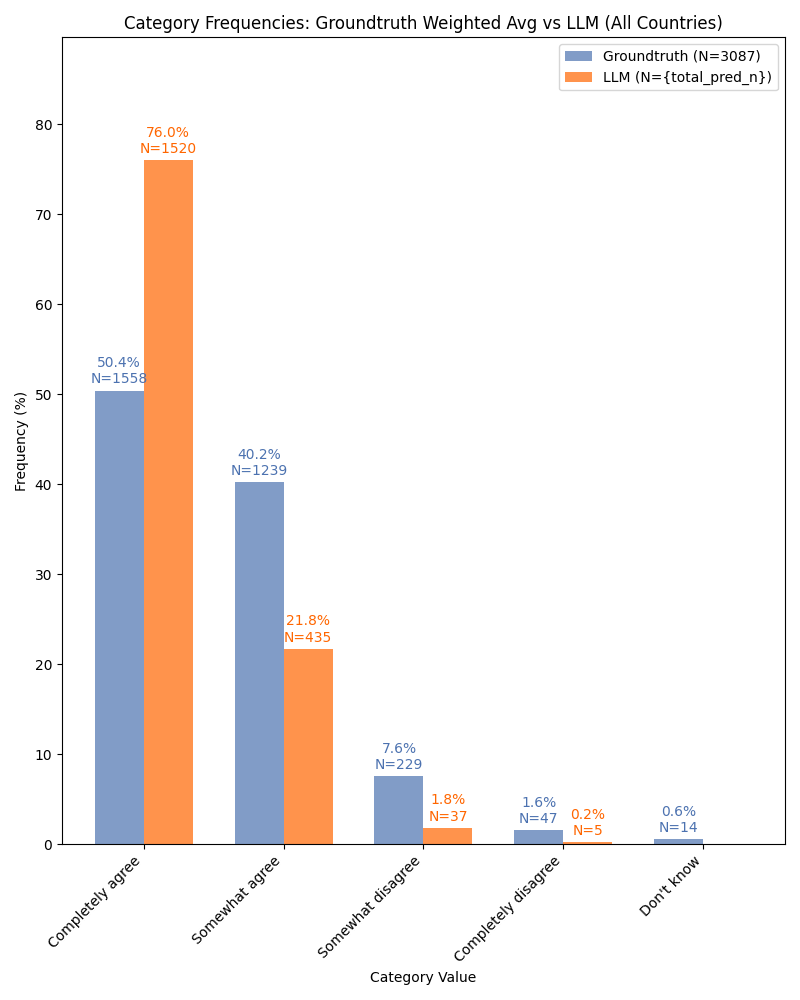}
        \caption*{(a) Demographics-only agents.}
        \label{fig:ex111_demog}
    \end{minipage}
    \hspace{0.04\textwidth}
    \begin{minipage}[t]{0.45\textwidth}
        \centering
        \includegraphics[width=\linewidth]{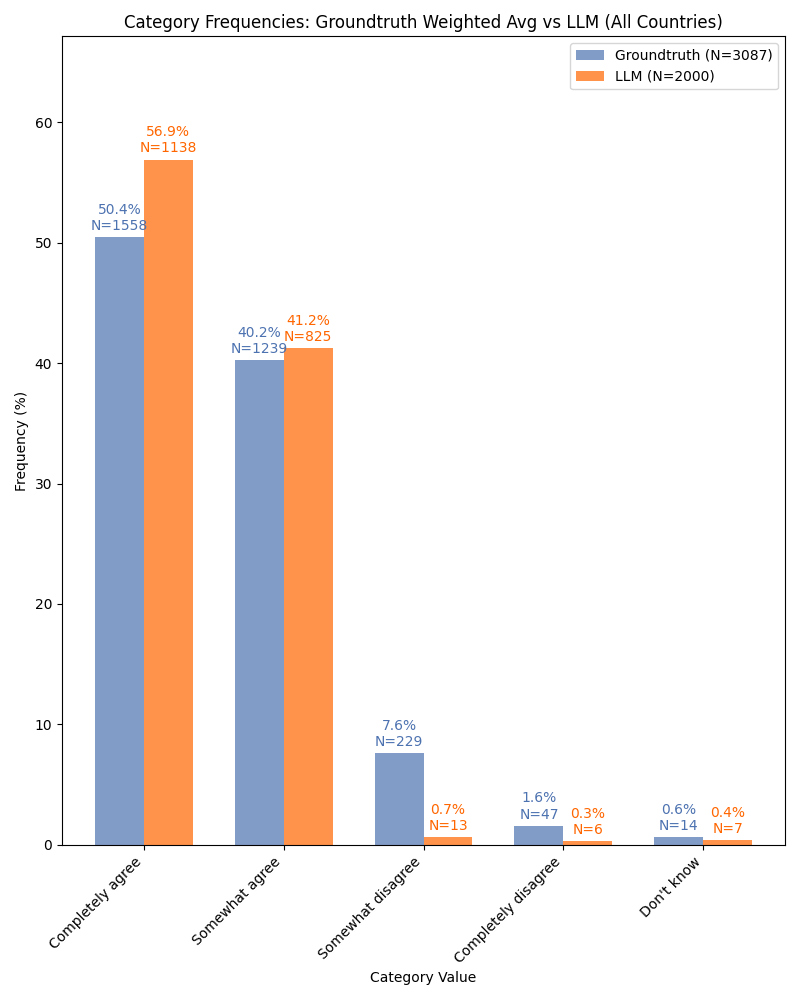}
        \caption*{(b) Survey-anchored agents.}
        \label{fig:ex111_full}
    \end{minipage}
    \caption{EUBAR-FRT01 - \enquote{\textit{To what extent do you agree or disagree with the following statements?} \textit{Before I buy something, I carefully consider whether I can afford it}}.
    Comparison of predicted response frequencies from demographics-only agents (Left) and survey-anchored agents (Right) against human responses.\newline
    Demographics-only agents exhibit response concentration bias toward the modal category, while survey-anchored agents improve category coverage. LLMs fail to reproduce the relative distribution within the disagreement categories, underestimating the proportion of “Somewhat disagree” responses compared to “Completely disagree.”}
    \label{fig:EUQ71_figures}
\end{figure*}

\paragraph{Central tendency bias}
Figure~\ref{fig:EUQ71_figures} shows results for risk tolerance (EUBAR-FRT01).
Consistent with the central tendency bias observed in SHARE, demographics-only agents exhibit \textbf{response concentration bias toward the modal category}, while survey-anchored agents demonstrate improved distributional alignment across response options. Both agents underestimate  the  distribution of answers within the disagreement categories.

Figure in Appendix \ref{fig:EUQ10_figures} (EUBAR-FTP01, \textit{if you lost your main source of income, how long could you continue to cover your living expenses?}) show same effect for both types of agents, again worse for demographics-only agents.

\begin{figure*}[]
    \centering
    \begin{minipage}[t]{0.45\textwidth}
        \centering
        \includegraphics[width=\linewidth]{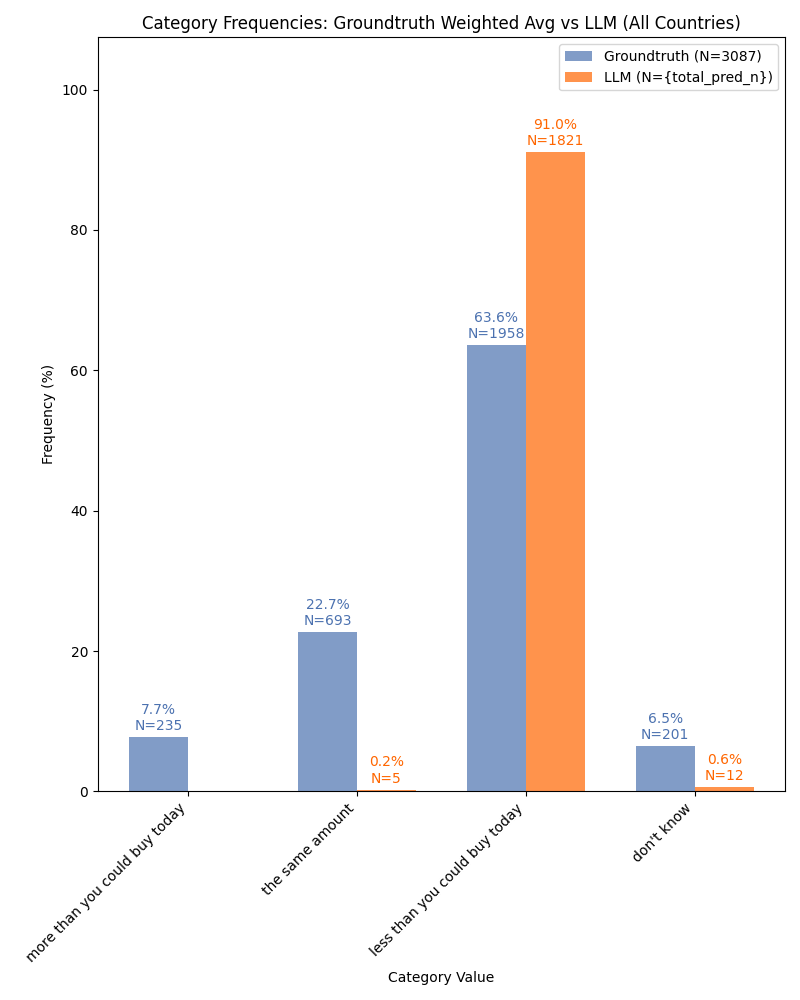}
        \caption*{(a) Demographics-only agents.}
    \end{minipage}
    \hspace{0.04\textwidth}
    \begin{minipage}[t]{0.45\textwidth}
        \centering
        \includegraphics[width=\linewidth]{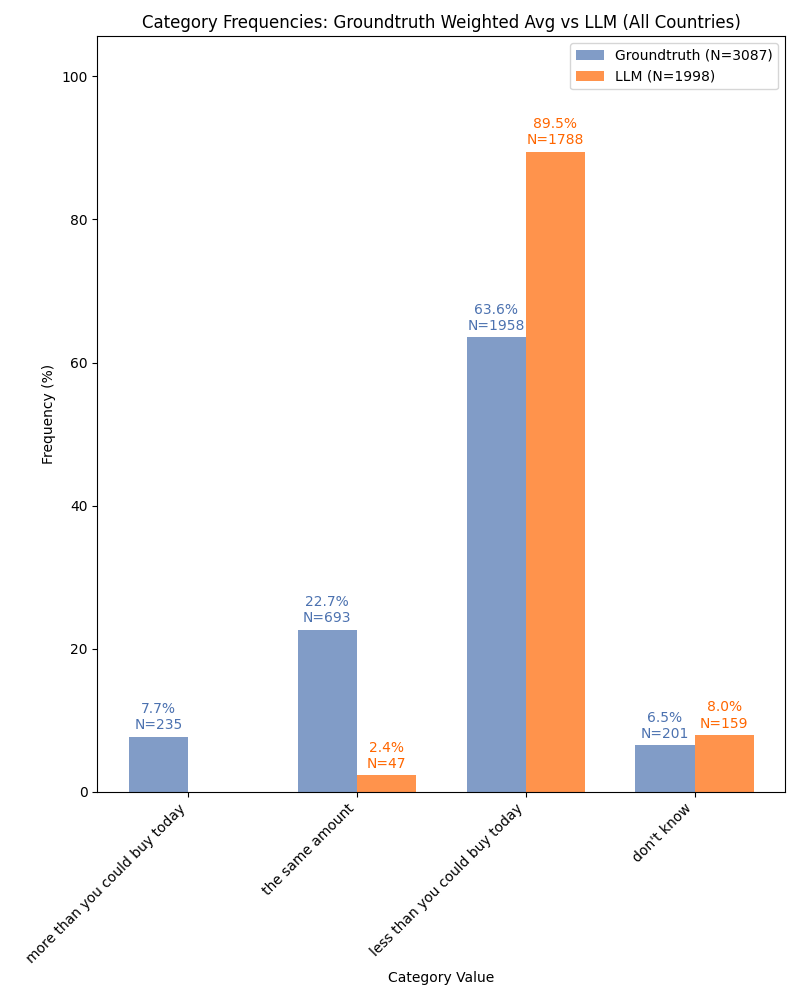}
        \caption*{(b) Survey-anchored agents.}
    \end{minipage}
    \caption{EUBAR-FK02 — \enquote{\textit{You are given a gift of €1,000 in one year and, over that year, inflation stays at 2\%. In one year’s time, with the €1,000 will you be able to buy}}. Correct answer: Less than today.\newline
    Comparison of predicted response frequencies from demographics-only agents (Left) and survey-anchored agents (Right) against human responses. Both exhibit hyperaccuracy, but survey-anchored agents reduce this effect and produce more diverse answers.}
    \label{fig:EUInflation_figures}
\end{figure*}

\paragraph{Hyper-accuracy}
 In the Financial Knowledge dimension (Figures ~\ref{fig:EUInflation_figures} and Appendix figure ~\ref{fig:EUInterest_figures}), agents outperform human respondents on both items, showing the already mentioned hyper-accuracy pattern. Survey-anchored agents reduce this effect and produce more diverse answers, but still suffer from hyper-accuracy.

\begin{figure*}[hbtp]
   \centering
    \begin{minipage}[t]{0.45\textwidth}
        \centering
        \includegraphics[width=\linewidth]{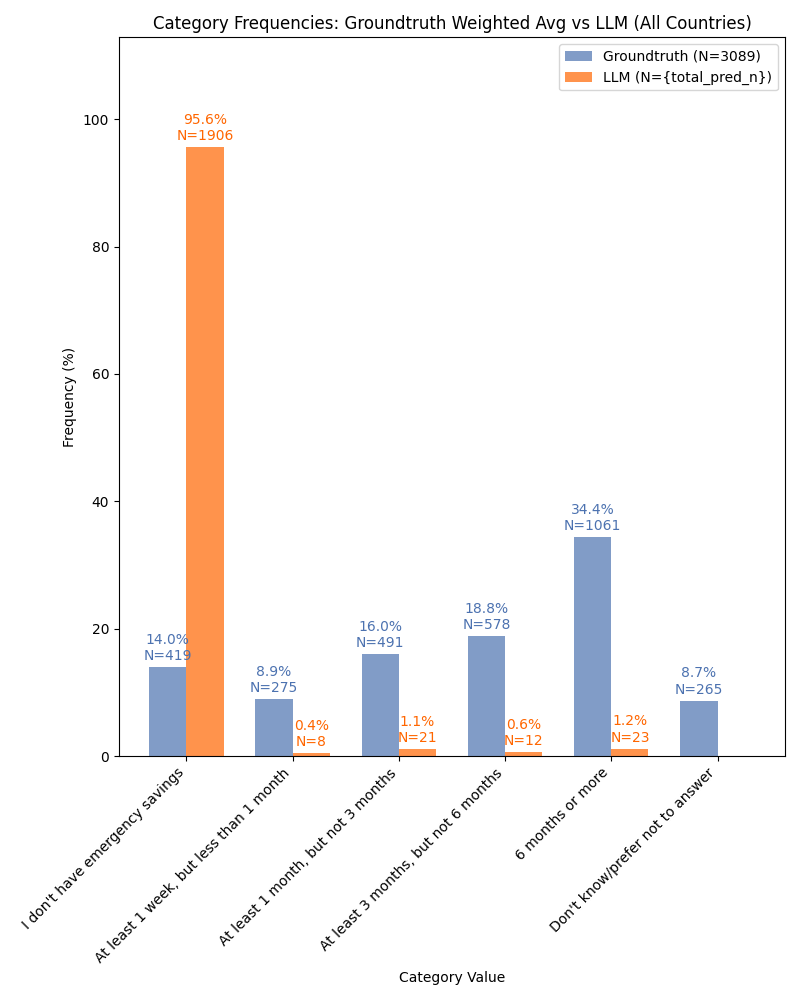}
        \caption*{(a) Demographics-only agents.}
    \end{minipage}
    \hspace{0.04\textwidth}
    \begin{minipage}[t]{0.45\textwidth}
        \centering
        \includegraphics[width=\linewidth]{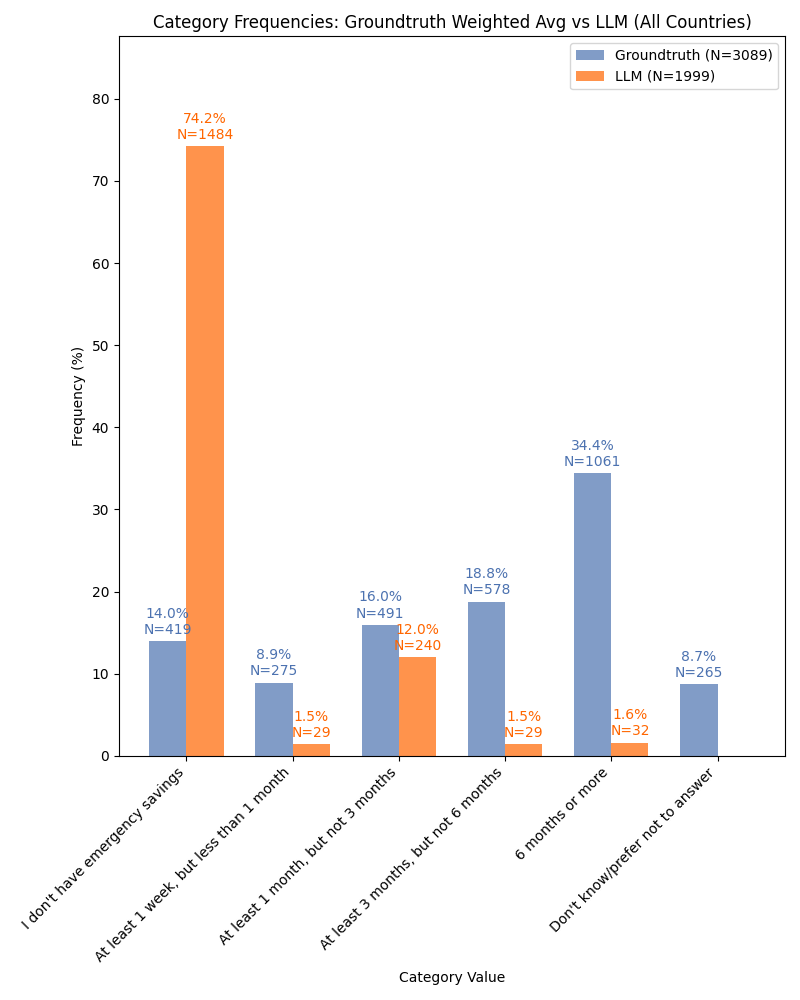}
        \caption*{(b) Survey-anchored agents.}
    \end{minipage}
    \caption{EUBAR-FTP01: \enquote{\textit{If you lost your main source of income today, how long could you continue to cover your living expenses, without borrowing any money or moving house?}} Demographics-only agents (Left) tend to focus on short-term duration categories suggesting limited temporal planning abilities. Adding context (Right) assist in transitioning to more mid-term responses, but still fail to mirror the exact proportions from the survey respondents.}
    \label{fig:EUQ8_figures}
\end{figure*}

\paragraph{Temporal reasoning limitations}
For EUBAR-FTP01 (\textit{\enquote{If you lost your main source of income, how long could you continue to cover your living expenses?}}), neither agent type captures the distribution of human responses (Figure~\ref{fig:EUQ8_figures}). This question requires reasoning about income and temporal horizons—distinguishing between one week, one month, and three months. Demographics-only agents tend to focus on short-term duration categories, while survey-anchored agents shift toward mid-term responses but still fail to mirror human proportions.

Prior work has documented that LLMs struggle with time-sensitive reasoning \cite{herel2024time,wallat2024temporal}. The TRAM benchmark \cite{wang2023tram} highlights temporal reasoning as a low-performing category for LLaMA-2 \cite{touvron2023llama}, the closest evaluated model to those used here. These findings suggest that difficulties with temporal relation comprehension underlie the weak alignment observed on time-horizon questions.

\subsection{Reproducing the hierarchical regression analysis from Jacobs-Lawson and Hershey}
\label{subsec:EXP3}

Beyond individual responses and country-level distributions, we test whether agents reproduce the \textit{internal consistency} of psychological constructs and the \textit{structural relationships} between them.

\subsubsection{Methodology}
\citet{jacobs2005influence} examined how three psychological factors---future time perspective (FTP), financial knowledge, and financial risk tolerance---influence retirement saving behaviors by surveying 270 working adults aged 25--45 in the United States. We constructed a comparable agent sample from the General Social Survey (GSS) \cite{marsden2016overview} of 2004, selecting US respondents with demographic characteristics that were similar to the original study population (see Appendix table \ref{tab:regression_analysis_population}).

\begin{table*}[]
\centering
\begin{tabular}{|p{4cm}|p{10cm}|}
\hline
\textbf{Dimension} & \textbf{Items} \\
\hline
\textbf{Knowledge of Financial Planning for Retirement (KFP)} &
1. I am very knowledgeable about financial planning for retirement. \newline
2. I know more than most people about retirement planning. \newline
3. I am very confident in my ability to do retirement planning. \newline
4. When I have a need for financial services, I know exactly where to obtain information on what to do. \newline
5. I am knowledgeable about how Social Security works. \newline
6. I am knowledgeable about how private investment plans work. \\
\hline
\textbf{Future Time Perspective (FTP)} &
1. I follow the advice to save for a rainy day. \newline
2. I enjoy thinking about how I will live years from now in the future. \newline
3. The distant future is too uncertain to plan for. (R) \newline
4. The future seems very vague and uncertain to me. (R) \newline
5. I pretty much live on a day-to-day basis. (R) \newline
6. I enjoy living for the moment and not knowing what tomorrow will bring. (R) \\
\hline
\textbf{Financial Risk Tolerance (FRT)} &
1. I am willing to risk financial losses. \newline
2. I prefer investments that have higher returns even though they are riskier. \newline
3. The overall growth potential of a retirement investment is more important than the level of risk of the investment. \newline
4. I am very willing to make risky investments to ensure financial stability in retirement. \newline
5. As a rule, I would never choose the safest investment when planning for retirement. \\
\hline
\textbf{Retirement Saving (RS)} &
1. Made meaningful contributions to a voluntary retirement savings plan. \newline
2. Relative to my peers, I have saved a great deal for retirement. \newline
3. Accumulated substantial savings for retirement. \newline
4. Made a conscious effort to save for retirement. \newline
5. Based on how I plan to live my life in retirement, I have saved accordingly. \\
\hline
\end{tabular}
\caption{Survey Questions from the Study by Jacobs-Lawson and Hershey. Participants answer between 1  (strongly disagree) and 7 (strongly agree). For each dimension, an average score was calculated. Items marked with (R) were reverse-coded prior to computing the averages.}
\label{tab:JLH_list_of_questions}
\end{table*}

The original study \cite{jacobs2005influence}  used hierarchical regression analysis to show that all three factors significantly predict saving behaviors both independently and through complex interactions, with the strongest effects occurring when individuals have high future time perspective combined with varying levels of knowledge and risk tolerance. Following their methodology, we built a population of agents that responded to each question on a 1--7 scale (1-strongly disagree, 7-strongly agree); dimension scores were computed by averaging items within each construct (reverse-coding where indicated). The list of questions used in the original research can be found in Table~\ref{tab:JLH_list_of_questions}.

We evaluate three agent conditions: \textit{demographics7-only agents} (hereafter ``Demo7'') prompted with seven demographic attributes (age, gender, income, education, marital status, employment status, and country); \textit{demographics3 agents} (hereafter ``Demo3'') using a reduced set of three attributes (age, gender, and country); and \textit{survey-anchored agents} that additionally receive responses to anchor items from the GSS (in average 465 items per agent). 


\subsubsection{Results: Distributional properties of answers by simulated agents}
\label{sec:distributional_properties}

Given that we do not have access to the participants’ responses from the original study (i.e., no ground truth is available), we examine the distributional properties of the answers by demographics-only and survey-anchored agents to characterize the nature of the synthetic populations they generate. This analysis does not require access to reference data and serves as a diagnostic of the generative process itself. Four complementary analyses are reported; all were computed across the four scales of the Jacobs-Lawson \& Hershey battery (FTP, KFP, FRT, RS).

\paragraph{Scale-level dispersion.} Table~\ref{tab:sd_comparison} reports means and standard deviations of agent-level scale scores. Demographics3 agents exhibit severely compressed variance across all scales (e.g., FTP: $\text{SD} = 0.16$; FRT: $\text{SD} = 0.28$ on a 1--7 scale), indicating near-uniform responding. Demographics7 agents show substantially greater dispersion (FTP: $\text{SD} = 0.52$; FRT: $\text{SD} = 0.84$), but remain narrower than survey-anchored agents for three of four scales (FTP: $0.52$ vs.\ $0.61$; KFP: $0.74$ vs.\ $0.90$; RS: $1.03$ vs.\ $1.23$). The exception is FRT, where Demographics7 produces wider dispersion ($0.84$ vs.\ $0.68$). This pattern of compressed scale-level variance in demographics-only agents is consistent with the central tendency bias identified in previous experiments (Section~\ref{subsec:EXP1} and Section \ref{subsec:EXP2}): the model defaults to prototypical response levels rather than generating the full range of individual differences.

\begin{figure}[t]
    \centering
     \includegraphics[width=\columnwidth]{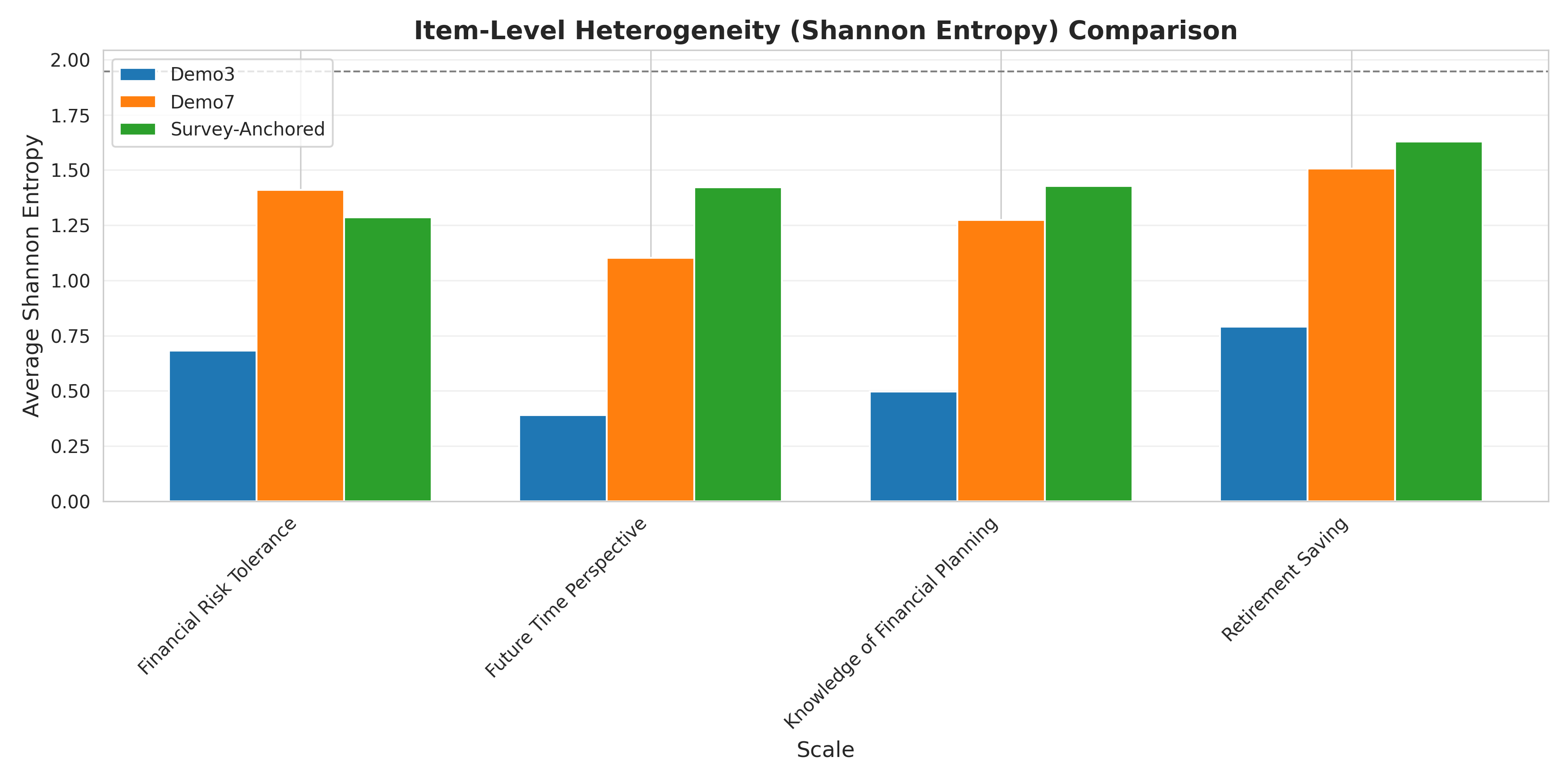}
     \caption{Average item-level Shannon entropy by condition and scale (maximum entropy for a 7-point scale: $H_{\max} = 1.95$). Survey-anchored agents show higher entropy for three out of four scales, indicating more dispersed item-level response distributions.}
     \label{fig:entropy_comparison}
 \end{figure}

\paragraph{Item-level entropy.} To examine response concentration at a finer granularity, we computed Shannon entropy $H = -\sum_{v=1}^{7} p_v \log_2 p_v$ for each item, where $p_v$ is the proportion of agents selecting response value $v$, and averaged across items within each scale ($H_{\max} = \log_2 7 = 1.95$ for a 7-point scale). As shown in figure~\ref{fig:entropy_comparison}, Demographics7 agents exhibit lower entropy than survey-anchored agents for three of four scales (FTP: $H = 1.10$ vs.\ $1.42$; KFP: $1.28$ vs.\ $1.43$; RS: $1.51$ vs.\ $1.63$), indicating more peaked item-level response distributions. The exception is again FRT ($1.41$ vs.\ $1.29$), consistent with that scale's wider dispersion under Demographics7. Demographics3 agents show extreme concentration (mean entropy $= 0.58$ across scales), confirming that minimal demographic conditioning produces highly stereotypical item-level responses .

\begin{figure}[t]
     \centering
     \includegraphics[width=\columnwidth]{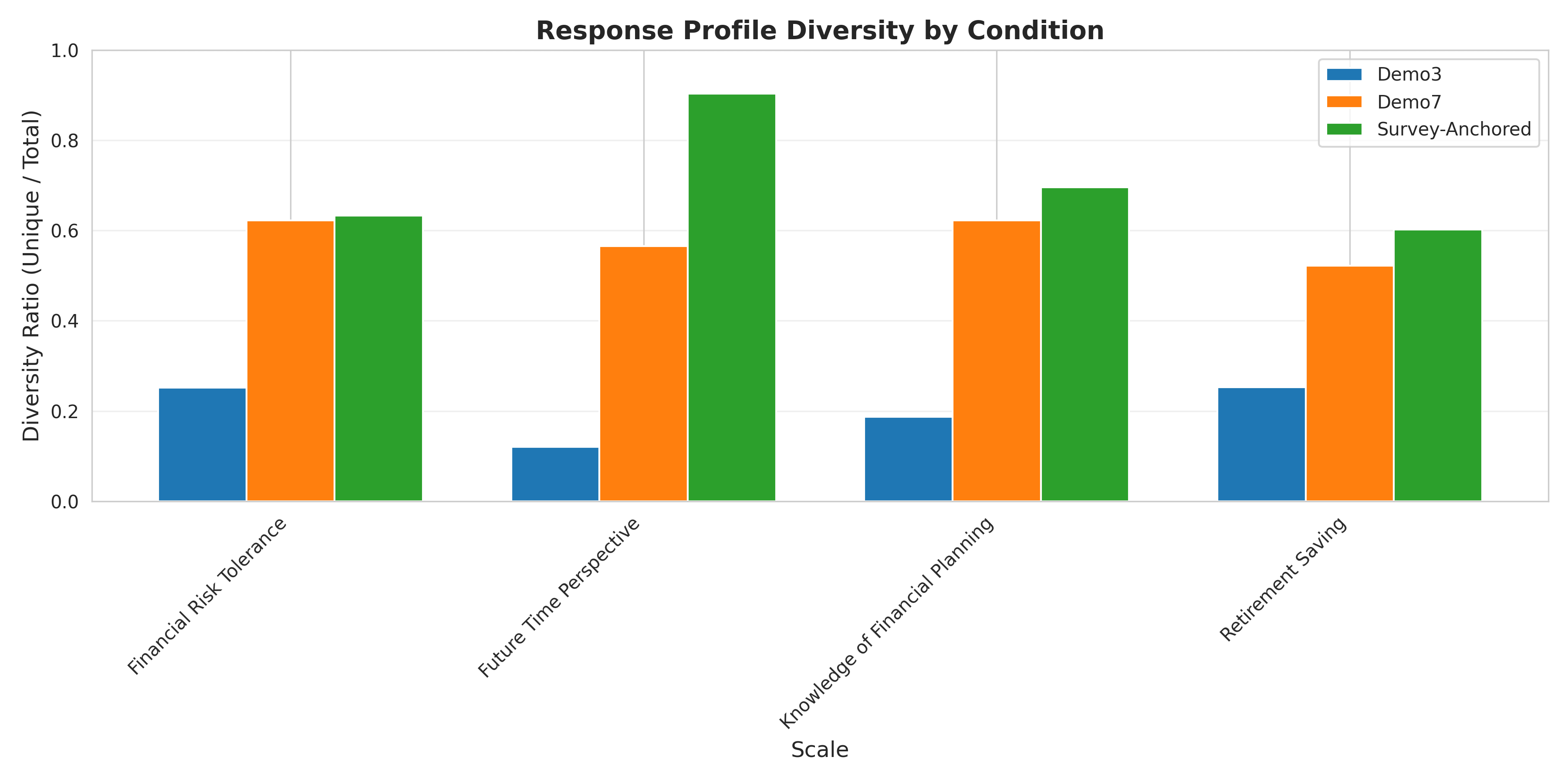}
     \caption{Response profile diversity ratio (unique profiles / total agents) by condition and scale. Survey-anchored agents produce consistently more unique item-level response patterns, particularly for FTP.}
     \label{fig:diversity_comparison}
 \end{figure}

\paragraph{Response profile diversity.} Each agent's response to a scale constitutes a $k$-dimensional integer vector (e.g., a 6-tuple for FTP with values in $\{1,\ldots,7\}$). We counted the number of distinct vectors per condition and computed the diversity ratio (unique profiles / total agents). \textit{Survey-anchored agents} yield higher diversity ratios than \textit{Demographics7 agents} across all four scales (Table~\ref{tab:profile_diversity}; Figure~\ref{fig:diversity_comparison}), though the gap varies: FTP shows the largest difference ($0.90$ vs.\ $0.57$), followed by RS ($0.60$ vs.\ $0.52$) and KFP ($0.70$ vs.\ $0.62$), while FRT is nearly matched ($0.63$ vs.\ $0.62$). The concentration of Demographics7 profiles is also reflected in top-profile coverage: for FTP, the 10 most frequent Demographics7 profiles account for 31.8\% of all responses versus 8.9\% for survey-anchored; for KFP, 26.7\% versus 17.8\% . This pattern mirrors the entropy results---the scales where survey-anchored agents show the greatest entropy advantage are also those with the largest diversity gap.

\begin{figure}[t]
     \centering
     \includegraphics[width=\columnwidth]{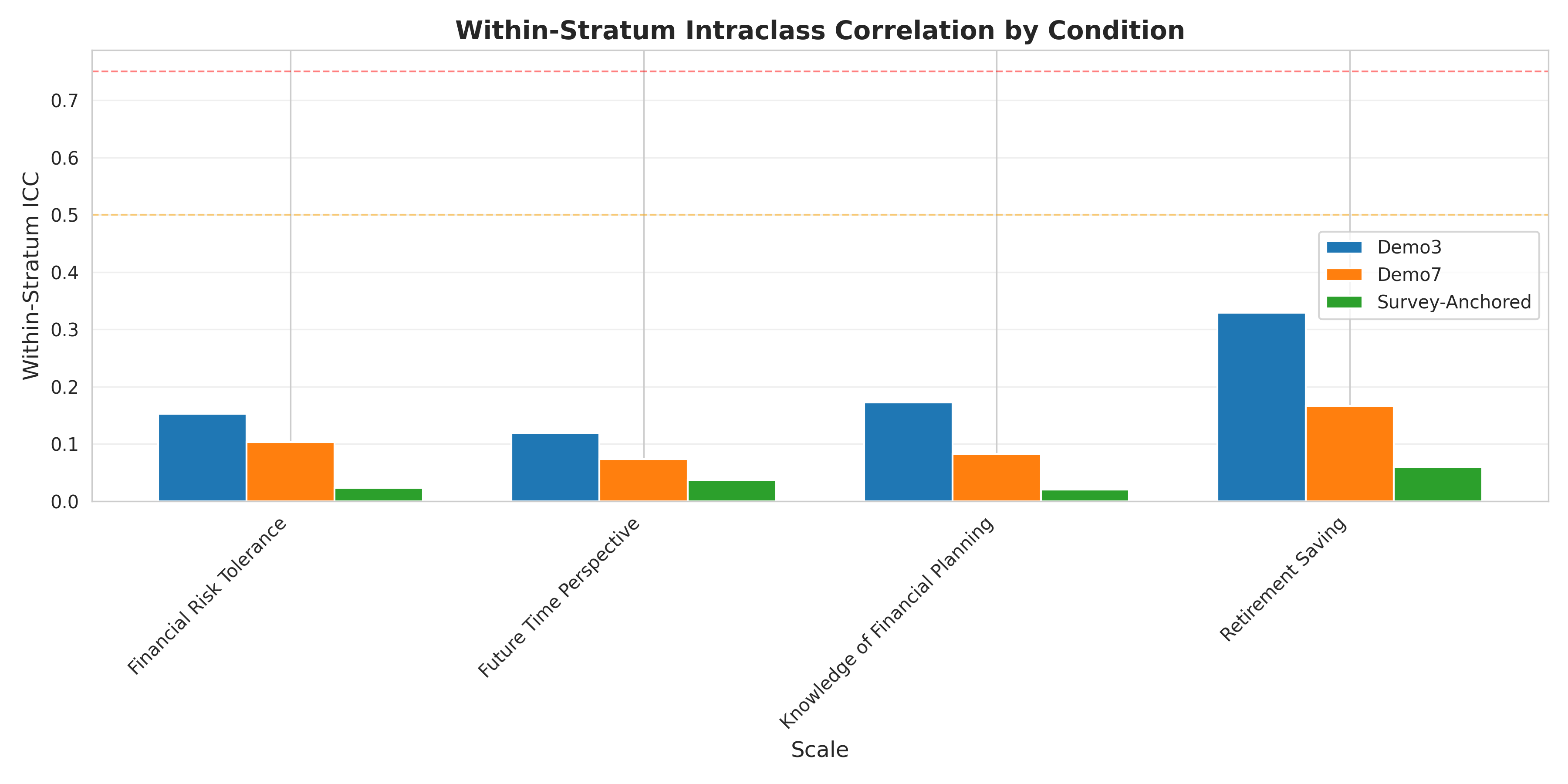}
     \caption{Within-stratum intraclass correlation (ICC) by condition and scale. Higher ICC indicates greater response homogeneity among agents sharing the same demographic stratum. Demographics-only agents show elevated ICC across all scales, consistent with demographic determinism.}
     \label{fig:icc_comparison}
 \end{figure}

\paragraph{Within-stratum intraclass correlation.} To assess the degree to which demographic attributes deterministically shape responses, agents were grouped into strata defined by age and gender categories. We computed the one-way random-effects intraclass correlation ICC(1) for scale mean scores, where $\text{ICC} = (\text{MS}_{\text{between}} - \text{MS}_{\text{within}}) / (\text{MS}_{\text{between}} + (n_g - 1)\,\text{MS}_{\text{within}})$ and $n_g$ is the average stratum size. Higher ICC values indicate that agents sharing a demographic stratum respond more similarly. Demographics7 agents show consistently higher ICC than survey-anchored agents across all scales (RS: $0.17$ vs.\ $0.06$; KFP: $0.08$ vs.\ $0.02$; FTP: $0.07$ vs.\ $0.04$; FRT: $0.10$ vs.\ $0.02$), indicating greater demographic determinism---a few demographic features rigidly constrain the response pattern. Survey-anchored agents, grounded in individual-level survey data, produce responses that vary more within demographic strata, as real humans do (Figure~\ref{fig:icc_comparison}).

\paragraph{Summary.} Across all four diagnostic analyses, demographics-only agents exhibit signatures of central tendency bias: compressed scale-level variance, lower item-level entropy, fewer unique response profiles, and higher within-stratum homogeneity. These findings converge with the results in sections \ref{subsec:EXP1} and \ref{subsec:EXP2} and indicate that the model generates internally plausible but undifferentiated response profiles when conditioned solely on demographic attributes. Survey-anchored agents mitigate these pathologies, producing more heterogeneous and individuated response patterns, though the degree of improvement is scale-dependent.


 \begin{table}[ht]
     \centering
     \renewcommand{\arraystretch}{1.3}
     \begin{tabular}{l|cc|cc|cc}
     \hline
     & \multicolumn{2}{c|}{\textbf{Demo7}} & \multicolumn{2}{c|}{\textbf{Survey-Anch.}} & \multicolumn{2}{c}{\textbf{Demo3}} \\
     \cmidrule{2-3}\cmidrule{4-5}\cmidrule{6-7}
     \textbf{Scale} & Mean & SD & Mean & SD & Mean & SD \\
     \hline
     FTP & 4.01 & 0.52 & 4.52 & 0.61 & 4.10 & 0.16 \\
     KFP & 3.51 & 0.74 & 3.42 & 0.90 & 3.98 & 0.23 \\
     FRT & 2.85 & 0.84 & 2.47 & 0.68 & 3.97 & 0.28 \\
     RS  & 2.96 & 1.03 & 3.09 & 1.23 & 3.82 & 0.45 \\
     \hline
     \end{tabular}
     \caption{Scale-level descriptive statistics across conditions. Demographics3 agents show severely compressed standard deviations across all scales. Demographics7 agents show narrower dispersion than survey-anchored agents for FTP, KFP, and RS; FRT is the exception.}
     \label{tab:sd_comparison}
\end{table}



\begin{table*}[ht]
    \centering
    \renewcommand{\arraystretch}{1.3}
    \begin{tabular}{l|ccc|ccc|ccc}
    \hline
    & \multicolumn{3}{c|}{\textbf{Demographics7}}
    & \multicolumn{3}{c|}{\textbf{Survey-Anchored}}
    & \multicolumn{3}{c}{\textbf{Demographics3}} \\
    \cmidrule{2-4} \cmidrule{5-7} \cmidrule{8-10}
    \textbf{Scale}
      & Unique & Div.\ ratio & Top-10
      & Unique & Div.\ ratio & Top-10
      & Unique & Div.\ ratio & Top-10 \\
    \hline
    FTP & 151 / 267 & 0.57 & 31.8\% & 244 / 270 & 0.90 &  8.9\% &  27 / 224 & 0.12 & 88.4\% \\
    KFP & 168 / 270 & 0.62 & 26.7\% & 188 / 270 & 0.70 & 17.8\% &  49 / 262 & 0.19 & 78.6\% \\
    FRT & 168 / 270 & 0.62 & 24.1\% & 169 / 267 & 0.63 & 27.0\% &  58 / 230 & 0.25 & 70.0\% \\
    RS  & 141 / 270 & 0.52 & 25.2\% & 162 / 269 & 0.60 & 21.6\% &  65 / 257 & 0.25 & 71.6\% \\
    \hline
    \end{tabular}
    \caption{Response profile diversity across conditions. \emph{Unique}: number of distinct item-level response vectors out of total agents. \emph{Div.\ ratio}: unique profiles divided by total agents. \emph{Top-10}: percentage of all agents accounted for by the 10 most frequent profiles. Higher diversity ratios and lower top-10 coverage indicate more individuated response generation. \newline Survey-anchored agents yield higher
diversity ratios than Demographics7 agents across all four scales}
    \label{tab:profile_diversity}
\end{table*}

\subsubsection{Results: Hierarchical regression analysis.}
Table~\ref{tab:regression_comparison_4col} presents the hierarchical regression results across conditions. All conditions recover Financial Planning Knowledge as the strongest predictor of retirement saving, consistent with the original study ($\beta = 0.51$). However, the conditions diverge in important ways.

\textit{Main effects (Level 1).} For Demo7 agents, the pattern of main effects closely approximates the original: Knowledge ($\beta = 0.60$), FTP ($\beta = 0.19$), and Risk Tolerance ($\beta = 0.16$) are all significant and in the expected direction, with relative magnitudes that mirror the original ($\beta = 0.51$, $0.25$, and $0.16$, respectively). Survey-anchored agents amplify the role of Knowledge ($\beta = 0.83$), retain a small but significant FTP effect ($\beta = 0.09$), but render Risk Tolerance non-significant ($\beta = 0.05$, $p = 0.13$). Demo3 agents are the only condition to produce a sign reversal: Risk Tolerance exhibits a negative relationship with saving ($\beta = -0.11$, $p = 0.05$), contradicting the original finding.

\textit{Interaction effects (Levels 2--3).} The original study's central finding is a significant three-way interaction among Knowledge, FTP, and Risk Tolerance ($\beta = -0.19$, $p < .01$), as well as a significant two-way interaction between Risk Tolerance and FTP ($\beta = 0.13$, $p < .01$). This is where conditions diverge most sharply. Demo7 agents fail to recover \textit{any} significant interaction: the three-way term is small and non-significant ($\beta = -0.05$, $p = 0.29$), and all two-way interactions are negligible. Demo3 agents show a similar pattern, with no significant interactions. Survey-anchored agents, in contrast, recover statistically significant effects at both the two-way and three-way levels: Risk $\times$ FTP ($\beta = -0.09$, $p = .03$), Knowledge $\times$ FTP ($\beta = 0.09$, $p = .01$), and the three-way interaction ($\beta = -0.09$, $p = .01$). Notably, the Knowledge $\times$ FTP interaction reaches significance for survey-anchored agents ($p = .01$) but was null in the original study ($p = 0.96$), representing a spurious interaction not present in the human data.

\textit{Model fit.} Both Demo7 ($R^2 = 0.81$) and survey-anchored agents ($R^2 = 0.85$) yield inflated $R^2$ values relative to the original ($R^2 = 0.59$). This suggests that LLM-generated responses are more deterministic than human responses, with less unexplained variance---consistent with the broader observation that LLMs produce overly systematic response patterns~\cite{aher2023using,argyle2023out}. Demo3 agents yield the lowest $R^2$ ($0.50$), reflecting the reduced predictive structure that accompanies degraded construct reliability.

The rest of the details of the comparison can be found in the Appendix \ref{app:additional_results_regression}. 

\paragraph{Summary.}
Survey-anchored agents are the only condition to recover statistically significant two-way and three-way interactions. However, important discrepancies remain: the dominance of Financial Knowledge is amplified ($\beta = 0.83$ vs.\ $0.51$), Risk Tolerance loses its main effect significance, and the simple slopes pattern is partially inverted relative to the original. These findings indicate that survey anchoring enables LLM agents to capture construct interdependencies that demographics alone cannot recover, while also highlighting the limits of current simulation approaches in reproducing the precise magnitude and direction of higher-order psychological effects.

\begin{table*}[h]
    \centering
    \tiny
    \renewcommand{\arraystretch}{1.2}
    \resizebox{\textwidth}{!}{%
    \begin{tabular}{l|ccc|ccc|ccc|ccc}
    \hline
    & \multicolumn{3}{c|}{\textbf{Original}} & \multicolumn{3}{c|}{\textbf{Demo7}} & \multicolumn{3}{c|}{\textbf{Demo3}} & \multicolumn{3}{c}{\textbf{Survey-anch. }} \\
    \cmidrule{2-4} \cmidrule{5-7} \cmidrule{8-10} \cmidrule{11-13}
    \textbf{Variable} & \textbf{$\beta$} & \textbf{t} & \textbf{p} & \textbf{$\beta$} & \textbf{t} & \textbf{p} & \textbf{$\beta$} & \textbf{t} & \textbf{p} & \textbf{$\beta$} & \textbf{t} & \textbf{p} \\
    \hline
    \textbf{Level 1} & & & & & & & & & & & & \\
    Fin. knowledge & 0.51 & 10.01 & \textbf{0.01} & 0.60 & 10.57 & \textbf{0.00} & 0.64 & 10.44 & \textbf{0.00} & 0.83 & 22.77 & \textbf{0.00} \\
    Future time persp. & 0.25 & 5.26 & \textbf{0.01} & 0.19 & 3.57 & \textbf{0.00} & 0.18 & 3.07 & \textbf{0.00} & 0.09 & 3.08 & \textbf{0.00} \\
    Risk tolerance & 0.16 & 3.53 & \textbf{0.01} & 0.16 & 3.66 & \textbf{0.00} & -0.11 & -1.97 & \textbf{0.05} & 0.05 & 1.52 & 0.13 \\
    \hline
    \textbf{Level 2} & & & & & & & & & & & & \\
    Knowl. × FTP & 0.00 & 0.05 & 0.96 & 0.05 & 0.65 & 0.52 & -0.05 & -0.73 & 0.47 & 0.09 & 2.53 & \textbf{0.01} \\
    Knowl. × Risk & -0.07 & -1.49 & 0.14 & 0.04 & 0.50 & 0.62 & 0.04 & 0.52 & 0.60 & -0.02 & -0.64 & 0.52 \\
    Risk × FTP & 0.13 & 2.72 & \textbf{0.01} & -0.02 & -0.18 & 0.86 & -0.01 & -0.21 & 0.83 & -0.09 & -2.22 & \textbf{0.03} \\
    \hline
    \textbf{Level 3} & & & & & & & & & & & & \\
    Knowl. × FTP × Risk & -0.19 & -4.00 & \textbf{0.01} & -0.05 & -1.06 & 0.29 & -0.08 & -1.00 & 0.32 & -0.09 & -2.47 & \textbf{0.01} \\
    \hline
    $R^2$ & 0.59 & & & 0.81 & & & 0.50 & & & 0.85 & & \\
    N & 270 & & & 267 & & & 174 & & & 266 & & \\
    \hline
    \end{tabular}%
    }
    \caption{Hierarchical regression results across conditions. Standardized betas ($\beta$) shown. \newline Only the survey-anchored agents recover the three-way interaction effects. Agents based solely on demographics reproduce the main effects, and survey-anchored agents do so as well, except in the case of Risk Tolerance. }    
    \label{tab:regression_comparison_4col}
\end{table*}


\section{Discussion}\label{sec:discu}

In this paper, we compare the performance of demographic agents versus survey-anchored agents in simulating human attitudes in the pension domain. Because the present work builds on a broad multidisciplinary dataset, a similar approach could also be leveraged to study other fields of relevance to European policymakers, such as health or economy. 

The results are promising as they pave the way for a future in which generative agents can accurately and reliably simulate human attitudes in social science domains. However, some work is still needed to reach that point. In our pension application, we evidence the risk of using demographics-only LLM agents to simulate human attitudes, as they can fail to capture human heterogeneity and therefore misrepresent differences among populations. This echoes previous arguments from traditional research on retirement planning (that do not involve generative agents): studying citizens' attitudes requires a deep understanding of both individual differences and country socio-political contexts to account for the great heterogeneity of pension systems in Europe \cite{rey2018influence}. In other words, using only individual demographic data for research on pensions is a well-known concern. Despite their attractivity in terms of availability and easiness of use, demographics-only AI agents should therefore not be taken for granted. 

On the other hand, even though we show that survey-anchored agents better capture human heterogeneity than demographics-only agents, it is worth noting that both exhibit systematic limitations such as central-tendency bias and over-reasoning on numeracy items. Therefore, we advise policymakers to be very cautious about the use of generative agents (particularly demographics-only agents) to derive trustworthy and reliable user insights, as well as to take political measures based on such insights in the near future.

\subsection{Limitations}
\label{sec:limit}

One fundamental limitation is the one mentioned by \cite{horton2023largelanguagemodelssimulated}: LLMs have been trained on a corpus that is not what general humans think, but a corpus produced by a highly selected pool of \enquote{humans creating public writing, and then selected again what they choose to say}. This raises the possibility that simulations may mirror the authors' biases instead of accurately representing human behavior within the population \cite{aher2023using}. Fine-tuning on survey data could mitigate this by injecting more representative human responses.

As outlined in the methodological framework, we did not utilize the full set of available information for each participant in SHARE. This choice was made to streamline the process, as some responses were distributed across multiple columns and required additional processing. Only the target variable was removed from the agents. Correlated variables could thus still be used by both the LLMs and the supervised machine learning baselines. Moreover, only a limited number of questions from SHARE or the Eurobarometer directly mapped to our focal constructs (FTP, FRT, FK), resulting in a reduced set of targets for evaluation.

Another limitation concerns temporal reliability. In survey and behavioral research, participant responses often vary over time. We did not assess this reliability but relied on a single set of pre-existing answers. 

We are also aware of limitations in prompting. LLMs are sensitive to response order and framing, and can sometimes be induced to choose outside options. We preserved the wording of SHARE items, except for numerical questions where we allowed the model to respond on a continuous range rather than discrete values. More fundamentally, \cite{gui_challenge_2023} show that ambiguous prompting can induce confounding in LLM-simulated experiments, as unspecified variables drift with the prompt content.

Finally, our use of \emph{local} LLMs was necessitated by the restricted nature of SHARE: no respondent information may leave the secure environment. Running locally improves data governance (no third-party transmission), auditability (versioned weights and prompts), and cost/throughput for large batch simulations. The trade-off is capability: frontier, cloud-hosted models typically exhibit stronger reasoning and calibration. Prior work also notes that larger models may better match human patterns yet amplify systematic distortions (e.g., numeracy hyper accuracy) and that "wisdom-of-crowds" ensembles of simulated respondents are not uniformly superior across tasks~\cite{aher2023using}.

\subsection{Future Work}
\label{sec:futurework}

Future work should extend beyond average predictive performance to examine subgroup-level accuracy. Errors may not be evenly distributed, and identifying which demographic or socioeconomic groups are misrepresented would connect this work to fairness and equity concerns.

In the short term, bias-mitigation strategies grounded in survey methodology could be explored to address the central-tendency bias and hyper-accuracy observed in demographics-only agents. In particular, post-hoc calibration of response distributions may help restore appropriate uncertainty and produce more realistic response patterns.

In the medium term, policy scenario testing could assess whether survey-anchored agents can predict responses to hypothetical policy interventions, such as pension reforms or retirement age changes. Validation against behavioral experiments or stated-preference studies would establish their potential value for policy evaluation and public opinion analysis.

It would also be valuable to test whether current findings generalize beyond the pension domain. Constructs like health, trust, or social support may elicit different reasoning patterns from agents, revealing new strengths or limitations.

Finally, the ability to deploy larger or fine-tuned models within secure environments would enable deeper tests of alignment and behavioral fidelity, particularly under privacy constraints.

\section{Conclusion}
\label{sec:conclu}

This paper compared two approaches for constructing LLM-based agents to simulate human survey responses: demographics-only agents, built from seven standard demographic variables, and survey-anchored agents, grounded in approximately 175 structured question–answer pairs per respondent. We evaluated both approaches across three levels of analysis — individual fidelity on 15 SHARE items, country-level distributional alignment on 5 Eurobarometer questions, and construct-level replication of a hierarchical regression from prior retirement research using 22 psychometric items — totaling 42 evaluation questions spanning the constructs of future time perspective, financial risk tolerance, and financial knowledge.

Our findings identify three systematic pathologies in demographics-only agents that survey-anchored agents partially mitigate. First, demographics-only agents exhibit pronounced central tendency bias, concentrating predictions toward population means and failing to reproduce the variance and tail behavior observed in human responses. Second, they display hyper-accuracy on objective financial literacy items, returning correct answers at near-ceiling rates and failing to replicate the error patterns and "don't know" responses characteristic of real respondents. Third, both agent types show limited sensitivity to temporal distinctions, though this limitation is more pronounced for demographics-only agents.

At the construct level, demographics-only agents recover main effects in a hierarchical regression of retirement saving on psychological predictors, but fail to reproduce three-way interactions that constitute one of the main findings of the original study. Survey-anchored agents are the only condition to recover significant interaction effects, though important discrepancies in effect magnitude and direction persist.

These results carry practical implications: demographics-only agents, despite their accessibility, risk generating internally consistent but undifferentiated synthetic populations that may mislead researchers and policymakers into overestimating consensus or underestimating heterogeneity. Survey-anchored agents offer a meaningful improvement, reducing TVD by an average of 29\% across our evaluation set, but they are not without limitations — they still exhibit residual central tendency bias, hyper-accuracy, and inflated model fit relative to human data.

Our evaluation is bounded by the retirement and financial planning domain, three European countries and US, and a single family of local LLMs. Whether these patterns generalize to other survey domains, cultural contexts, or model architectures remains an open question. Nevertheless, the consistency of our findings across individual, country, and study-level analyses suggests 
that caution should be exercised when relying solely on demographic-based agents, as they may overlook nuanced individual differences and contextual factors, underscoring the need for more comprehensive approaches in future research.


\backmatter

\section*{Declarations}

\textbf{Funding} This research did not receive any grant from funding agencies.\\
\textbf{Conflict of interest} Michael Bernstein discloses that he is a co-founder of Simile AI, Inc., a company developing generative agent simulations.\\
\textbf{Competing interests} The authors declare that they have no competing interests.\\
\textbf {Ethics approval and consent to participate} The SHARE project is continuously reviewed by both primarily responsible ethics committees (University of Mannheim and Max Planck Society, Germany) as well as national ethics committees in participating countries. The reviews refer to all facets of the project, including study design and the informed consent. The reviews confirm the project to comply with important legal norms and international ethical standards. The owner of the SHARE study is the "Survey of Health, Ageing and Retirement in Europe – European Research Infrastructure Consortium (SHARE-ERIC)". After registration persons with a scientific background may use the data in the context of "scientific-use files". The secondary analysis performed here underwent no further ethical approval. Further information on the ethical approval can be found on the SHARE project web page \url{http://www.share-project.org}.\\
\textbf {Consent for publication} The authors have provided their consent to publish their research article.\\
\textbf {Data availability} This paper uses data from SHARE Wave 9 (DOI: 10.6103/SHARE.w9.900), see Börsch-Supan et al. (2013) for methodological details. The SHARE data collection has been funded by the European Commission, DG RTD through FP5 (QLK6-CT-2001-00360), FP6 (SHARE-I3: RII-CT-2006-062193, COMPARE: CIT5-CT-2005-028857, SHARELIFE: CIT4-CT-2006-028812), FP7 (SHARE-PREP: GA N°211909, SHARE-LEAP: GA N°227822, SHARE M4: GA N°261982, DASISH: GA N°283646) and Horizon 2020 (SHARE-DEV3: GA N°676536, SHARE-COHESION: GA N°870628, SERISS: GA N°654221, SSHOC: GA N°823782, SHARE-COVID19: GA N°101015924) and by DG Employment, Social Affairs \& Inclusion through VS 2015/0195, VS 2016/0135, VS 2018/0285, VS 2019/0332, VS 2020/0313, SHARE-EUCOV: GA N°101052589 and EUCOVII: GA N°101102412. Additional funding from the German Federal Ministry of Research, Technology and Space (01UW1301, 01UW1801, 01UW2202), the Max Planck Society for the Advancement of Science, the U.S. National Institute on Aging (U01\_AG09740-13S2, P01\_AG005842, P01\_AG08291, P30\_AG12815, R21\_AG025169, Y1-AG-4553-01, IAG\_BSR06-11, OGHA\_04-064, BSR12-04, R01\_AG052527-02, R01\_AG056329-02, R01\_AG063944, HHSN271201300071C, RAG052527A) and from various national funding sources is gratefully acknowledged (see \url{www.share-eric.eu}).
\\

\onecolumn
\begin{appendices}
\renewcommand{\thetable}{A\arabic{table}}
\renewcommand{\thefigure}{A\arabic{figure}}
\section{}\label{secA1}

\subsection{List of sections in the SHARE dataset}

\begin{table}[h]
\centering
\begin{minipage}{0.45\textwidth}
\centering
\begin{tabular}{|l|l|}
\hline
\textbf{Code} & \textbf{Description} \\
\hline
AC & Activities \\
AS & Assets \\
BR & Behavioral risk \\
CF & Cognitive function \\
CH & Children \\
CO & Consumption \\
DN & Demographics and networks \\
EP & Employment and pensions \\
EX & Expectations \\
FT & Financial transfers \\
GS & Gripstrength \\
\hline
\end{tabular}
\end{minipage}
\hspace*{0.05\textwidth}
\begin{minipage}{0.45\textwidth}
\centering
\begin{tabular}{|l|l|}
\hline
\textbf{Code} & \textbf{Description} \\
\hline
HC & Health care \\
HH & Household income \\
HO & Housing \\
IT & Computer Use \\
IV & Interviewer \\
LI & Linking \\
MH & Mental health \\
PH & Physical health \\
SN & Social Networks \\
SP & Social support \\
TE & Time Expenditures \\
XT & End of life interview \\
\hline
\end{tabular}
\end{minipage}
\caption{Different sections covered by the SHARE dataset. Only Interviewer and Linking were excluded.}
\label{tab:SHARE_sections}
\end{table}

\subsection{System prompt used when prompting the LLM}

\begin{tcolorbox}[colback=gray!5, colframe=gray!80, title=System Prompt]
\enquote{You are an expert behavioral analyst and survey researcher. Your task is to analyze a set of survey questions and the corresponding answers provided by a single respondent. Based on the patterns, tone, preferences, and reasoning evident in their responses, infer how this same person would likely answer a new, unseen question. Your predictions should be thoughtful, consistent with the respondent’s previous answers, and reflect their likely perspective, values, and communication style.}
\end{tcolorbox}

\subsection{Example of [SURVEY INPUT], the question/answers from each participant}

\begin{tcolorbox}[colback=gray!5, colframe=gray!80, title=[SURVEY INPUT]: List of questions/answers from a partipant made available to the LLM]
\{\enquote{country}: \enquote{Germany},
\enquote{Age}: \enquote{54},
\enquote{On a scale from 0 to 10 where 0 means completely dissatisfied and 10 means completely satisfied, how satisfied are you with your life?}: 8.0,
\enquote{How often do you think your age prevents you from doing the things you would like to do? }: \enquote{Sometimes},
\enquote{How often do you feel that what happens to you is out of your control?}: \enquote{Sometimes},
\enquote{How often do you feel left out of things? }: \enquote{Often},
\enquote{How often do you think that you can do the things that you want to do?}: \enquote{Often},
\enquote{How often do you think that family responsibilities prevent you from doing what you want to do? }: \enquote{Rarely},
... 
(Continues until having an average of 175 question/answers for each survey-anchored agent)
\end{tcolorbox}

\subsection{List of items from SHARE for demographics-only agents}
\label{app:demographics7_share_items}
The following items from SHARE were used to build the demographics-only agents with 7 demographics.

\begin{table}[ht]
    \centering
    \caption{SHARE items used to build Demographics-only agents}
    \label{tab:demographics7}
    \begin{tabular}{|l|p{10cm}|}
    \hline
        \textbf{Demographic} & \textbf{SHARE Question} \\
    \hline
        Country & country \\
    \hline
        Age & Age \\
    \hline
        Gender & Note sex of respondent from observation (ask if unsure) \\
    \hline
        Employment Status &  In general, which of the following best describes your current employment situation? \\
    \hline
        Marital Status & What is your marital status? \\
    \hline
        Income (Financial Situation) & Thinking of your household's total monthly income, would you say that your household is able to make ends meet... \\
    \hline
        Years of Full-Time Education & How many years have you been in full-time education? \\
    \hline
    \end{tabular}
\end{table}

\subsection{Implementation details of SHARE survey-anchored agents}

Given the extensive size of the SHARE Dataset, with over 3000 columns for each individual, we automated the extraction of question/answers from the questionnaires and tables. Some questions had country or language specific mapping (for example the education levels) so we translated those corresponding codes (see table \ref{tab:SHARE_special_coding} for exhaustive list of those variables). In the cases where the data was missing we omitted the question from the list for that participant. Due to the added complexity in automating these cases, some of the questions were discarded when the answers were mapped into different columns (for example question br005 - (\enquote{\textit{What did you smoke before you stopped?}}) provided the answers in columns br005d1 (cigarettes), br005d2 (pipe), br005d3 (cigars or cigarillos), br005d4 (e-cigarettes)). This limitation can be improved in a future version of the synthetic agents.

From an initial list of 3436 possible variables the resulting synthetic agents included 175 question/answers in average. Not all the synthetic agents had the same question/answers as participants refused to answer some questions or did not apply to their context.

\begin{table}[h]
\centering
\begin{tabular}{|l|l|}
\hline
\textbf{Variable Code} & \textbf{Description} \\
\hline
DN004 & Country of Birth \\
DN005 & Other Country \\
DN006 & Year to Country \\
DN007 & Citizenship \\
DN502 & When Became Citizen \\
DN503 & Nationality Since Birth \\
DN010 & Highest Education \\
DN021 & Highest Education (Partial) \\
DN051 & Highest Education of Parent \\
DN052 & Other Highest Education of Parent \\
\hline
\end{tabular}
\caption{Variables that required specific country or language mapping}
\label{tab:SHARE_special_coding}
\end{table}

\subsection{Robustness experiments}
\begin{figure*}[]
    \centering
    \begin{minipage}[b]{0.45\textwidth}
        \centering
        \includegraphics[width=\linewidth]{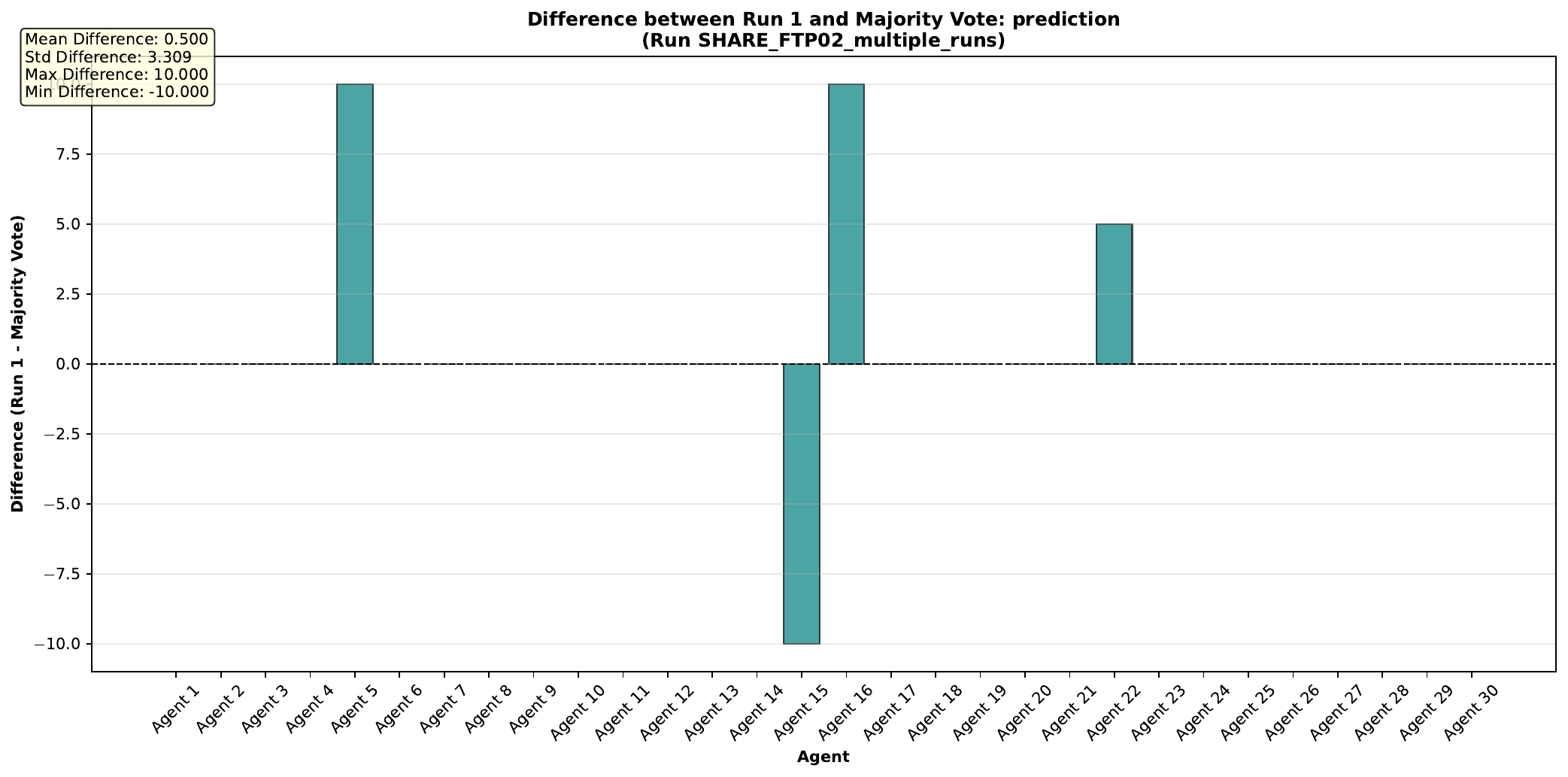}
        \caption*{(a) Numerical question.}
        \label{fig:robustness_numerical}
    \end{minipage}
    \hspace{0.04\textwidth}
    \begin{minipage}[b]{0.45\textwidth}
        \centering
        \includegraphics[width=\linewidth]{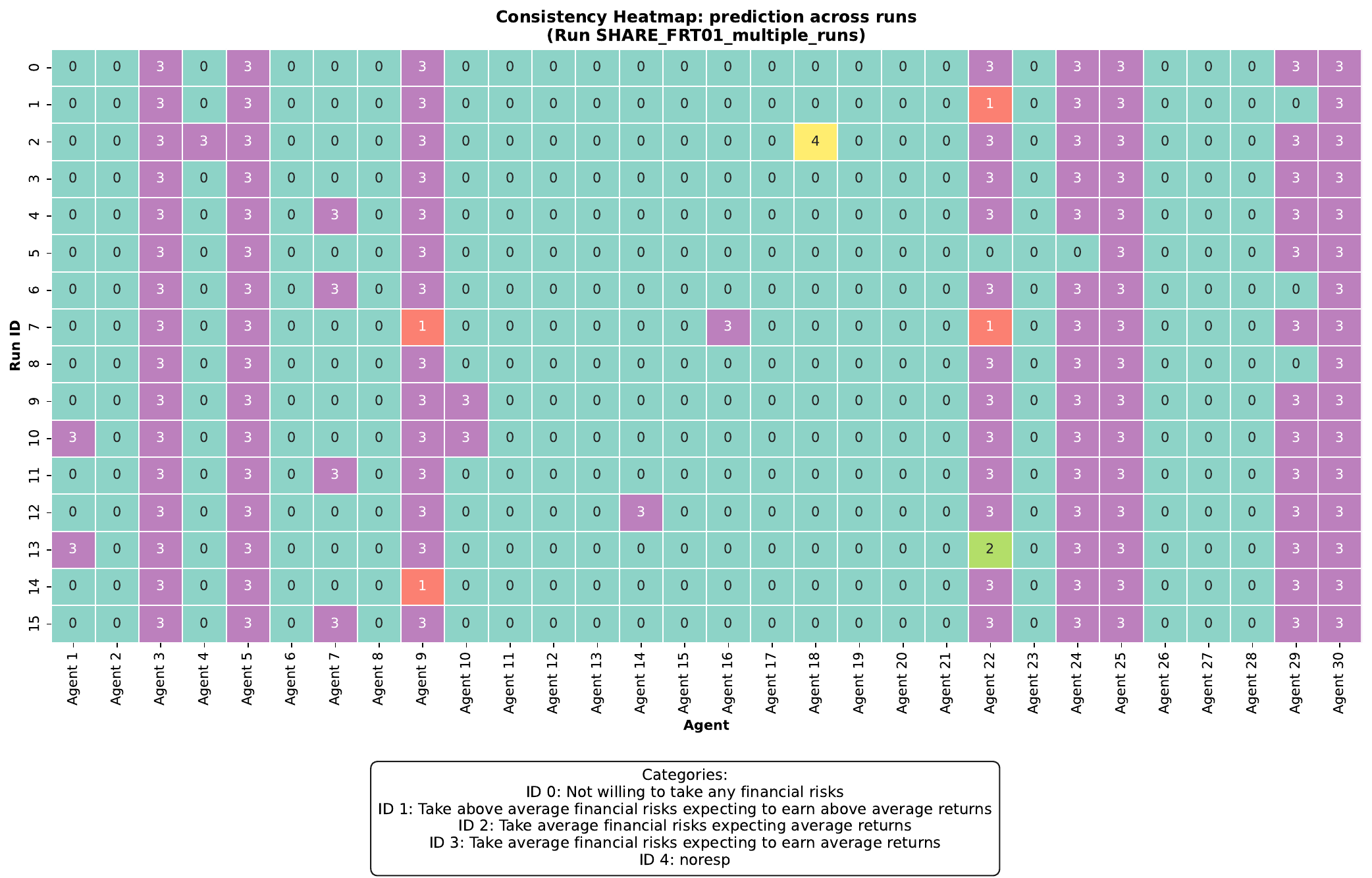}
        \caption*{(b) Categorical question.}
        \label{fig:robustness_categorical}
    \end{minipage}
    
    \caption{Robustness tests. Same question asked 15 times to same agent. Test on 30 agents (x-axis). Results on numerical (a) show the difference between single run and majority voting (y-axis difference between single run and majority voting) does not exceed 10\% in any case and is 0 for most of the agents. For categorical (b) show results of the 15 runs for the 30 agents (15 runs on y-axis). Majority voting would not have affected outcomes significantly.}
    \label{fig:robustness}
\end{figure*}

For each question and participant, we queried the LLM a single time. We verified that conducting multiple
runs with majority voting would not have materially changed the outcomes, although it would have considerably
increased the duration of the experiments. We performed robustness tests on 30 agents by running the same question 15 times. Figure \ref{fig:robustness} shows the results for both numerical and categorical values. Both sets of results demonstrate a high level of consistency.
\label{app:robustness_experiments}

\subsection{List of SHARE questions excluded to avoid information leakage}
\label{appendix:excluded_questions_information_leakage}
To prevent information leakage, SHARE questions that were semantically overlapping with the evaluation questions were excluded from the survey-anchored agent profiles. Consequently, agents did not have access to the evaluated items or near-duplicate formulations during anchoring, ensuring that performance differences cannot be explained by cross-dataset question overlap.

\begin{table}[b]
\centering
\begin{tabular}{|l|p{10cm}|}
\hline
\textbf{SHARE Code} & \textbf{Question Text} \\
\hline
cf011\_ & Next I would like to ask you some questions which assess how people use numbers in everyday life. \\
cf012\_ & If the chance of getting a disease is 10 percent, how many people out of 1000 (one thousand) would be expected to get the disease? \\
cf013\_ & In a sale, a shop is selling all items at half price. Before the sale, a sofa costs 300 [FLCurr]. How much will it cost in the sale? \\
cf014\_ & A second hand car dealer is selling a car for 6,000 [FLCurr]. This is two-thirds of what it costs new. How much did the car cost new? \\
cf015\_ & Let's say you have 2000 [FLCurr] in a savings account. The account earns ten per cent interest each year. How much would you have in the account at the end of two years? \\
cf108\_ & Now let's try some subtraction of numbers. One hundred minus 7 equals what? \\
cf109\_ & And 7 from that \\
cf110\_ & And 7 from that \\
cf111\_ & And 7 from that \\
cf112\_ & And 7 from that \\
\hline
\end{tabular}
\caption{Questions from SHARE omitted from survey-anchored agents for the EUROBAROMETER target questions}
\end{table}

\subsection{Results on additional local language models: Gemma3:12b and LLaMa 3.1:8b}
\label{app:resulst_gemma3_llama3}
\begin{figure}[H]
    \centering

    \begin{minipage}[b]{0.45\textwidth}
        \centering
        \includegraphics[width=\textwidth]{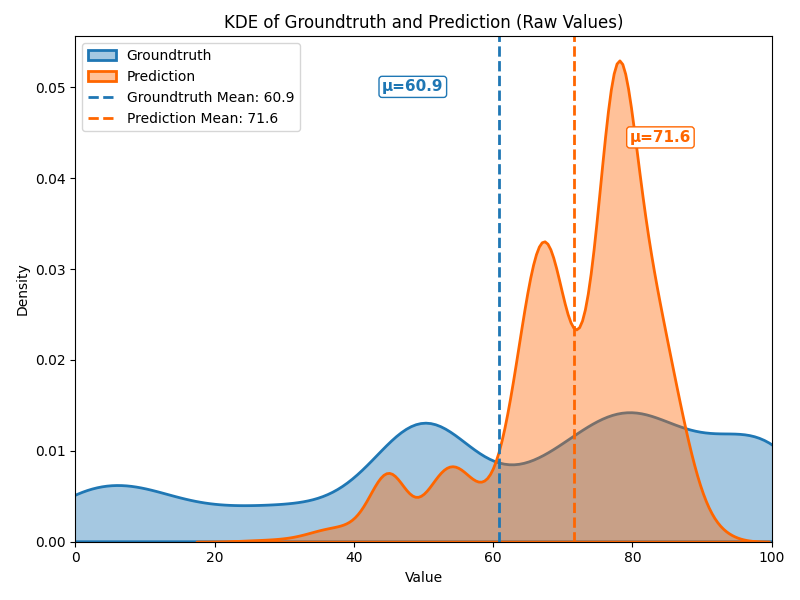}
        \textbf{Demographic-only agents (country/age/gender)}
        \label{fig:gemma3_ex009_demog_kde}
    \end{minipage}
    \hfill
    \begin{minipage}[b]{0.45\textwidth}
        \centering
        \includegraphics[width=\textwidth]{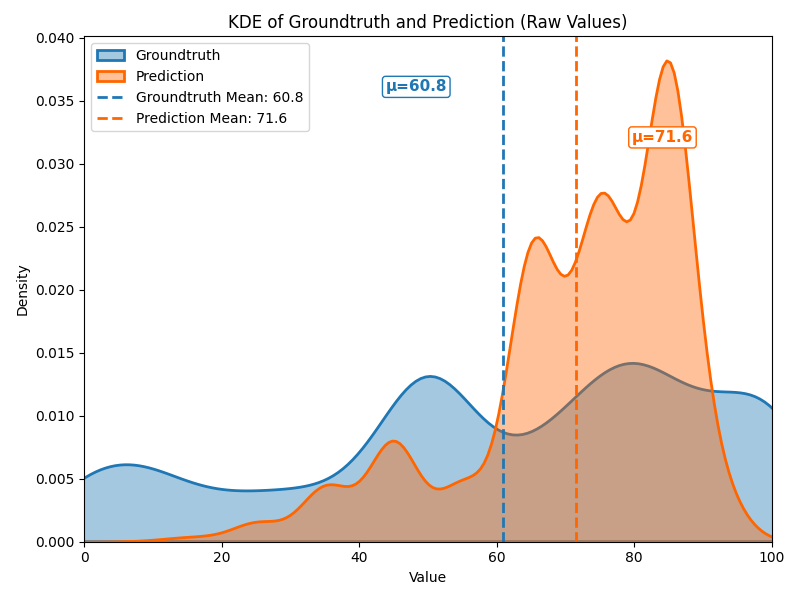}
        \textbf{Survey-anchored agents}
        \label{fig:gemma3_ex009_full_kde}
    \end{minipage}

    \vspace{0.5cm}

    \begin{minipage}[b]{0.45\textwidth}
        \centering
        \includegraphics[width=\textwidth]{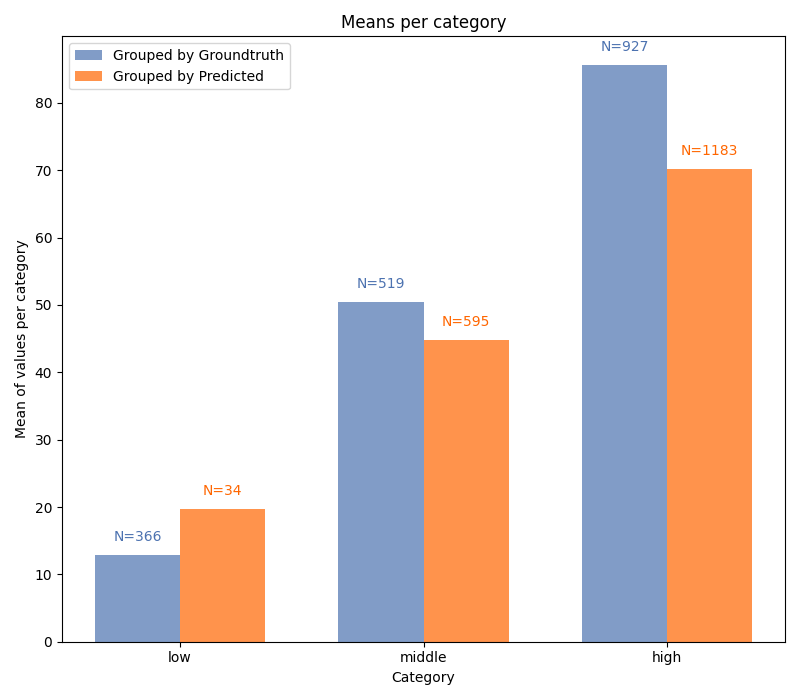}
        \textbf{Survey-anchored agents: LLM categorization fidelity}
        \label{fig:gemma3_ex009_full_barplot}
    \end{minipage}
    \hfill
    \begin{minipage}[b]{0.45\textwidth}
        \centering
        \includegraphics[width=\textwidth]{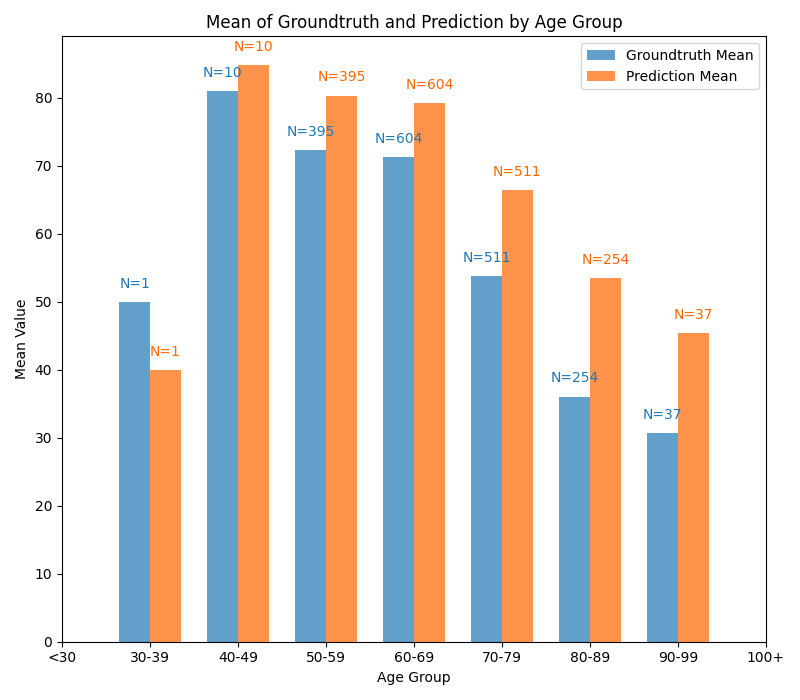}
        \textbf{Survey-anchored agents: Mean prediction by age group}
        \label{fig:gemma3_ex009_full_byages}
    \end{minipage}

    \caption{Gemma3:12b-SHARE-FTP01 — \enquote{\textit{What are the chances that you will live to age XX or more?}}. Similar findings to Qwen3.\newline
    Kernel density estimates (top row) show minimal distributional differences between agent types for this question. For LLM categorization (bottom left), category-level means are recovered more accurately in the center, though extremes are over/underestimated. Finally, LLMs correctly reproduce the monotonic decline in subjective life expectancy with increasing age (bottom right).}
    \label{fig:gemma3_ex009_figures}
\end{figure}

\begin{figure}[H]
    \centering

    \begin{minipage}[b]{0.45\textwidth}
        \centering
        \includegraphics[width=\linewidth]{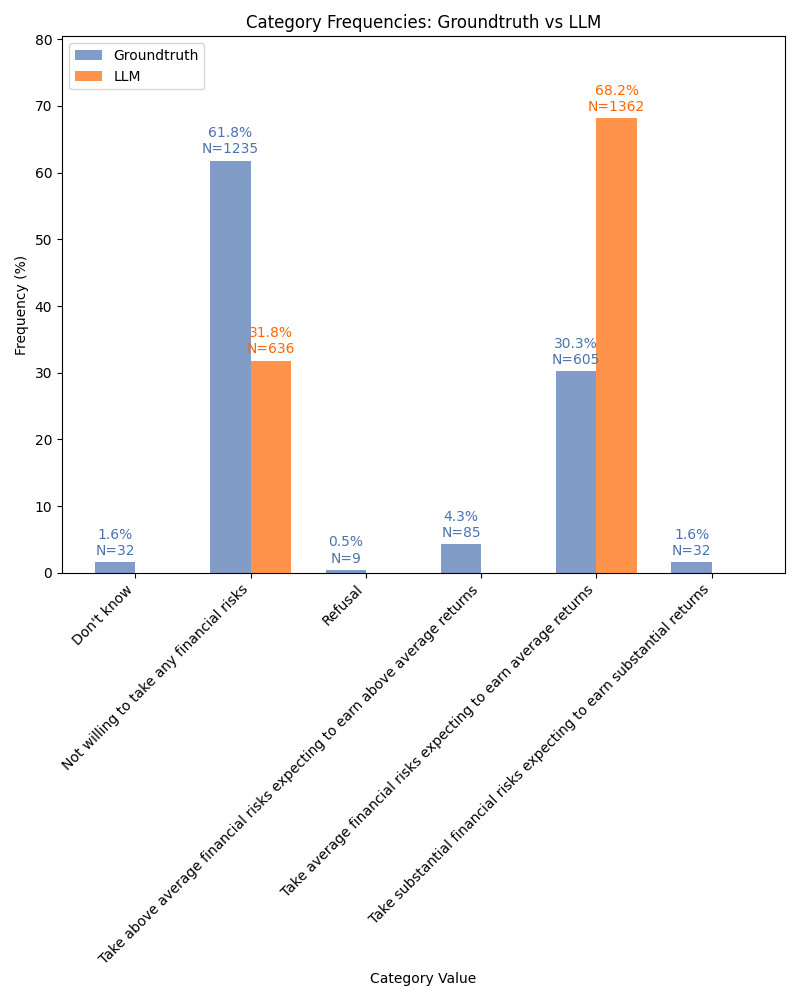}
        \textbf{Demographics-only agents (country/age/gender)}
        \label{fig:gemma3_ex110_demog}
    \end{minipage}
    \hspace{0.04\textwidth}
    \begin{minipage}[b]{0.45\textwidth}
        \centering
        \includegraphics[width=\linewidth]{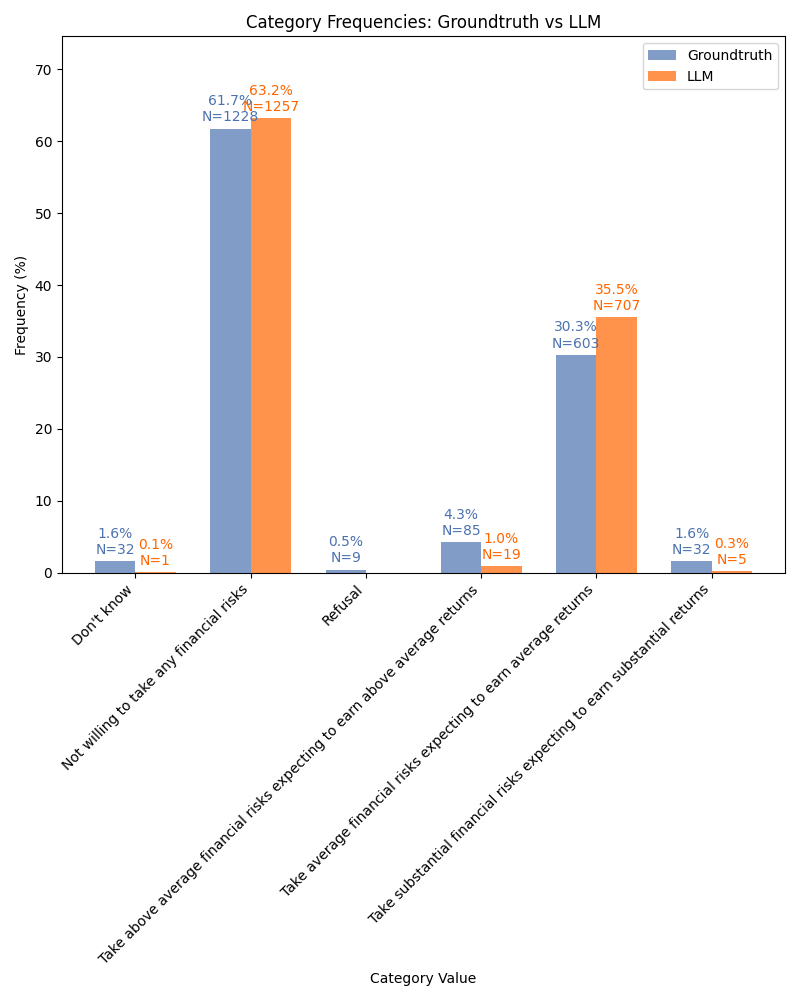}
        \textbf{Survey-anchored agents}
        \label{fig:gemma3_ex110_full}
    \end{minipage}

    \caption{Gemma3:12b-SHARE-FRT-01 — \enquote{\textit{Which amount of risk are you willing to take when you save or make investments?}}.\newline
    Gemma3:12b replicates key findings from Qwen3 14b. Demographics-only agents (left) fail to replicate the most frequent answers from human participants. Survey-anchored agents (right) replicate the most frequent human responses.}
    \label{fig:gemma3_ex110_figures}
\end{figure}

\begin{figure}[H]
    \centering

    \begin{minipage}[b]{0.45\textwidth}
        \centering
        \includegraphics[width=\textwidth]{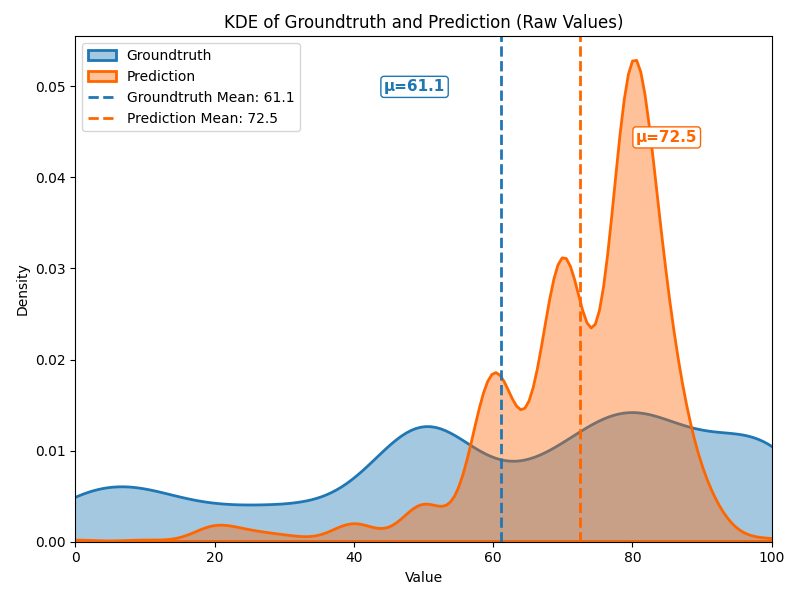}
        \textbf{Demographics-only agents (country/age/gender)}
        \label{fig:llama31_8b_ex009_demog_kde}
    \end{minipage}
    \hfill
    \begin{minipage}[b]{0.45\textwidth}
        \centering
        \includegraphics[width=\textwidth]{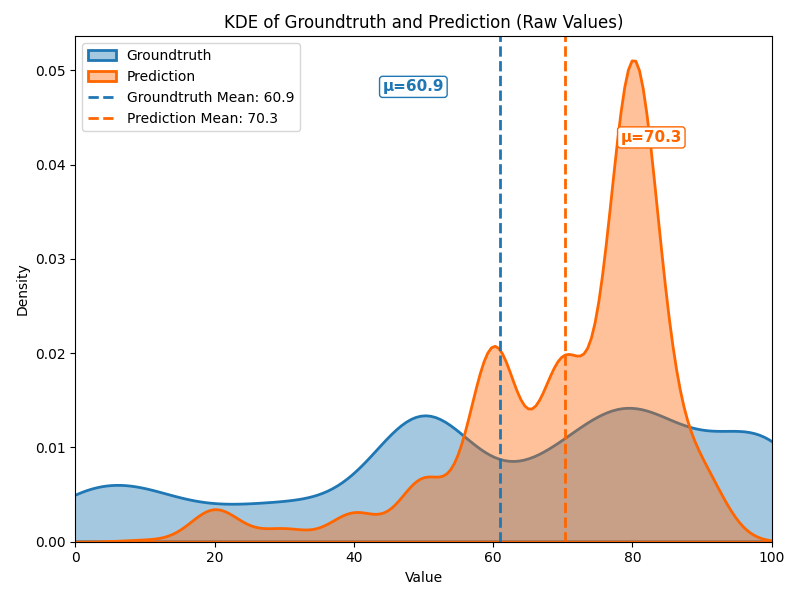}
        \textbf{Survey-anchored agents}
        \label{fig:llama31_8b_ex009_full_kde}
    \end{minipage}

    \vspace{0.5cm}

    \begin{minipage}[b]{0.45\textwidth}
        \centering
        \includegraphics[width=\textwidth]{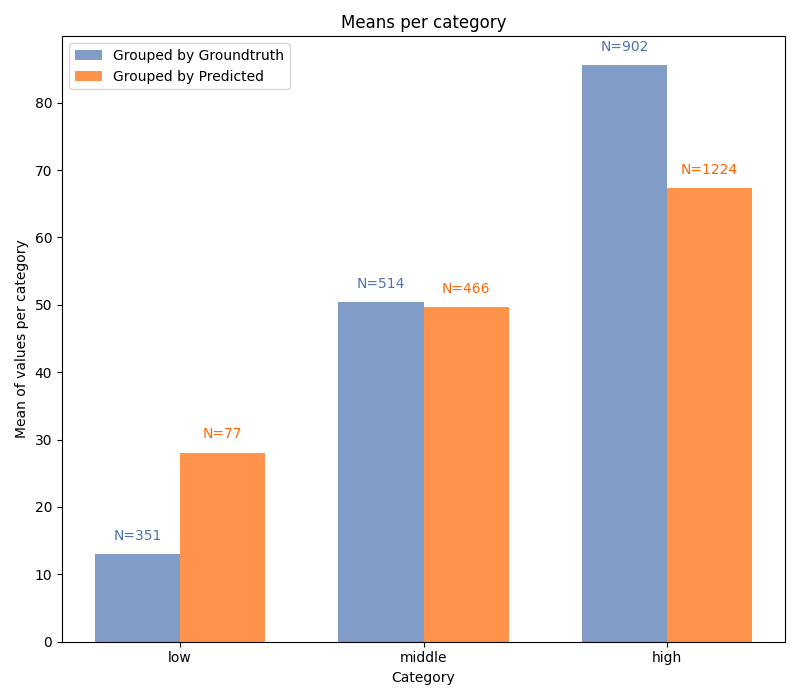}
        \textbf{Survey-anchored agents: LLM categorization fidelity}
        \label{fig:llama31_8b_ex009_full_barplot}
    \end{minipage}
    \hfill
    \begin{minipage}[b]{0.45\textwidth}
        \centering
        \includegraphics[width=\textwidth]{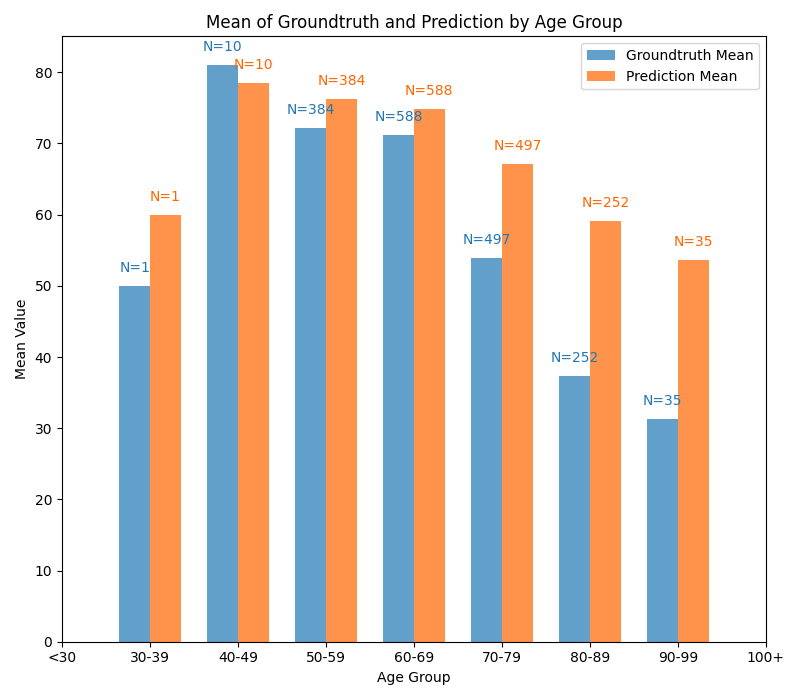}
        \textbf{Survey-anchored agents: Mean prediction by age group}
        \label{fig:llama31_8b_ex009_full_byages}
    \end{minipage}

    \caption{Llama3.1:8b-SHARE-FTP01 — \enquote{\textit{What are the chances that you will live to age XX or more?}}.\newline
    Kernel density estimates (top row) show minimal distributional differences between agent types for this question. For LLM categorization (bottom left), category-level means are recovered more accurately in the center, though extremes are over/underestimated. Finally, LLMs correctly reproduce the monotonic decline in subjective life expectancy with increasing age (bottom right).}
    \label{fig:llama31_8b_ex009_figures}
\end{figure}

\subsection{Calculation of Total Variation Distance}
\label{app:tvd}
The Total Variation Distance (TVD) \cite{gibbs2002choosing} is a statistical distance metric used to quantify the difference between two probability distributions, $P$ and $Q$, over the same measurable space $\mathcal{X}$. It is defined as half the $\ell_1$ distance between the probability measures and is always bounded between 0 and 1. A TVD of 0 indicates the distributions are identical, while a TVD of 1 indicates they have disjoint support.

\subsubsection{TVD for Categorical Variables (Discrete Distributions)}

When comparing two discrete probability distributions, $P$ and $Q$, over a finite or countably infinite set of categories $\mathcal{X} = \{x_1, x_2, \ldots\}$, the TVD is calculated by summing the absolute differences of the probability mass functions, $P(X=x)$ and $Q(X=x)$, across all categories and dividing by two:
\begin{equation}
d_{\text{TV}}(P, Q) = \frac{1}{2} \sum_{x \in \mathcal{X}} \left|P(X=x) - Q(X=x)\right|
\label{eq:tvd_discrete}
\end{equation}

\subsubsection{TVD for Numerical Variables (Continuous Distributions)}

When comparing two continuous probability distributions, $P$ and $Q$, with probability density functions (PDFs) $p(x)$ and $q(x)$, the TVD is calculated by integrating the absolute difference of the PDFs over the entire domain, $\mathcal{X}$, and dividing by two:
\begin{equation}
d_{\text{TV}}(P, Q) = \frac{1}{2} \int_{\mathcal{X}} \left|p(x) - q(x)\right| \, dx
\label{eq:tvd_continuous}
\end{equation}
This continuous form represents half the area between the two density curves. In both cases, an equivalent, foundational definition for the TVD is the maximum difference in probability that the two distributions assign to \textit{any} event $A \subseteq \mathcal{X}$: $d_{\text{TV}}(P, Q) = \sup_{A \subseteq \mathcal{X}} |P(A) - Q(A)|$.

\subsubsection{Discretized Total Variation Distance (TVD) Approximation}
\label{app:tvd_discretized}

To compute the Total Variation Distance (TVD) between two continuous (numerical) distributions, $P$ (ground truth) and $Q$ (prediction), a discrete approximation method is employed. This involves partitioning the shared data range into $K$ bins to convert the continuous distributions into discrete probability mass functions (PMFs).

\begin{enumerate}
    \item \textbf{Discretization:} A common range encompassing both grountruth and predictions, $\mathcal{R}$, is divided into $K$ equally-sized bins, $\mathcal{B} = \{b_1, b_2, \ldots, b_{K}\}$. In our implementation, $K=50$ bins were used.
    \item \textbf{PMF Generation:} Histograms for the ground truth and prediction datasets are generated using the shared bins $\mathcal{B}$. These histograms are normalized such that the sum of the bin heights equals $1$, yielding the discrete PMFs $P(b_i)$ and $Q(b_i)$.
\end{enumerate}

The TVD is then calculated using the discrete formula (Equation~\ref{eq:tvd_discrete_approx}), which measures half the sum of the absolute differences between the two PMFs across all bins:

\begin{equation}
d_{\text{TV}}(P, Q) \approx \frac{1}{2} \sum_{i=1}^{K} \left|P(b_i) - Q(b_i)\right|
\label{eq:tvd_discrete_approx}
\end{equation}

\subsection{Approximating the averages of population subgroups}
\label{appendix:Validation_LLM_Categorization}
\begin{figure}[t]
    \centering
    \includegraphics[width=0.66\textwidth]{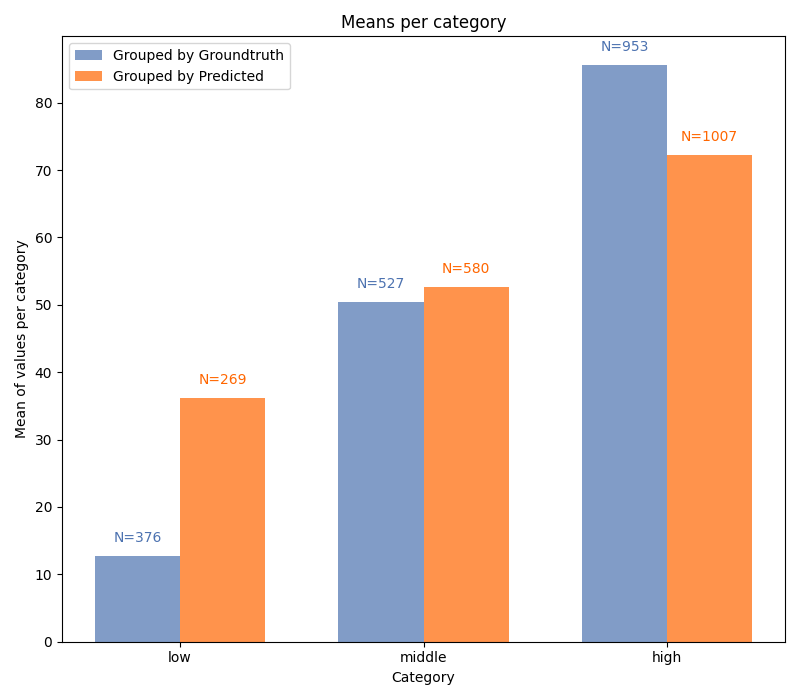}
    \caption{\small Answers were categorized into three groups (low (0 to 33.33), middle (33.34 to 66.66), high (66.67 to 100) using either ground truth responses (blue) or LLM predictions (orange), with category means computed from actual human values in both cases}
    \label{fig:appendix_ex009_demog_barplot}
\end{figure}

For numerical questions only, participant answers were categorized into three groups using either ground truth responses (blue) or LLM predictions (orange), with category means computed
from actual human values in both cases (see Figure \ref{fig:appendix_ex009_demog_barplot})
This validation process compares the grouping of human participant answers based on \textbf{ground truth (GT)} responses versus \textbf{Large Language Model (LLM)} predictions. The core goal is to determine if the LLM's categorization method successfully preserves the statistical properties (means) of the actual human values within each group.

\subsection*{Categorization of participant answers}

Participants are categorized into $K=3$ distinct groups ($\mathcal{G}_1, \mathcal{G}_2, \mathcal{G}_3$) using two separate methods:

\begin{enumerate}
    \item \textbf{Ground Truth Categorization (GT):} Participants are assigned to groups based on their actual human responses. Let $P_{i}^{\text{GT}}$ be the set of participants assigned to group $\mathcal{G}_i$ using this method.
    \item \textbf{LLM Prediction Categorization (LLM):} Participants are assigned to groups based on the categories predicted by the LLM. Let $P_{i}^{\text{LLM}}$ be the set of participants assigned to group $\mathcal{G}_i$ using this method.
\end{enumerate}
In our case, the groups are "low, middle, high" and depend on the value of the answer [0-33.33],[33.34-66.66],[66.67-100]).

\subsection*{Mean Computation}

For both categorization methods, the group mean is computed using the actual human value ($\mathcal{V}_j$) associated with each participant $j$.

Let $\mathcal{V}_j$ be the actual human value for participant $j$. The computed mean for group $\mathcal{G}_i$ for each method is defined as:

\begin{itemize}
    \item \textbf{Ground Truth Mean ($\mu_{i}^{\text{GT}}$):}
    $$
    \mu_{i}^{\text{GT}} = \frac{1}{|P_{i}^{\text{GT}}|} \sum_{j \in P_{i}^{\text{GT}}} \mathcal{V}_j
    $$
    \item \textbf{LLM Prediction Mean ($\mu_{i}^{\text{LLM}}$):}
    $$
    \mu_{i}^{\text{LLM}} = \frac{1}{|P_{i}^{\text{LLM}}|} \sum_{j \in P_{i}^{\text{LLM}}} \mathcal{V}_j
    $$
\end{itemize}
Where $|P_{i}|$ denotes the number of participants in the respective set.

\subsection*{Validation Criteria: Preservation}

The validation is successful if the group means are \textbf{preserved}, implying the categorization methods yield statistically similar group averages for the inherent human values. This condition is met when the difference between the means for each group $i$ is minimal:

$$
|\mu_{i}^{\text{GT}} - \mu_{i}^{\text{LLM}}| \approx 0 \quad \text{for } i \in \{1, 2, 3\}
$$

Visually, this is represented by comparing the bar heights for GT (blue) and LLM (orange) across all three groups ($\mathcal{G}_1, \mathcal{G}_2, \mathcal{G}_3$).

\subsection{Hyperparameters used during the experiments}

\begin{table}[htpb]
    \centering
    \begin{tabular}{|l|l|c|}
        \hline
        Category & Hyperparameter & Value \\ \hline
        Model & architecture & qwen3 \\ \hline
        Model & parameters & 14.8N \\ \hline
        Model & context length & 40960 \\ \hline
        Model & embedding length & 5120 \\ \hline
        Model & quantization & $Q4_{K_M}$ \\ \hline
        Parameters & repeat penalty & 1 \\ \hline
        Parameters & temperature & 0.6 \\ \hline
        Parameters & topk & 20 \\ \hline
        Parameters & topp & 0.95 \\ \hline
        Parameters & repeatpenalty & 1 \\ \hline
        Parameters & think & true \\ \hline
        Parameters & numctx & 3000 to 8000 \\ \hline
    \end{tabular}
    \caption{Hyperparameters used during the experiments}
    \label{tab:hyperparams}
\end{table}

\subsection{Supervised learning pipeline}
The Supervised Learning method is implemented as a pipeline that employs either a RandomForestClassifier or a RandomForestRegressor. The pre-processing and post-processing steps for every target value include:

\begin{itemize}
    \item Select the rows for the countries of interest (Spain, France, Germany).
    \item Select only the rows with non-null value in the target variable.
    \item Exclude the rows where the value of the target variable is \textit{Refusal/Don't know}.
    \item Remove columns with too many missing values (\textgreater 30\%).
    \item Implement one-hot encoding for categorical columns.
    \item Split the dataset into train, validation, and test sets (60\%, 20\%, 20\%).
    \item Run Grid Search for RandomForest hyperparameters. See Table~\ref{tab:rf_hyperparams}.
    \item Report metrics on the test set for the best model.
\end{itemize}

\begin{table}[h!]
\centering
\begin{tabular}{|l|l|}
\hline
\textbf{Random Forest Parameter} & \textbf{Values} \\ \hline
\texttt{n\_estimators} & 5,\; 10,\; 20,\; 50 \\ \hline
\texttt{max\_depth} & 3,\; 5 ,\; 7 \\ \hline
\texttt{min\_samples\_split} & 10,\; 20,\; 50 \\ \hline
\texttt{min\_samples\_leaf} & 5,\; 10,\; 20 \\ \hline
\end{tabular}
\caption{Hyperparameter grid for random forest tuning using Grid Search}
\label{tab:rf_hyperparams}
\end{table}

\newpage
\subsection{Characteristics of the population of agents used in the regression analysis}
\label{tab:regression_analysis_population}

\begin{table*}[h!]
\centering
\label{tab:demographics_comparison}
\begin{tabular}{lcc}
\hline
\textbf{Characteristic} & \textbf{Original Study (2005) by \cite{jacobs2005influence}} & \textbf{Agents Population (GSS 2004)} \\
\hline
\multicolumn{3}{l}{\textit{Sample Size}} \\
Total participants & 270 working adults & 270 \\
Sample source & National household panel & GSS 2004 survey \\
\hline
\multicolumn{3}{l}{\textit{Age Distribution}} \\
Age range (years) & 25--45 & 25--45 \\
Mean age (SD) & 36.2 (6.18) & 35.5 (4.68) \\
\hline
\multicolumn{3}{l}{\textit{Gender Distribution}} \\
Male & 154 (57.0\%) & 154 (57.0\%) \\
Female & 116 (43.0\%) & 116 (43.0\%) \\
\hline
\multicolumn{3}{l}{\textit{Education}} \\
Median years of schooling & 14.0 & 14.0 \\
\hline
\end{tabular}
\begin{flushleft}
\small
\caption{Comparison of Demographics: Original Study vs Agents Population. The original study recruited 270 working adults from a national household panel as part of a larger study on psychological determinants of retirement planning. The agents population 
was created through stratified sampling from the General Social Survey 2004 dataset to match the original study's 
demographic targets.}
\end{flushleft}
\label{tab:original_study_vs_agent_populations}
\end{table*}

\newpage

\subsection{Supplementary EUROBAROMETER questions}

\begin{figure*}[htpb]
   \centering
    \begin{minipage}[t]{0.45\textwidth}
        \centering
        \includegraphics[width=\linewidth]{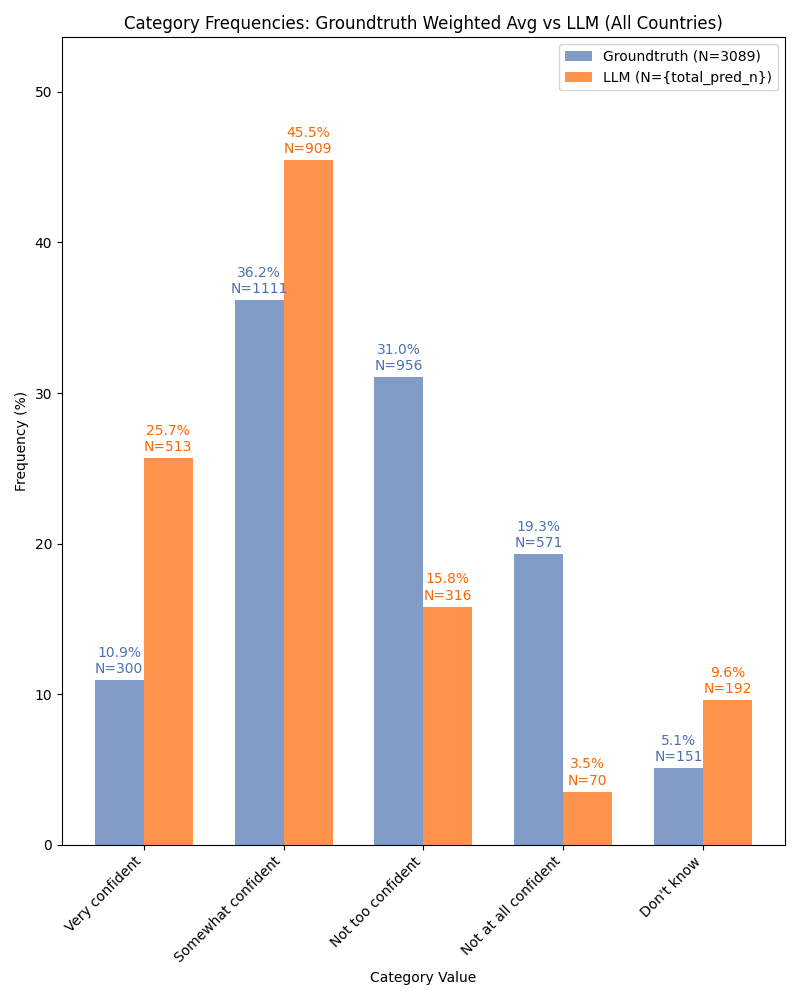}
        \caption*{(a) Demographics-only agents.}
    \end{minipage}
    \hspace{0.04\textwidth}
    \begin{minipage}[t]{0.45\textwidth}
        \centering
        \includegraphics[width=\linewidth]{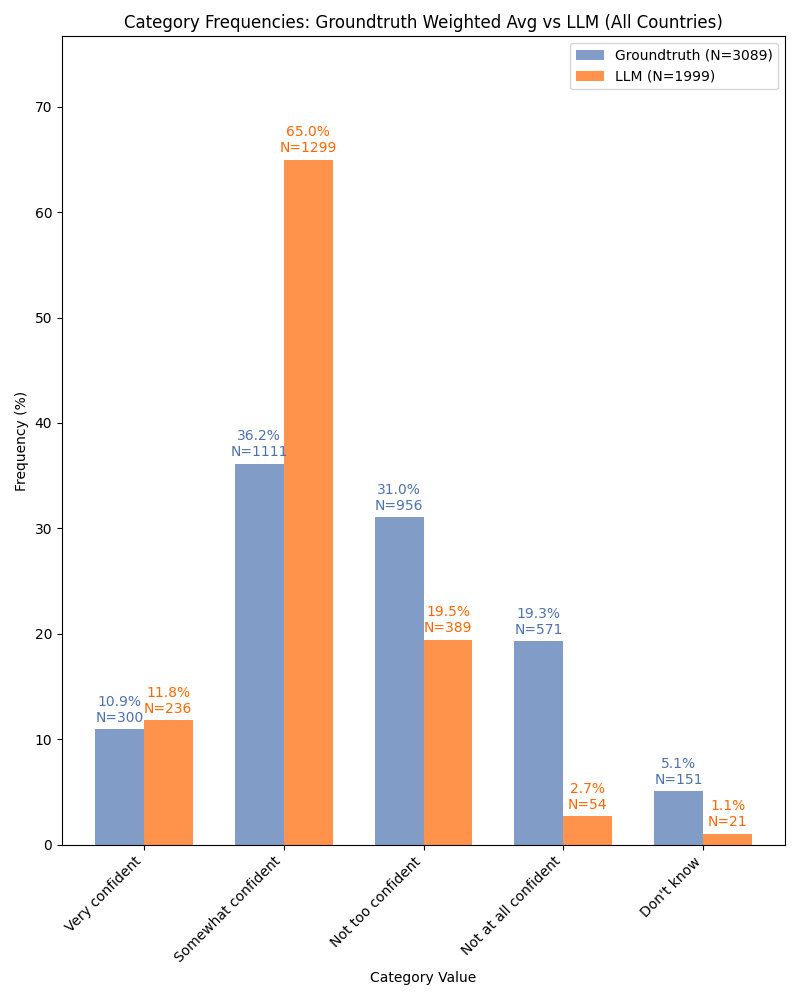}
        \caption*{(b) Survey-anchored agents.}
    \end{minipage}
    \caption{EUBAR-FTP02 - \enquote{\textit{Overall, how confident are you that you will have enough money to live comfortably throughout your retirement years?}}
    Both types of agents show a tendency to concentrate their responses around the most frequent category. In this setting, the bias is stronger for survey-anchored agents.}
    \label{fig:EUQ10_figures}
\end{figure*}

\begin{figure}[htpb]
   \centering
    \begin{minipage}[t]{0.45\textwidth}
        \centering
        \includegraphics[width=\linewidth]{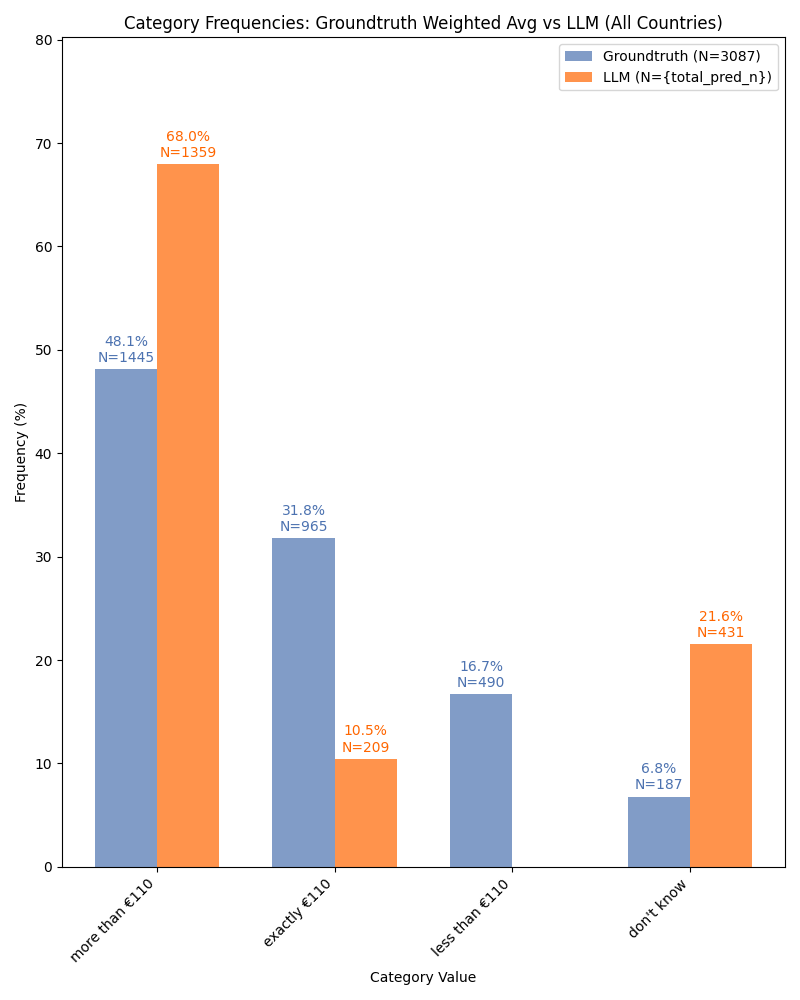}
        \caption*{(a) Demographics-only agents.}
    \end{minipage}
    \hspace{0.04\textwidth}
    \begin{minipage}[t]{0.45\textwidth}
        \centering
        \includegraphics[width=\linewidth]{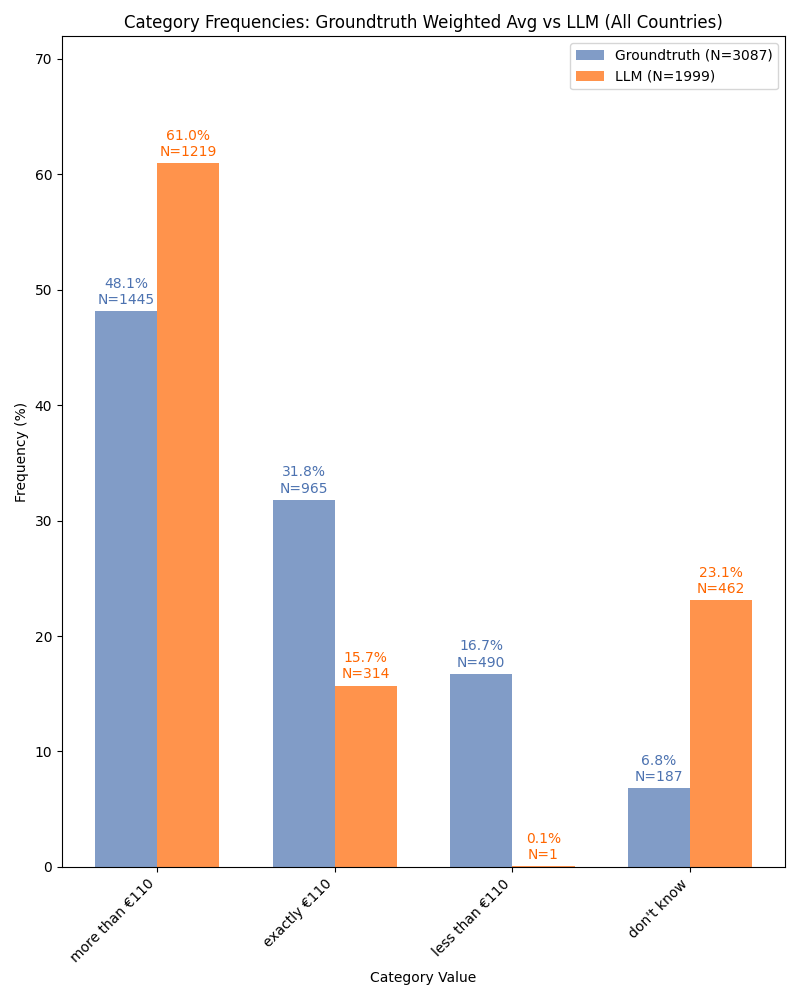}
        \caption*{(b) Survey-anchored agents.}
    \end{minipage}

    \caption{EUBAR-FK01 — \enquote{\textit{Imagine that someone puts €100 into a savings account with interest rate of 2\% per year. How much would there be in five years?}}. Correct answer: More than 110€. Answers from full agents.\newline
    Comparison of predicted response frequencies from demographics-only agents (Left) and survey-anchored agents (Right) against human responses. Both exhibit hyperaccuracy, but survey-anchored agents reduce this effect and produce more diverse answers.}
    \label{fig:EUInterest_figures}
\end{figure}

\newpage

\newpage

\subsection{Additional results from the replication of the regression analysis by Jacobs-Lawson and Hershey}
\label{app:additional_results_regression}

\paragraph{Internal consistency (Cronbach's alpha).}

\citet{jacobs2005influence} computed Cronbach's alpha to assess the internal consistency reliability of the multi-item scales measuring the study’s key constructs (acceptable threshold: $\alpha > 0.7$~\cite{saad1999testing}). 
Table~\ref{tab:cronbach_comparison} reveals an initially counterintuitive pattern: Demographics7 agents produce Cronbach's $\alpha$ values that closely match those reported by \citet{jacobs2005influence}---for instance, $\alpha = 0.71$ vs.\ $0.75$ (FTP), $0.91$ vs.\ $0.94$ (KFP), $0.81$ vs.\ $0.83$ (FRT), and $0.94$ vs.\ $0.93$ (RS)---whereas survey-anchored agents yield \emph{lower} $\alpha$ for FTP ($0.55$) and FRT ($0.60$). At face value, this could suggest that demographics-only conditioning suffices for psychometrically coherent simulation.

\begin{table*}[ht]
\centering
\renewcommand{\arraystretch}{1.4}
\resizebox{\textwidth}{!}{%
\begin{tabular}{|l|c|c|c|c|}
\hline
\textbf{Dimension} & \textbf{Jacobs-Lawson \& Hershey (2005)} N=270 & \textbf{Demographics7  Agents} N=267 & \textbf{Demographics3 Agents} N=174 & \textbf{Survey-Anchored Agents} N=266 \\
\hline
Future Time Perspective & 0.75 & 0.71 & 0.24 & 0.55 \\
Knowledge of Financial Planning for Retirement & 0.94 & 0.91 & 0.63 & 0.92 \\
Financial Risk Tolerance & 0.83 & 0.81 & 0.25 & 0.60 \\
Retirement Saving Indicator & 0.93 & 0.94 & 0.80 & 0.95 \\
\hline
\end{tabular}%
}
\caption{Internal consistency
reliability of the multi-item scales measuring the
study’s key constructs. Demographics7 (age, country, gender, marital status, education years, income, occupation) agents yield Cronbach's alpha values that align with those reported in Jacobs-Lawson \& Hershey (2005). In contrast, Demographics3 (age,country, gender) agents fail to exhibit the expected internal coherence. Survey-anchored agents reproduce acceptable values in Financial literacy and Retirement Savings scores.}
\label{tab:cronbach_comparison}
\end{table*}


\begin{table*}[ht]
    \centering
    \renewcommand{\arraystretch}{1.3}
    \resizebox{\textwidth}{!}{%
    \begin{tabular}{l|cccc|cccc|cccc}
    \hline
    & \multicolumn{4}{c|}{\textbf{Demographics7} ($N=267$--$270$)}
    & \multicolumn{4}{c|}{\textbf{Survey-Anchored} ($N=267$--$270$)}
    & \multicolumn{4}{c}{\textbf{Demographics3} ($N=224$--$262$)} \\
    \cmidrule{2-5} \cmidrule{6-9} \cmidrule{10-13}
    \textbf{Scale}
      & $\alpha$ & $\bar{r}$ & $\bar{\sigma}^2_{\text{item}}$ & $\sigma^2_{\text{scale}}$
      & $\alpha$ & $\bar{r}$ & $\bar{\sigma}^2_{\text{item}}$ & $\sigma^2_{\text{scale}}$
      & $\alpha$ & $\bar{r}$ & $\bar{\sigma}^2_{\text{item}}$ & $\sigma^2_{\text{scale}}$ \\
    \hline
    FTP ($k=6$) & 0.71 & 0.25 & 0.65 & 0.27 & 0.54 & 0.15 & 1.21 & 0.37 & 0.20 & 0.04 & 0.13 & 0.03 \\
    KFP ($k=6$) & 0.91 & 0.62 & 0.80 & 0.55 & 0.92 & 0.64 & 1.15 & 0.81 & 0.61 & 0.22 & 0.16 & 0.05 \\
    FRT ($k=5$) & 0.81 & 0.48 & 1.26 & 0.71 & 0.61 & 0.29 & 1.19 & 0.47 & 0.16 & 0.05 & 0.35 & 0.08 \\
    RS  ($k=5$) & 0.94 & 0.75 & 1.32 & 1.05 & 0.95 & 0.78 & 1.85 & 1.52 & 0.78 & 0.42 & 0.38 & 0.20 \\
    \hline
    \end{tabular}%
    }
    \caption{Cronbach's $\alpha$ decomposition across conditions. $\bar{r}$: mean inter-item correlation; $\bar{\sigma}^2_{\text{item}}$: average item-level variance; $\sigma^2_{\text{scale}}$: variance of agent-level mean scores; $k$: number of items. Demographics7 achieves high $\alpha$ for FTP and FRT through elevated $\bar{r}$ despite lower $\bar{\sigma}^2_{\text{item}}$ than survey-anchored agents, indicating stereotypical co-variation rather than genuine response heterogeneity. See Appendix~\ref{app:alpha_decomposition} for derivation.}
    \label{tab:alpha_decomposition}
\end{table*}
However, decomposing $\alpha$ into its constituent components reveals that this surface-level match reflects qualitatively different generative mechanisms. Cronbach's $\alpha$ is a joint function of the number of items, mean inter-item correlation ($\bar{r}$), and item-level variance \citep{cortina1993coefficient} (see Appendix~\ref{app:alpha_decomposition} for derivation). As shown in Table~\ref{tab:alpha_decomposition}, Demographics7 agents achieve high $\alpha$ through \emph{elevated inter-item correlations} paired with \emph{compressed item-level variance}: for FTP, $\bar{r} = 0.25$ with $\bar{\sigma}^2_{\text{item}} = 0.65$, compared to $\bar{r} = 0.15$ with $\bar{\sigma}^2_{\text{item}} = 1.21$ for survey-anchored agents. A similar pattern holds for FRT ($\bar{r} = 0.48$ vs.\ $0.29$; $\bar{\sigma}^2_{\text{item}} = 1.26$ vs.\ $1.19$). This indicates that Demographics7 items co-vary in lockstep across a narrower range---a signature of stereotypical response generation rather than genuine psychometric coherence.

\begin{figure*}[htbp]
    \centering
    \setlength{\tabcolsep}{0pt}
    \setlength{\abovecaptionskip}{2pt}
    \setlength{\belowcaptionskip}{2pt}

    \newcommand{\rowvspace}{\vspace{0.3em}}

    \begin{minipage}{0.85\textwidth}
        \centering
        \resizebox{0.95\textwidth}{!}{
            \begin{tikzpicture}
\begin{scope}[xshift=0cm]
\begin{axis}[
    width=7cm, height=6cm,
    xlabel={Centered Risk Tolerance Score},
    ylabel={Retirement Savings Score},
    title={{\textbf{High Future Time Perspective}}},
    title style={font=\bfseries, at={(0.5,1.02)}, anchor=south},
    xmin=-3, xmax=3, ymin=0, ymax=7,
    grid=major, grid style={gray!30, line width=0.5pt},
    legend pos=south east,
    legend style={font=\footnotesize, draw=black, fill=white, fill opacity=0.9},
    every axis plot/.append style={line width=1.5pt},
    tick label style={font=\footnotesize},
    label style={font=\footnotesize}
]
\addplot[black, solid, domain=-3:3] {4.5 + 0.16*x};
\addlegendentry{High Knowledge}
\addplot[black, dashed, domain=-3:3] {2.2 + 0.55*x};
\addlegendentry{Low Knowledge}

\node[font=\tiny, fill=white, draw=black, rounded corners=1pt, inner sep=2pt] 
    at (axis cs:-1.5,5.1) {$\beta = .16^*$};
\node[font=\tiny, fill=white, draw=black, rounded corners=1pt, inner sep=2pt] 
    at (axis cs:1.5,3.9) {$\beta = .55^*$};
\end{axis}
\end{scope}

\begin{scope}[xshift=8.5cm]
\begin{axis}[
    width=7cm, height=6cm,
    xlabel={Centered Risk Tolerance Score},
    ylabel={Retirement Savings Score},
    title={{\textbf{Low Future Time Perspective}}},
    title style={font=\bfseries, at={(0.5,1.02)}, anchor=south},
    xmin=-3, xmax=3, ymin=0, ymax=7,
    grid=major, grid style={gray!30, line width=0.5pt},
    legend pos=south east,
    legend style={font=\footnotesize, draw=black, fill=white, fill opacity=0.9},
    every axis plot/.append style={line width=1.5pt},
    tick label style={font=\footnotesize},
    label style={font=\footnotesize}
]
\addplot[black, solid, domain=-3:3] {3.8 + 0.20*x};
\addlegendentry{High Knowledge}
\addplot[black, dashed, domain=-3:3] {2.6 + 0.05*x};
\addlegendentry{Low Knowledge}

\node[font=\tiny, fill=white, draw=black, rounded corners=1pt, inner sep=2pt] 
    at (axis cs:1.5,4.7) {$\beta = .20^{\dagger}$};
\node[font=\tiny, fill=white, draw=black, rounded corners=1pt, inner sep=2pt] 
    at (axis cs:0,2.4) {$\beta = .05$};
\end{axis}
\end{scope}
\end{tikzpicture}
\label{fig:simple_slopes}
        }
        \caption*{(a) Original study}
        \label{fig:JLH_original_study}
    \end{minipage}

    \rowvspace
    \begin{minipage}{0.85\textwidth}
        \centering
        \includegraphics[width=0.95\textwidth]{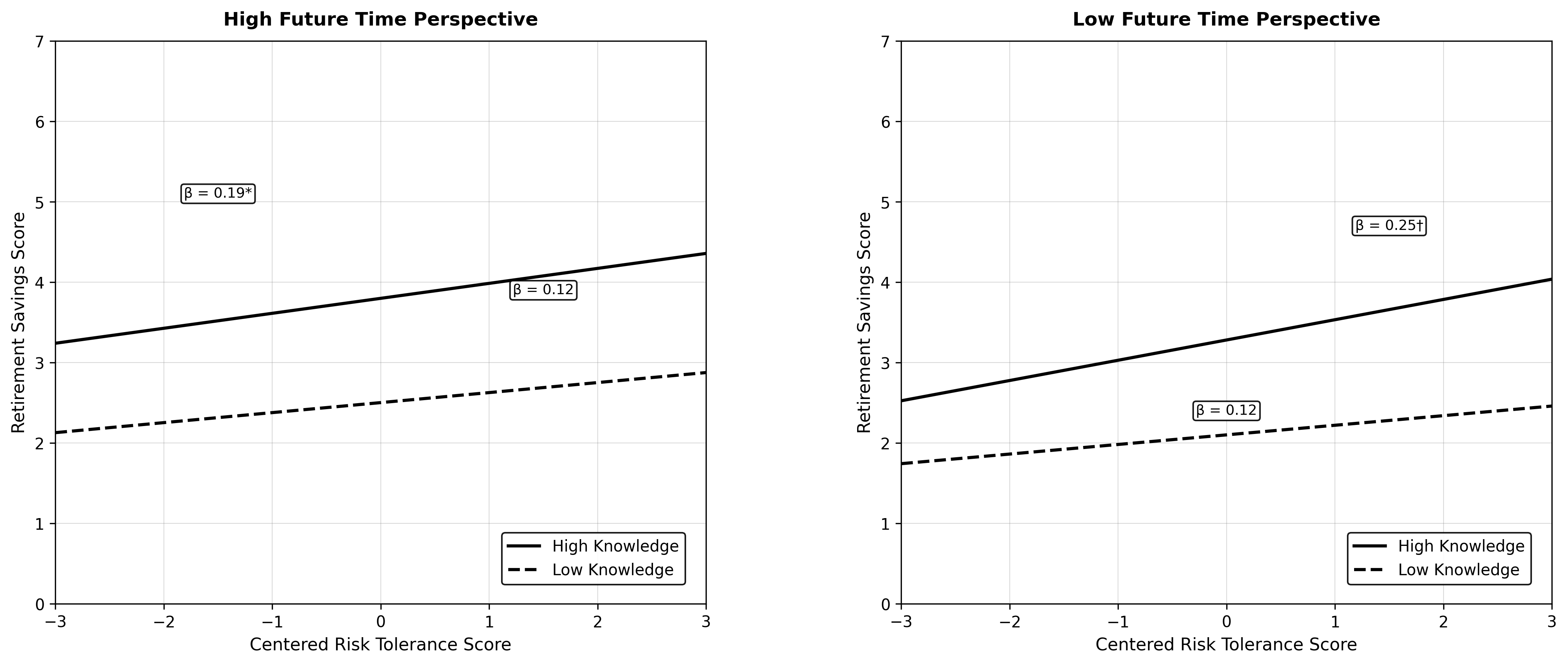}
        \caption*{(c) Demographics7-only agents}
        \label{fig:Reg_analysis_full}
    \end{minipage}

    \rowvspace

    \begin{minipage}{0.85\textwidth}
        \centering
        \includegraphics[width=0.95\textwidth]{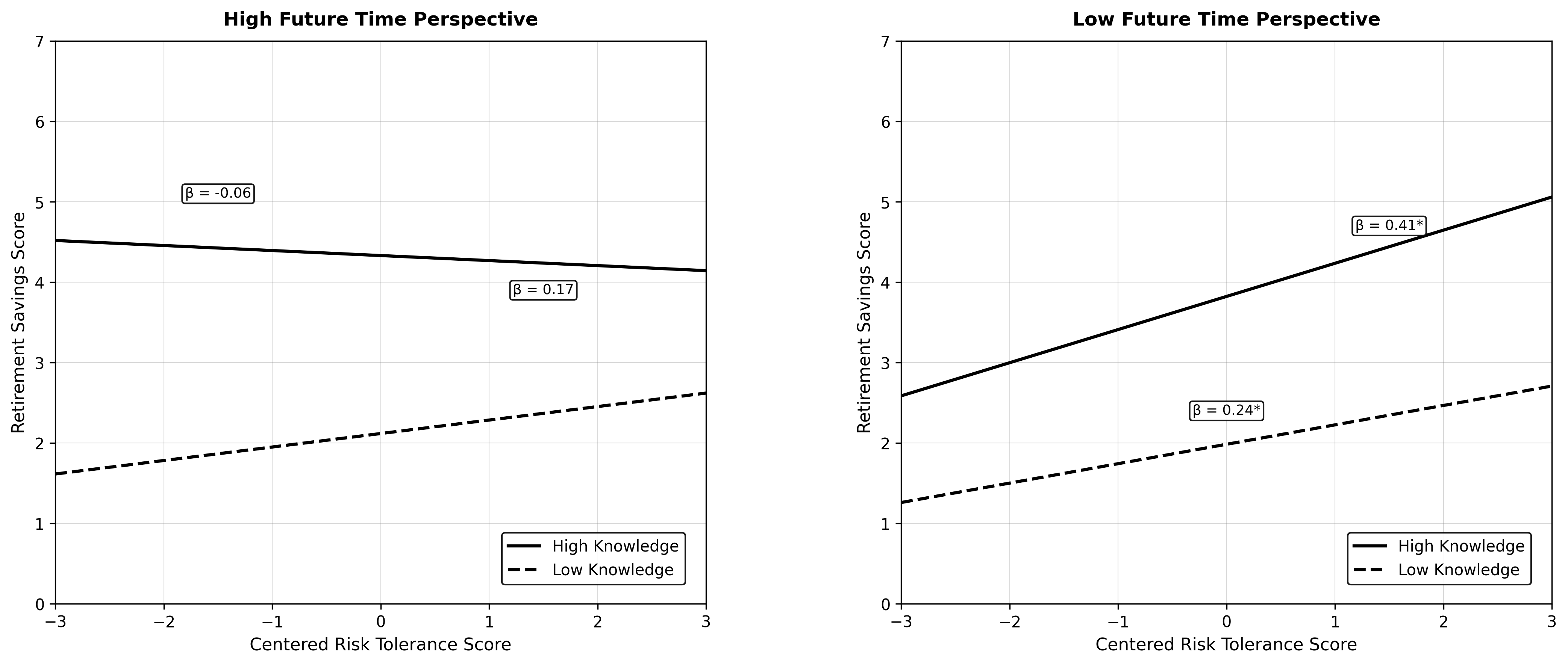}
        \caption*{(d) Survey-anchored agents}
        \label{fig:gss_v2_reg_analysis}
    \end{minipage}

    \caption{Demographics-only agents show non-significant effect.
    Significant positive relationship between Risk Tolerance and Savings behavior for Survey Anchored with Low Future Time Perspective and High Knowledge (solid line) is observed as the original study but not with the same magnitude.}
    \label{fig:simple_slope_analysis}
\end{figure*}

\paragraph{Simple slopes analysis.}
Following \citet{jacobs2005influence}, we decomposed the three-way interaction by regressing saving on risk tolerance at combinations of high and low levels of FTP and financial knowledge, using the methodology of \citet{aiken1991multiple,cohen2013applied}. Figure~\ref{fig:simple_slope_analysis} presents the simple slopes for each condition alongside the original study.

In the original study (Figure~\ref{fig:simple_slope_analysis}a), the strongest relationship between risk tolerance and saving emerges for individuals high in FTP and low in knowledge ($\beta = 0.55$, $p < .05$), followed by those high in both FTP and knowledge ($\beta = 0.16$, $p < .05$). Among low-FTP individuals, the relationship is marginally significant for those with high knowledge ($\beta = 0.20$, $p < .10$) and non-significant for those with low knowledge ($\beta = 0.05$). This pattern suggests that a strong future orientation amplifies the role of risk tolerance in driving saving behavior, particularly when financial knowledge is limited.

Demo7 agents (Figure~\ref{fig:simple_slope_analysis}b) show the following patterns: all slopes are positive, and one reaches significance at the High FTP / High Knowledge combination ($\beta = 0.19$, $p < .05$), with a marginal effect at Low FTP / High Knowledge ($\beta = 0.25$, $p < .10$). While the direction of effects is consistent with the original, the specific pattern diverges: the original found the strongest slope for High FTP / Low Knowledge ($\beta = 0.55$), whereas Demo7 agents show relatively flat, similar-magnitude slopes across conditions ($\beta = 0.12$ to $0.25$), failing to capture the differential moderation that is the hallmark of the three-way interaction.

Survey-anchored agents (Figure~\ref{fig:simple_slope_analysis}c) produce the most differentiated pattern. For individuals with low FTP, risk tolerance significantly predicts saving at both high ($\beta = 0.41$, $p < .05$) and low ($\beta = 0.24$, $p < .05$) knowledge levels. For those with high FTP, effects are weaker and non-significant: a slight negative slope for high knowledge ($\beta = -0.06$) and a positive but non-significant slope for low knowledge ($\beta = 0.17$). This pattern partially inverts the original finding, where the strongest effects appeared under high FTP rather than low FTP.

\subsection{Cronbach's Alpha Decomposition}
\label{app:alpha_decomposition}

Cronbach's $\alpha$ \citep{cronbach1951coefficient} is the most widely used estimator of internal consistency reliability. For a scale with $k$ items, it is defined as:
\begin{equation}
\alpha = \frac{k}{k-1} \left( 1 - \frac{\sum_{i=1}^{k} \sigma^2_i}{\sigma^2_X} \right)
\label{eq:alpha_raw}
\end{equation}
where $\sigma^2_i$ is the variance of item $i$ and $\sigma^2_X = \sum_{i} \sigma^2_i + 2\sum_{i<j} \sigma_{ij}$ is the total scale variance (i.e., the variance of the sum of all items).

\paragraph{Standardized formulation.} An algebraically equivalent expression rewrites $\alpha$ in terms of the mean inter-item correlation $\bar{r}$ \citep{cortina1993coefficient}:
\begin{equation}
\alpha_{\text{std}} = \frac{k \, \bar{r}}{1 + (k-1)\,\bar{r}}
\label{eq:alpha_standardized}
\end{equation}
where $\bar{r} = \frac{2}{k(k-1)} \sum_{i<j} r_{ij}$ is the average Pearson correlation across all $\binom{k}{2}$ item pairs. This formulation makes explicit that, holding the number of items constant, $\alpha$ is entirely determined by how strongly items correlate with one another on average.

\paragraph{Decomposition rationale.} Equations~\ref{eq:alpha_raw} and~\ref{eq:alpha_standardized} reveal that two datasets can yield the same $\alpha$ through different mechanisms \citep{sijtsma2009use}. A high $\alpha$ can arise from (a)~high inter-item correlations $\bar{r}$ with modest item variances, or (b)~moderate $\bar{r}$ with large item variances. In the context of LLM-based simulation, this distinction is diagnostic: if a simulated population achieves high $\alpha$ primarily through elevated $\bar{r}$ while exhibiting compressed $\bar{\sigma}^2_{\text{item}}$ relative to a reference population, this indicates that the model generates stereotypically coherent profiles---items move in lockstep---without reproducing the full range of individual differences observed in human respondents.

\paragraph{Reported quantities.} For each scale and condition, we report four quantities (Table~\ref{tab:alpha_decomposition}):

\begin{enumerate}
    \item \textbf{Cronbach's $\alpha$}, computed via Equation~\ref{eq:alpha_raw}.
    \item \textbf{Mean inter-item correlation} $\bar{r}$, computed from the Pearson correlation matrix of items.
    \item \textbf{Average item variance} $\bar{\sigma}^2_{\text{item}} = \frac{1}{k}\sum_{i=1}^{k} \sigma^2_i$, capturing the dispersion of responses within individual items.
    \item \textbf{Scale variance} $\sigma^2_{\text{scale}}$, defined as the variance of agent-level mean scores, capturing between-agent differentiation at the construct level.
\end{enumerate}

Comparing $\bar{r}$ and $\bar{\sigma}^2_{\text{item}}$ across conditions disambiguates whether matched $\alpha$ values reflect comparable generative processes. A condition with high $\bar{r}$ but low $\bar{\sigma}^2_{\text{item}}$ achieves internal consistency through rigid co-variation over a narrow response range. A condition with moderate $\bar{r}$ but high $\bar{\sigma}^2_{\text{item}}$ achieves consistency while preserving the response heterogeneity observed in human samples.

\subsection{Results to questions beyond retirement attitudes}

\begin{figure}[t]
    \centering
    \begin{minipage}[t]{0.32\textwidth}
        \centering
        \includegraphics[width=\textwidth]{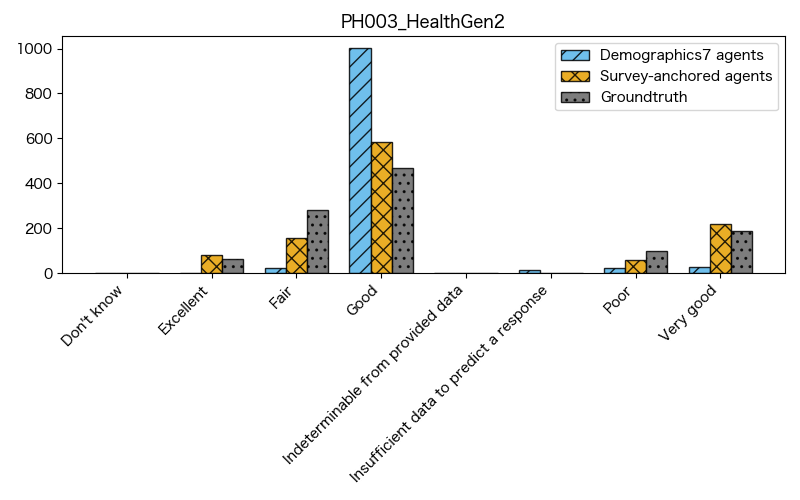}
        \small\textit{PH003\_HealthGen2: Would you say your health is?}
    \end{minipage}\hfill
    \begin{minipage}[t]{0.32\textwidth}
        \centering
        \includegraphics[width=\textwidth]{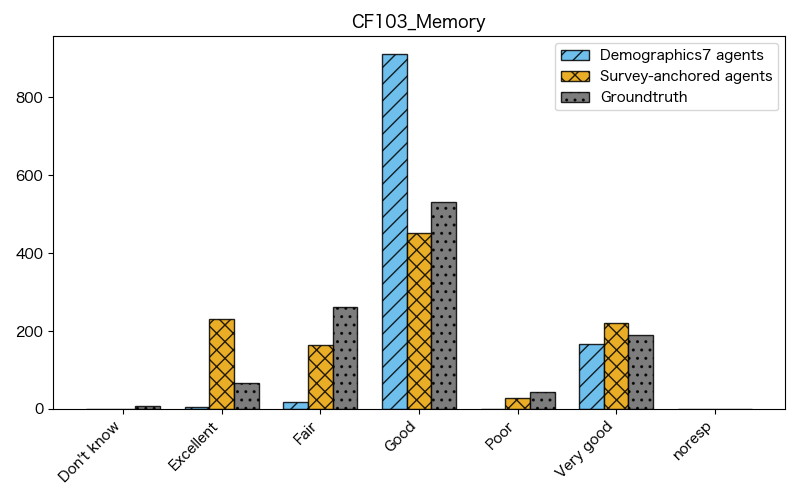}
        \small\textit{CF103\_Memory: How would you rate your memory at the present time?}
    \end{minipage}\hfill
    \begin{minipage}[t]{0.32\textwidth}
        \centering
        \includegraphics[width=\textwidth]{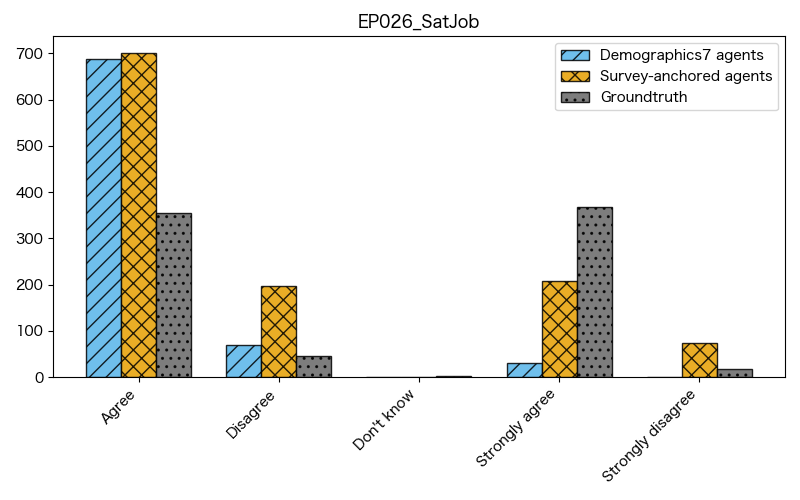}
        \small\textit{EP026\_SatJob: All things considered, I am satisfied with my job.}
    \end{minipage}
    
    \vspace{0.5cm}
    
    \begin{minipage}[t]{0.32\textwidth}
        \centering
        \includegraphics[width=\textwidth]{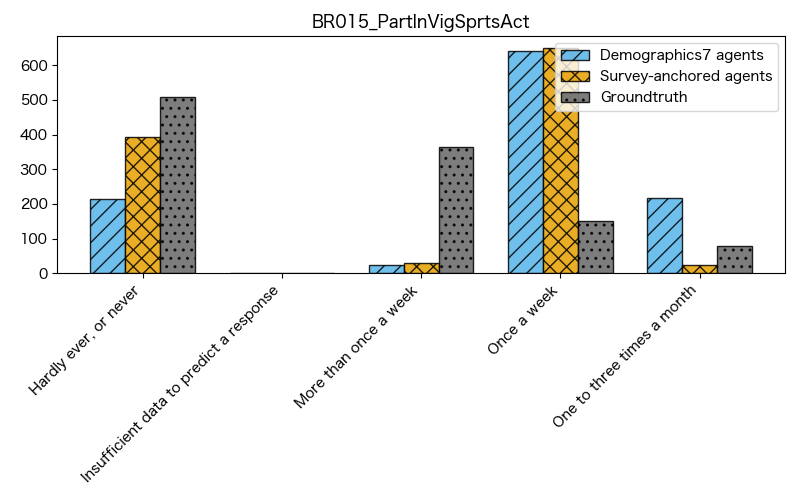}
        \small\textit{BR015\_PartInVigSprtsAct: How often do you engage in vigorous physical activity?}
    \end{minipage}\hfill
    \begin{minipage}[t]{0.32\textwidth}
        \centering
        \includegraphics[width=\textwidth]{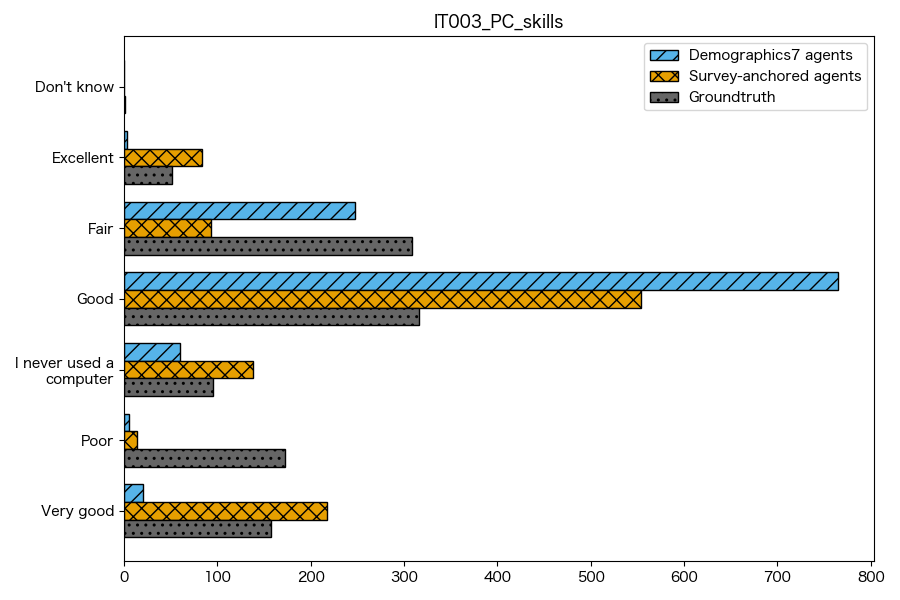}
        \small\textit{IT003\_PC\_skills: How would you rate your computer skills?}
    \end{minipage}\hfill
    \begin{minipage}[t]{0.32\textwidth}
        \centering
        \includegraphics[width=\textwidth]{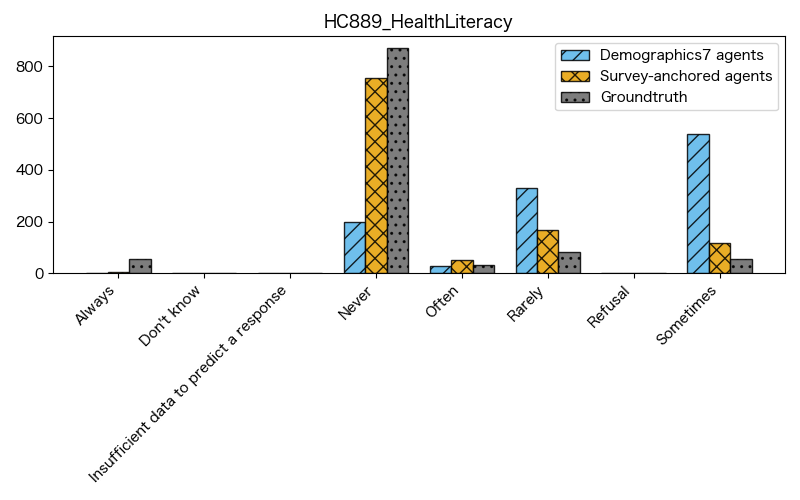}
        \small\textit{HC889\_HealthLiteracy: How often do you need help reading medical instructions?}
    \end{minipage}
    
    \vspace{0.5cm}
    
    \begin{minipage}[t]{0.32\textwidth}
        \centering
        \includegraphics[width=\textwidth]{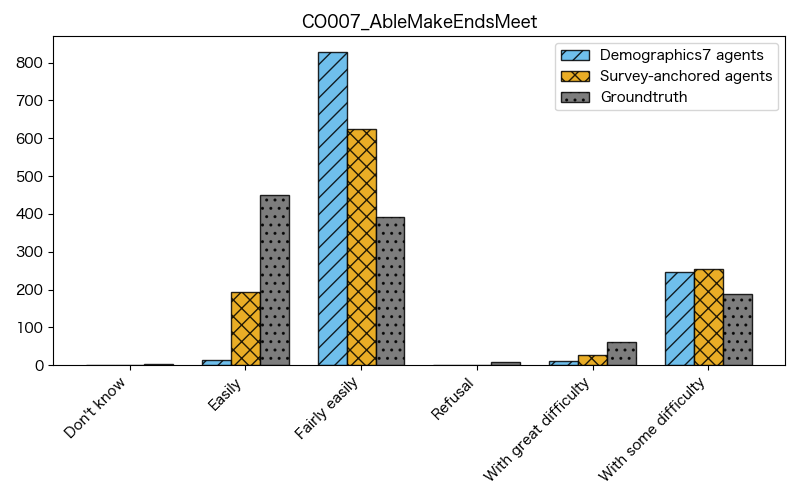}
        \small\textit{CO007\_AbleMakeEndsMeet: Is your household able to make ends meet?}
    \end{minipage}\hfill
    \begin{minipage}[t]{0.32\textwidth}
        \centering
        \includegraphics[width=\textwidth]{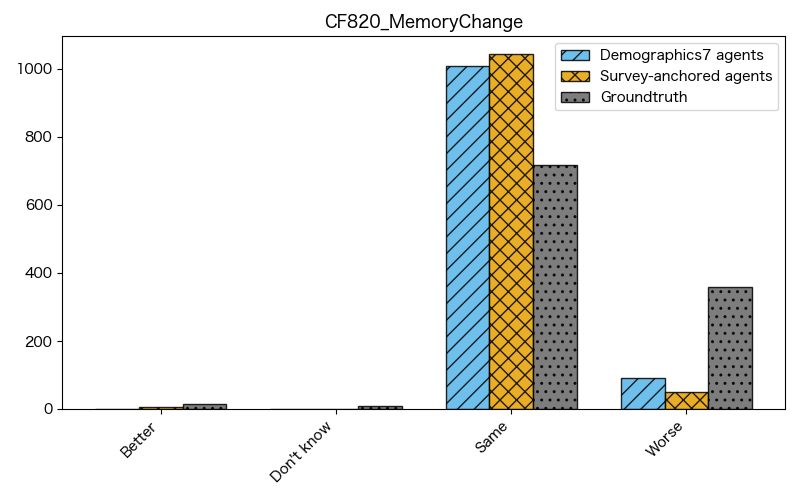}
        \small\textit{CF820\_MemoryChange: Compared to last interview, is your memory better, same, or worse?}
    \end{minipage}\hfill
    \begin{minipage}[t]{0.32\textwidth}
        \centering
        \includegraphics[width=\textwidth]{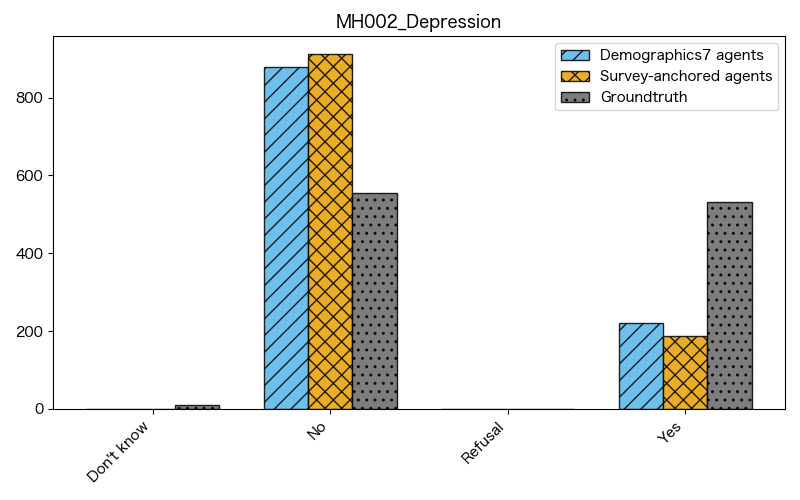}
        \small\textit{MH002\_Depression: In the last month, have you been sad or depressed?}
    \end{minipage}
    
    \vspace{0.5cm}
    
    \begin{minipage}[t]{0.32\textwidth}
        \centering
        \includegraphics[width=\textwidth]{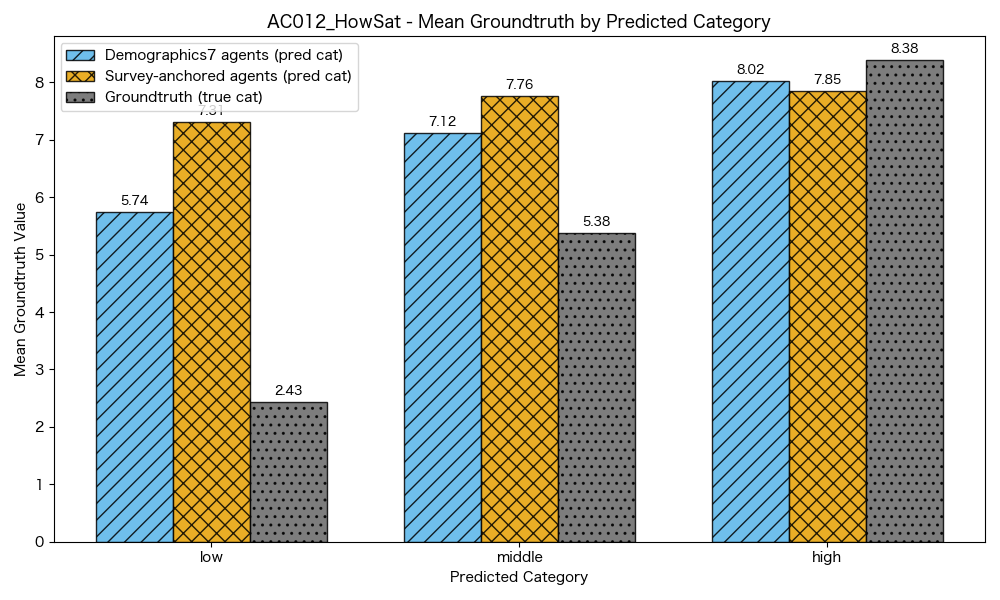}
        \small\textit{AC012\_HowSat: On a scale from 0 to 10, how satisfied are you with your life? (mean by category)}
    \end{minipage}\hfill
    \begin{minipage}[t]{0.32\textwidth}
        \centering
        \includegraphics[width=\textwidth]{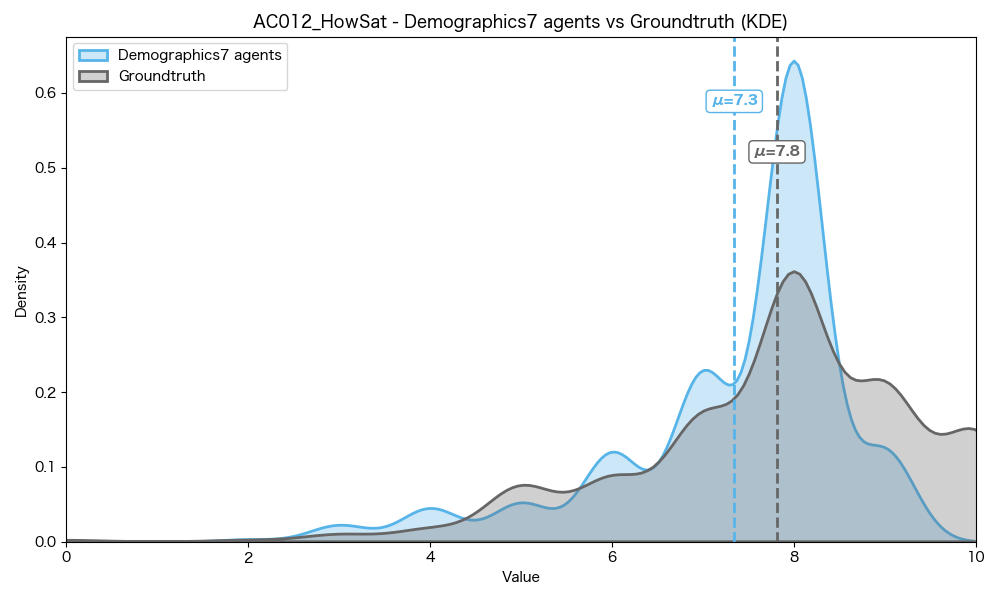}
        \small\textit{AC012\_HowSat: On a scale from 0 to 10, how satisfied are you with your life? (demographics vs GT)}
    \end{minipage}\hfill
    \begin{minipage}[t]{0.32\textwidth}
        \centering
        \includegraphics[width=\textwidth]{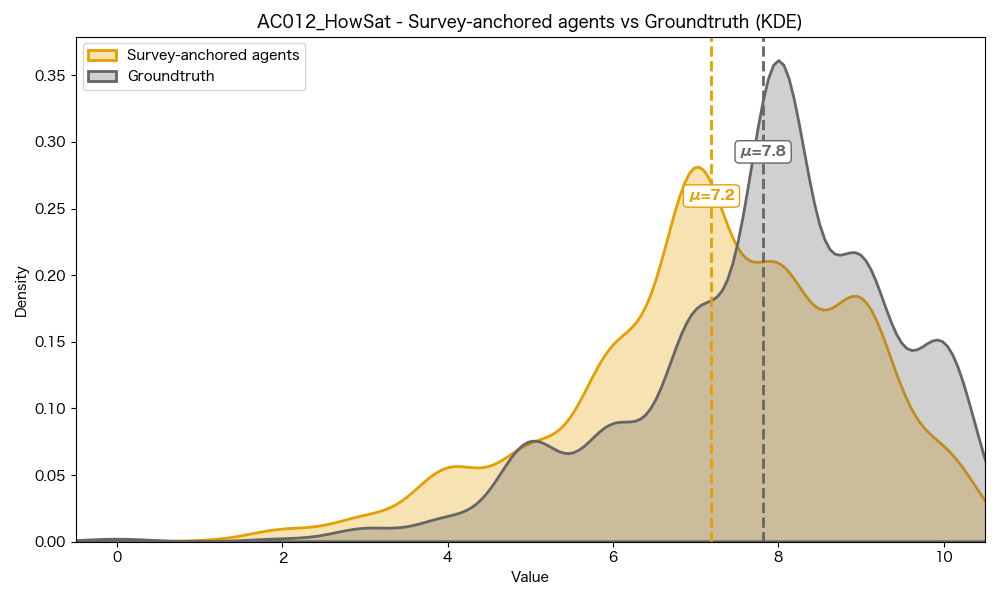}
        \small\textit{AC012\_HowSat: On a scale from 0 to 10, how satisfied are you with your life? (survey vs GT)}
    \end{minipage}
    
    \caption{Comparison of agent predictions vs. groundtruth (GT) for variables beyond retirement attitudes. Demographics-only agents (blue) and survey-anchored agents (orange) are compared against real participant responses (grey, grountruth). Rows 1-3 show categorical variables, row 4 shows AC012 numerical variable life satisfaction with three different visualizations.}
    \label{fig:phase2_all}
\end{figure}

\end{appendices}
\clearpage

\bibliography{sn-bibliography}

\end{document}